\documentstyle[12pt,french,fancyheadings]{book}

\catcode`\@=12
\newcommand{\be}{\begin{equation}}
\newcommand{\ee}{\end{equation}}
\newcommand{\bn}{\begin{eqnarray}}
\newcommand{\en}{\end{eqnarray}}
\newcommand{\bnn}{\begin{eqnarray*}}
\newcommand{\enn}{\end{eqnarray*}}
\newcommand{\n}{\noindent}

\begin{document}
\thispagestyle{empty}
\mbox{}\vskip -3.7cm
\centerline{ \Large \sc UNIVERSIT\'E de NICE-SOPHIA
ANTIPOLIS}
\medskip \centerline{\Large \sc U.F.R Facult\'e des Sciences}
\vskip 2 cm
\centerline{\bf \Huge \sc T\,\,\,\,\,H\,\,\,\,\,\`E\,\,\,\,\,S\,\,\,\,\,E}
\medskip \centerline{\bf \LARGE en Sciences Physiques}
\medskip \centerline{\bf \Large (Physique Th\'eorique)}
\vskip 2.3 cm
\centerline{\bf \LARGE --------------------------------------------------}
\centerline{\bf \LARGE SOLUTIONS STATIONNAIRES}
\medskip \centerline{\bf \LARGE EN TH\'EORIE DE KALUZA-KLEIN}
\centerline{\bf \LARGE --------------------------------------------------}
\vskip 2.3 cm
\centerline{\large Pr\'esent\'ee par}
\vskip 0.7 cm
\centerline{\bf \large \sc Mustapha AZREG-A\"{I}NOU}
\vskip 0.7 cm
\centerline{\large Pour obtenir le Grade de Docteur Es-Sciences}
\vskip 0.7 cm
\centerline{\large Soutenue le 07 Novembre 1995 \`a l'Institut Non
  Lin\'eaire de Nice (INLN)}
\vskip 0.7 cm
\centerline{\large devant la Commission d'examen}
\vskip 1 cm

\noindent{\bf \small F. ROCCA}\hspace{27.9mm}{\it Pr\'esident} (Universit\'e de
Nice)\\
{\bf \small R. KERNER}\hspace{24.1mm}{\it Rapporteur} (Universit\'e de Pierre
et Marie Curie)\\
{\bf \small J. MADORE}\hspace{23.8mm}{\it Rapporteur} (Universit\'e de Paris
Sud)\\
{\bf \small G. CL\'EMENT}\hspace{21.2mm}{\it Examinateur} (Universit\'e
de Pierre et Marie Curie)\\
{\bf \small C. DUVAL}\hspace{27.9mm}{\it Examinateur} (Universit\'e
d'Aix-Marseille II)\\
{\bf \small J. P. PROVOST}\hspace{17.7mm}{\it Examinateur} (Universit\'e de
Nice)

\newpage
\thispagestyle{empty}
\centerline{\bf \LARGE R\,e\,m\,e\,r\,c\,i\,e\,m\,e\,n\,t\,s}
\vskip 2.5cm
La r\'ealisation de cette th\`ese, propos\'ee et dirig\'ee par
Monsieur G\'erard CL\'EMENT, Charg\'e de Recherche (CNRS) au Laboratoire
de Gravitation et Cosmologie Relativistes (GCR), Universit\'e Pierre
et Marie Curie (UPMC), Paris, a trouv\'e sa fin gr\^{a}ce \`a ses conseils,
suggestions et critiques dont je tiens \`a le remercier vivement.\\

J'exprime ma reconnaissance \`a Monsieur Fran\c{c}ois ROCCA,
Professeur, Institut Non Lin\'eaire de Nice (INLN), Universit\'e de Nice-Sophia
Antipolis (UNSA), pour l'honneur qu'il m'a fait en acceptant la pr\'esidence
du jury de th\`ese.\\

Je tiens aussi \`a remercier Messieurs Christian DUVAL, Professeur,
Centre de Physique Th\'eorique, Universit\'e d'Aix-Marseille II, Richard
KERNER, Professeur (GCR, UPMC), John MADORE, Directeur de Recherche
(CNRS), Laboratoire de Physique Th\'eorique et Hautes \'Energies,
Universit\'e de Paris Sud, Orsay, et Jean-Pierre PROVOST, Professeur
(INLN, UNSA), d'avoir bien voulu \mbox{accepter} de participer au jury
de cette th\`ese.\\

Je remercie \'egalement Monsieur G\'erard IOOSS, Professeur (UNSA) et
\mbox{Directeur} de l'INLN o\`u ce travail a \'et\'e r\'ealis\'e.

\newpage
\thispagestyle{empty}
\mbox{}\vskip 3cm
\mbox{}\hspace{8.5cm}\`A ma m\`ere\\
\mbox{}\hspace{11.0cm}{\bf \em Guemra}

\newpage

\pagenumbering{Roman}

\tableofcontents

\newpage
\pagenumbering{arabic}

\chapter*{Introduction}
\markboth{Introduction}{}
\addcontentsline{toc}{chapter}{Introduction}
Un premier pas vers l'unification de la gravitation et de
l'\'electromagn\'etisme \'etait purement {\em g\'eom\'etrique}. Les
id\'ees de Kaluza \cite{Kaluza} et de Klein \cite{Klein} ont conduit \`a
\'edifier, sur le mod\`ele g\'eom\'etrique de la th\'eorie de la
relativit\'e g\'en\'erale d'Einstein \cite{Weinberg}, un mod\`ele
semblable (\`a cinq dimensions) o\`u les lois de la gravitation et de
l'\'electromagn\'etisme fusionnent naturellement. Autrement dit, la {\em
courbure en un point de l'espace-temps \`a cinq dimensions y engendre
ce qu'on per\c{c}oit \^{e}tre les forces gravitationnelle et
\'electromagn\'etique}. D'autres th\'eories ult\'erieures d'unification
ont emprunt\'e l'id\'ee d'un {\em espace-temps plus large}
(\`a plus de quatre dimensions).

Ce qu'on appelle commun\'ement th\'eorie de Kaluza-Klein est la version
des id\'ees de Kaluza et de Klein d\'evelopp\'ees et clarifi\'ees par
Jordan et Thiry \cite{Jordan}: c'est une relativit\'e g\'en\'erale
d'Einstein ordinaire \`a cinq dimensions, {\em la dimension suppl\'ementaire
\'etant compactifi\'ee en un cercle} (section \ref{sec-Klein});
c'est l\`a l'id\'ee de base. En effet, si l'on suppose de plus
l'existence d'une {\em isom\'etrie} d\'efinie par un vecteur de
Killing parall\`ele \`a la cinqui\`eme direction, certains rapports
entre les composantes suppl\'ementaires de la m\'etrique se
transforment, lors des translations parall\`eles au vecteur de Killing
et ind\'ependantes de la coordonn\'ee suppl\'ementaire, comme le font
les potentiels \'electromagn\'etiques lors d'une transformation de
jauge et seront donc identifi\'es avec ces derniers; seule une
composante suppl\'ementaire de la m\'etrique reste invariante et
correspond au champ scalaire de la th\'eorie. Les plus importants
r\'esultats de la th\'eorie de Kaluza-Klein sont les suivants:\\

${\bullet}$ La charge \'electrique (4-dimensionnelle) est proportionnelle
\`a l'impulsion suivant la cinqui\`eme direction. Une particule neutre
est donc au repos sur le cercle de Klein \cite{Klein}.\\

${\bullet}$ La quantification de la charge est justifi\'ee comme \'etant
celle de la cinqui\`eme composante du moment d'impulsion \cite{Klein}.\\

${\bullet}$ La constante de structure fine acquiert une
interpr\'etation g\'eom\'etrique et peut \^{e}tre exprim\'ee en fonction
du rapport de la longueur de Planck et du rayon de Klein (voir
\'equation (\ref{eq fine})).\\

${\bullet}$ Contrairement \`a la relativit\'e g\'en\'erale d'Einstein,
celle de Kaluza-Klein admet des solutions solitons \cite{Chodos,Sorkin,Gross}.
\\

${\bullet}$ L'unification peut inclure d'autres forces si on continue
\`a augmenter la dimension de l'espace-temps \cite{KK, KS}.\\

La th\'eorie de Kaluza-Klein, sans terme de source, admet diverses
solutions {\em stationnaires}
\cite{Leutwyler,Dobiasch,Chodos,Sorkin,Gross,Clement131,Xi,Clement137,
Belinsky,Fer} \`a sym\'etrie {\em sph\'erique, axiale et m\^{e}me
sans sym\'etrie}, parmi
lesquelles des solutions {\em r\'eguli\`eres et d'\'energie finie}
\cite{Dobiasch,Chodos,Sorkin,Gross,Clement131}. Ces derni\`eres
peuvent \^{e}tre interpr\'et\'ees comme des mod\`eles de particules
(\'etendues) d'une part \`a cause de leurs propri\'et\'es physiques
\cite{Clement137, Clement491} (citons entre autres le spin et le
moment dipolaire magn\'etique avec {\em un rapport gyromagn\'etique
relativiste correct \'egal \`a $g=2$} pour les solutions wormholes)
et d'autre part parce qu'elles manifestent un comportement assez
semblable à celui des particules \'el\'ementaires quand elles sont
trait\'ees comme cibles dans des probl\`emes de diffusion
\cite{theseA, AAGC1}.

Les plus surprenantes des solutions r\'eguli\`eres sont les
monop\^{o}les de Kaluza-Klein d\'ecouverts par Sorkin \cite{Sorkin}
puis ind\'ependamment par Gross et Perry \cite{Gross}. Ces derniers
ont construit explicitement le monop\^{o}le magn\'etique, puis le
dip\^{o}le magn\'etique (sans source) par
combinaison d'un monop\^{o}le et d'un anti-monop\^{o}le, et ont
discut\'e leurs propri\'et\'es physiques et topologiques sans
toutefois approfondir l'\'etude des g\'eod\'esiques de leurs
g\'eom\'etries; la possibilit\'e de construction de solutions
multi-p\^{o}les \'etait aussi mentionn\'ee.

Dans \cite{theseA, AAGC1} on a consid\'er\'e la g\'eom\'etrie du wormhole
(charg\'e sans masse) de Kaluza-Klein, et pr\'esent\'e une \'etude
d\'etaill\'ee des
g\'eod\'esiques de cette solution; elle nous
a permis d'aborder ensuite l'\'etude de la diffusion {\em classique}
des particules
\mbox{\'el\'ementaires} (neutres ou charg\'ees) \`a laquelle a \'et\'e
consacr\'e le chapitre 6 de \cite{theseA}. Dans l'hypoth\`ese
d'une diffusion sans recul, on a
montr\'e que les particules neutres sont diffus\'ees par le wormhole, et
obtenu des sections efficaces de diffusion non \mbox{rigouresement} nulles,
mais de l'ordre de $6\times 10^{-68}\,\mbox{cm}^{2}$; cet effet est donc
{\em inobservable}. Pour les particules charg\'ees, la
section efficace de diffusion ne diff\`ere essentiellement, dans le
cas non-relativiste, de celle de Rutherford que par l'effet de {\em gloire}
observ\'e vers l'arri\`ere. Un effet analogue est observ\'e par application des
lois de la relativit\'e restreinte au mouvement ``classique"
des particules charg\'ees
\'el\'ementaires, et introduit des {\em corrections} \`a la
formule de Rutherford \cite{theseA}. Dans une approche
semi-classique \cite{AAGC1}, on montre que ces
corrections ne sont, pour les particules sans spin,
qu'un artefact de l'approximation classique.

La diffusion par des solutions non r\'eguli\`eres a \'et\'e aussi
prise en compte. Dans \cite{Karim} on a consid\'er\'e les dyons de Kaluza-Klein
\cite{Xi} de charges \'electrique et magn\'etique {\em \'egales}, o\`u une
\'etude semblable a \'et\'e men\'ee; nous nous contenterons de citer
les r\'esultats de cette \'etude concernant la diffusion. Les sections
efficaces de
diffusion des photons, {\em proportionnelles} au carr\'e de la longueur
caract\'eristique du dyon, ont \'et\'e consid\'er\'ees comme \'etant
totalement {\em n\'egligeables}. Pour des valeurs du param\`etre d'impact
relativement grandes par rapport \`a la longueur caract\'eristique du
dyon, la section efficace de diffusion vers {\em l'avant} des particules
charg\'ees est du {\em type de Rutherford}\,; c'est la somme de deux
contributions dues \`a la diffusion par deux monop\^{o}les \'electrique
et magn\'etique.

Mais l'interpr\'etation --des solutions r\'eguli\`eres ou solitons-- en
termes de mod\`eles de particules \'etendues ne sera compl\`ete que si
{\em la condition de stabilit\'e est remplie}. Mentionnons, que si
les monop\^{o}le et dip\^{o}le de
Kaluza-Klein sont stables pour des raisons topologiques
\cite{Gross,Belinskii},
A. Tomimatsu \cite{Tom} fut le premier et le seul, \`a notre
connaissance,
qui s'est int\'eress\'e \`a l'\'etude de la stabilit\'e des solutions
statiques de Kaluza-Klein. En \'etudiant (voir section 2.4)
la stabilit\'e de la classe
des solutions statiques pour lesquelles le 5-tenseur m\'etrique est
diagonal, il a conclu \`a la stabilit\'e de la {\em seule solution de
Schwarzschild} contre les excitations monopolaires; les autres solutions
\'etant instables. Dans cette th\`ese nous reprenons en d\'etail, dans
le cas g\'en\'eral o\`u la 5-m\'etrique n'est pas diagonale, l'\'etude
de la stabilit\'e des solutions statiques de Kaluza-Klein, avec des
m\'ethodes l\'eg\`erement diff\'erentes de celles de Tomimatsu. Nous
arriverons \`a la conclusion qu'un certain nombre de solutions non
diagonales sont stables \cite{AAGC2}, ainsi que certaines solutions
diagonales non \'etudi\'ees par Tomimatsu.

Les solutions stationnaires de Kaluza-Klein les plus \'etudi\'ees dans
la litt\'erature sont les solutions \`a sym\'etrie
sph\'erique. L'\'etude des solutions \`a sym\'etrie cylindrique a
\'et\'e \'ebauch\'ee dans \cite{Fer}, mais une \'etude syst\'ematique
reste \`a faire. Ce cas de la sym\'etrie cylindrique est pourtant
physiquement int\'eressant, de telles solutions peuvent
s'interpr\'eter comme des mod\`eles de cordes cosmiques, analogues aux
mod\`eles de cordes cosmiques neutres ou supraconductrices dans un
espace-temps \`a 4 dimensions \'etudi\'es durant la derni\`ere
d\'ecennie par de nombreux auteurs \cite{Vilenkin,Hiscock,Gott,Linet}
et \cite{Witten}. Signalons d'autre part que la th\'eorie de
Kaluza-Klein peut \^{e}tre g\'en\'eralis\'ee par l'adjonction du terme
de Gauss-Bonnet \cite{Madore,Zwiebach, Zumino,WheelerB273,Wiltshire}
\`a l'action d'Einstein-Hilbert \`a 5 dimensions. Seules les solutions
\`a sym\'etrie sph\'erique de cette th\'eorie g\'en\'eralis\'ee ont
\'et\'e \'etudi\'ees jusqu'ici. L'autre grand th\`eme de cette th\`ese
sera donc l'\'etude des solutions \`a sym\'etrie cylindrique de la
th\'eorie de Kaluza-Klein et de son extension par Gauss-Bonnet, avec
application \`a un mod\`ele de corde cosmique supraconductrice.

Le contenu du pr\'esent m\'emoire est r\'eparti sur 3 chapitres. Dans
les 4 premi\`eres sections du chapitre 1 nous introduisons la
th\'eorie de Kaluza-Klein sous sa derni\`ere version due \`a Jordan et
Thiry, et dans la section 1.5 nous g\'en\'eralisons l'action
d'Einstein-Hilbert pour tenir compte du terme de Gauss-Bonnet,
quadratique par rapport au tenseur de Riemann, et qui conduit aussi,
apr\`es d\'erivation, \`a des \'equations du second ordre. Nous avons
consacr\'e la derni\`ere section du chapitre 1 \`a l'introduction de
la d\'ecomposition $n+p$ de la $d$-m\'etrique ($d=n+p$),
\`a l'aide de laquelle les composantes du tenseur de Riemann sont
exprim\'ees en fonction d'\'el\'ements des $n$-m\'etrique et $p$-m\'etrique.

Les deux chapitres suivants sont ind\'ependants. Dans le chapitre 2,
nous commen\c{c}ons par donner et classer les solutions 2-statiques \`a
sym\'etrie sph\'erique de la th\'eorie de Kaluza-Klein. Puis nous
\'ecrivons et s\'eparons les \'equations gouvernant les petites
oscillations monopolaires de la 5-m\'etrique. Nous utilisons ensuite
ces \'equations pour discuter la stabilit\'e classe par classe et cas
par cas. Nous concluons ce chapitre par la section 2.4 o\`u un parall\`ele
avec l'\'etude de Tomimatsu est pr\'esent\'e.

La recherche de solutions cordes cosmiques est entam\'ee dans le
chapitre 3. Dans la section 3.2, nous pr\'esentons une \'etude
syst\'ematique des solutions exactes \mbox{4-statiques} \`a sym\'etrie
cylindrique de la th\'eorie de Kaluza-Klein, et nous discutons les solutions
du type corde cosmique. Dans la section 3.3, une \'etude semblable est
men\'ee pour une classe de solutions en tenant compte du terme de
Gauss-Bonnet dans l'action d'Einstein-Hilbert; les solutions cordes
cosmiques obtenues sont aussi solutions des \'equations de Kaluza-Klein
sans terme de
Gauss-Bonnet, et par cons\'equent elles ne d\'ependent pas
explicitement de sa pr\'esence dans l'action g\'en\'eralis\'ee.
Nous obtenons, dans la section 3.4, une solution
perturbative de la th\'eorie g\'en\'eralis\'ee, dont nous montrons
ensuite qu'elle est exacte; cette solution peut s'interpr\`eter comme
une corde cosmique
supraconductrice de la th\'eorie de Kaluza-Klein, avec le terme de
Gauss-Bonnet comme principale source. Une \'etude qualitative des
g\'eod\'esiques de la g\'eom\'etrie de cette solution est conduite
dans la section 3.5. Nous concluons dans la section 3.6.

Enfin l'Annexe 1 rassemble les principales formules utilis\'ees dans
ce m\'emoire. Les deux autres Annexes 2 et 3 sont consacr\'es \`a des
d\'emonstrations sp\'ecifiques se rapportant aux 2 sous-sections 3.3.1
et 3.3.2 de la section 3.3. Dans l'Annexe 4 on int\`egre l'\'equation
de Killing pour la m\'etrique corde cosmique supraconductrice de la
sous-section 3.4.3.

\newpage

\chapter{La th\'eorie de Kaluza-Klein et\\
         son extension de Gauss-Bonnet}

\thispagestyle{plain}
\markboth{La th\'eorie de K-K et son extension de G-B}{}
\section{L'id\'ee de Kaluza}
En tentant d'unifier les seules deux forces bien \'etablies \`a
l'\'epoque, \`a savoir la gravit\'e et l'\'electromagn\'etisme, Kaluza
postule en 1921 \cite{Kaluza} l'existence d'une cinqui\`eme dimension
suppl\'ementaire pour l'espace-temps. Il consid\`ere une th\'eorie d'Einstein
de la gravit\'e dans un espace \`a cinq dimensions muni d'une
m\'etrique sym\'etrique

\be
ds^{2} = g_{AB}(x^{C})\,dx^{A}\,dx^{B}
\ee

\noindent (o\`u $A, B, C, \dots = 1, 2, 3, 4, 5$) de signature
$(-\,-\,-\,+\,-)$; la dimension suppl\'ementaire $x^{5}$ est suppos\'ee donc
du genre espace. Puis il proc\`ede \`a la d\'ecomposition 4+1

\be
\label{eq d 4+1}
g_{AB} = \left( \begin{array}{cc}
                \bar{g}_{\mu\nu}+{\lambda}^2\,A_{\mu}\,A_{\nu}\,g_{55} &
                \lambda\,A_{\mu}\,g_{55}\\
                \lambda\,A_{\nu}\,g_{55} & g_{55}
                \end{array} \right)
\ee

\noindent (o\`u $\mu, \nu, \dots = 1, 2, 3, 4$) avec

\bn
\label{eq potentiel}
A_{\mu} & \equiv & {\lambda}^{-1}\,\frac{g_{\mu5}}{g_{55}}\\
\label{eq projete}
\bar{g}_{\mu\nu} & \equiv & g_{\mu\nu}-\frac{g_{\mu5}\,g_{\nu5}}{g_{55}}.
\en

\noindent La m\^{e}me d\'ecomposition appliqu\'ee aux \'equations
d'Einstein dans le vide permet de retrouver (voir (\ref{eq KK})), dans
l'hypoth\`ese o\`u $g_{AB}$ ne d\'epend pas de $x^{5}$, les
\'equations d'Einstein pour $\bar{g}_{\mu\nu}(x^{\rho})$,
les \'equations de Maxwell pour $A_{\mu}(x^{\rho})$ et l'\'equation
de Klein-Gordon sans masse pour $g_{55}(x^{\rho})$.

\section{L'id\'ee de Klein}
\label{sec-Klein}
\setcounter{equation}{0}
Klein \cite{Klein} cherche \`a justifier deux choses: l'hypoth\`ese de
l'ind\'ependance de la m\'etrique par rapport \`a la dimension
suppl\'ementaire $x^{5}$ (hypoth\`ese non justifi\'ee de Kaluza) et le
fait que cette dimension soit inobservable. Pour ce faire, il postule
que l'espace-temps a la topologie du produit $V^{4}\,\times\,S^{1}$,
o\`u $V^{4}$ est topologiquement \'equivalent \`a un espace minkowskien
\`a quatre dimensions $M^{4}$, et $S^{1}$ est un cercle de rayon $a$
param\'etris\'e par $x^{5}$:
\be
0 \leq x^{5} \leq 2\,\pi\,a.
\ee
\noindent Il suppose l'existence d'une isom\'etrie d\'efinie par un
vecteur de Killing de genre espace, ce qui signifie que l'espace-temps est
homog\`ene dans la cinqui\`eme direction (il n'y a pas moyen de
d\'eterminer la position dans la cinqui\`eme dimension). Il suppose de
plus que le rayon $a$ est si petit que la cinqui\`eme dimension est
inobservable; il d\'eterminera ensuite par des consid\'erations {\em
quantiques} (voir section 1.6 plus loin) la valeur de ce rayon

\be
\label{eq rayon}
a = 2\left(\frac{\hbar\,G}{\alpha\,c^{3}}\right)^{1/2} \simeq 3,7.10^{-32}\,cm
\ee

\noindent o\`u $\alpha$ est la constante de structure fine. Le rayon
$a$ est bien de l'ordre de la longueur de Planck $\ell_{P}$,

\be
\label{eq fine}
a = \sqrt{\frac{4}{\alpha}}\,\ell_{P}
\ee

\noindent avec $\ell_{P} \simeq 1,6.10^{-33}\,cm$\,.

\section{Sym\'etries de {\boldmath $V^{4}\,\times\,S^{1}$}}
\label{sec-V.S}
\setcounter{equation}{0}
Les transformations de coordonn\'ees qui conservent la topologie
de $V^{4}\,\times\,S^{1}$ sont:\\\\
${\bullet}$ le groupe des transformations g\'en\'erales \`a quatre dimensions
de $V^{4}$

\be
x^{\mu} \rightarrow x^{\prime\,\mu}(x^{\nu})
\ee

\noindent sous l'action duquel, les composantes
$\bar{g}_{\mu\nu}$ de la matrice (\ref{eq d 4+1}) se transforment comme des
tenseurs d'ordre 2, les composantes $A_{\mu}$ comme des tenseurs d'ordre
1 (des vecteurs) et $g_{55}$ comme un tenseur d'ordre 0 (un scalaire)\\\\
${\bullet}$ et le groupe de rotations $U(1)$ du cercle $S^{1}$

\be
x^{5} \rightarrow x^{\prime\,5} = x^{5} + f(x^{\mu})
\ee

\noindent sous l'action duquel, les composantes de la matrice
(\ref{eq d 4+1}) se transforment comme
$$
\bar{g}^{\,\prime}_{\mu\nu} = \bar{g}_{\mu\nu}
$$
\be
A^{\prime}_{\mu} = A_{\mu}-{\lambda}^{-1}\,\partial_{\mu}f
\ee
$$
g^{\prime}_{55} = g_{55}
$$

\noindent o\`u on voit appara\^{\i}tre manifestement la transformation
de jauge pour les potentiels $A_{\mu}$.

\section{\'Equations de la th\'eorie de Kaluza-Klein}
\label{sec-equationKK}
\setcounter{equation}{0}
Les \'equations de la th\'eorie de Kaluza-Klein (dans le vide)
s'obtiennent en extr\'emisant l'action d'Einstein-Hilbert (on omettra
le terme cosmologique)

\be
S = -\frac{1}{16\,\pi\,G_{5}}\,\int d^{\,5}x\,\sqrt{g}\,R
\ee

\noindent (o\`u $R$ est le scalaire de courbure \`a cinq dimensions
(\ref{eq scalaire}), $g \equiv \mbox{det}g_{AB}$ et $G_{5}$ la constante
de gravitation \`a cinq dimensions) pour des variations arbitraires
des $g_{AB}$. Elles sont donn\'ees par

\be
\label{eq E}
G_{AB} = 0
\ee

\noindent o\`u $G_{AB} \equiv R_{AB} - (1/2)g_{AB}\,R$ est le
tenseur d'Einstein \`a divergence covariante nulle

\be
G^{AB}\,_{;\,A} = 0
\ee

\noindent (o\`u $(;)$ d\'enote la d\'eriv\'ee covariante \`a cinq
dimensions  (\ref{eq dc})). Les \'equations (\ref{eq E}) sont \'equivalentes
aux \'equations suivantes
$$
R^{\mu\nu} = 0 \;\; ; \;\; R^{\mu}\,_{5} = 0 \;\; ; \;\; R_{55} = 0
$$
qui s'\'ecrivent en introduisant l'ansatz (\ref{eq d 4+1}) \cite{KK}
$$
\bar{R}_{\mu}\,^{\nu} =
\frac{{\lambda}^{2}}{2}\,g_{55}\,F_{\mu\rho}\,F^{\nu\rho} +
|g_{55}|^{-1/2}\,\bar{D}_{\mu}\,\bar{D}^{\nu}|g_{55}|^{1/2}
$$
\be
\label{eq KK}
\bar{D}_{\nu}\left((g_{55})^{3/2}\,F^{\mu\nu}\right) = 0
\ee
$$
\Box|g_{55}|^{1/2} = -
\frac{{\lambda}^{2}}{4}\,|g_{55}|^{3/2}\,F_{\mu\nu}\,F^{\mu\nu}.
$$
Dans (\ref{eq KK}), $\bar{D}_{\mu}$ et $\bar{R}_{\mu}\,^{\nu}$ sont
respectivement la d\'eriv\'ee covariante et le tenseur de Ricci
d\'efinis par rapport \`a $\bar{g}_{\mu\nu}$, dont l'inverse $\bar{g}^{\mu\nu}$
sert \`a \'elever les indices $\mu, \nu, \ldots$, $F_{\mu\nu} \equiv
\partial_{\mu}A_{\nu} - \partial_{\nu}A_{\mu}$ est le champ
\'electromagn\'etique associ\'e au quadripotentiel $A_{\mu}$, et
$\Box$ est l'op\'erateur dalembertien,
$$
\Box{\bullet} \equiv \frac{1}{\sqrt{-\bar{g}}}\,
\partial_{\nu}\left(\sqrt{- \bar{g}}\,
\bar{g}^{\,\mu\nu}\,\partial_{\mu}{\bullet}\right)
$$
avec $\bar{g} \equiv \mbox{det}\bar{g}_{\mu\nu}$. Sous cette forme, les
\'equations (\ref{eq KK}), g\'en\'eralisent de fa\c{c}on \'evidente
les \'equations d'Einstein-Maxwell de la th\'eorie de la gravitation coupl\'ee
\`a l'\'electromagn\'etisme. En effet, si on suppose $g_{55} = Cste$,
les deux premi\`eres \'equations (\ref{eq KK}) se r\'eduisent aux
\'equations d'Einstein-Maxwell moyennant le choix

\be
\label{eq lambda}
{\lambda}^{2} = 2\,\kappa = 16\,\pi\,G\,c^{-2}
\ee

\noindent (dans le cas $g_{55} = -1$). Si au contraire, suivant Jordan
et Thiry \cite{Jordan}, on traite sur le m\^{e}me plan les trois
\'equations (\ref{eq KK}), on obtient une th\'eorie unifi\'ee de la
gravitation et de l'\'electromagn\'etisme coupl\'es au champ
scalaire $g_{55}$, le choix (\ref{eq lambda}) restant toujours valable
dans le cas o\`u le champ scalaire se comporte \`a grande distance des
sources comme $g_{55} \rightarrow -1$.

\section{Terme de Gauss-Bonnet}
\setcounter{equation}{0}
Dans les espaces-temps \`a quatre dimensions, l'action d'Einstein-Hilbert
est l'uni-\\que action g\'eom\'etrique conduisant \`a des \'equations
diff\'erentielles du second ordre pour la m\'etrique $g_{AB}$. Les \'equations
d'Einstein sont donc l'unique choix si on exige que les \'equations du
champ gravitationnel \`a quatre dimensions soient du second ordre; de
plus elles sont lin\'eaires par rapport aux d\'eriv\'ees secondes de la
m\'etrique. Dans les th\'eories de Kaluza-Klein \`a plus de quatre
dimensions il y a d'autres choix possibles si on \'ecarte la condition
{\em d'\^{e}tre lin\'eaires par rapport aux d\'eriv\'ees secondes de
$g_{AB}$}, en introduisant dans l'action d'Einstein-Hilbert des termes de
courbure d'ordre sup\'erieur dont la variation par rapport \`a
$g_{AB}$ donne des tenseurs d'ordre 2 qui s'ajoutent au tenseur
d'Einstein $G_{AB}$ (comparer (\ref{eq E}) et (\ref{eq GB})). En effet,
Lovelock \cite{Love} a montr\'e qu'il existe en g\'en\'eral d'autres {\em
tenseurs sym\'etriques du second rang} --autres que le tenseur
m\'etrique et le tenseur d'Einstein-- {\em \`a divergence
covariante nulle et qui contiennent jusqu'\`a la d\'eriv\'ee seconde de
la m\'etrique $g_{AB}$}\,.

\`A cinq dimensions, il y a un seul tenseur de ce type: c'est le
tenseur de Lanczos \cite{Madore} que l'on obtient en variant, par rapport
\`a la m\'etrique, le terme correspondant dans l'action appel\'e terme de
Gauss-Bonnet. Si on sait de plus que les th\'eories de Lovelock de la
gravit\'e ne comportent pas de fant\^{o}mes \cite{Zwiebach, Zumino} et
qu'elles conservent la parit\'e \cite{WheelerB273}, on voit qu'il n'y
a pas  de raison physique de restreindre, dans les th\'eories de
Kaluza-Klein, l'action au seul terme d'Einstein-Hilbert.

Dans tout le reste de ce chapitre, on continuera \`a omettre le terme
cosmologique mais on tiendra compte du terme de Gauss-Bonnet. L'action
g\'en\'eralis\'ee dans le vide s'\'ecrit donc

\be
\label{eq action}
S_{G} = -\frac{1}{16\,\pi\,G_{5}}\,\int d^{\,5}x\,\sqrt{g}\,\cal L
\ee

\noindent avec
\be
\label{eq densite}
{\cal L} \equiv R + \frac{\gamma}{2}\,{\cal L}_{{\cal GB}}
\ee

\noindent o\`u $\gamma$ est une constante et ${\cal L}_{{\cal GB}}$ est le
terme de Gauss-Bonnet donn\'e par

\be
{\cal L}_{{\cal GB}} \equiv R^{ABCD}\,R_{ABCD} - 4\,R^{AB}\,R_{AB} + R^{2}
\ee

\noindent (o\`u $R_{ABCD}$ est le tenseur de Riemann (\ref{eq
Riemann}) et $R_{AB}$ le tenseur de Ricci (\ref{eq Ricci})).
Les \'equations du champ s'obtiennent en variant l'action par rapport
\`a $g_{AB}$; elles s'\'ecrivent

\be
\label{eq GB}
G_{AB} + \gamma\,L_{AB} = 0
\ee

\noindent o\`u $L_{AB}$ est le tenseur de Lanczos (sym\'etrique) tel que

\be
L^{AB}\,_{;\,A} = 0.
\ee

\noindent Il est d\'efini par

\be
\label{eq L}
L_{AB} \equiv R_{A}\,^{CDE}\,R_{BCDE} - 2\,R^{CD}\,R_{ACBD} -
2\,R_{AC}\,R_{B}\,^{C} + R\,R_{AB} - \frac{1}{4}\,{g_{AB}}\,
{\cal L}_{{\cal GB}}\,.
\ee

\noindent En prenant la trace de (\ref{eq L}) on trouve

\be
L^{A}\,_{A} = {\cal L}_{{\cal GB}} - \frac{5}{4}\,{\cal L}_{{\cal GB}} =
-\frac{1}{4}\,{\cal L}_{{\cal GB}}.
\ee

\noindent En reportant ensuite dans la trace de (\ref{eq GB}) on obtient

\be
\label{eq 6R}
\gamma\,{\cal L}_{{\cal GB}} = -6\,R
\ee
puis en reportant dans (\ref{eq densite}) on trouve enfin

\be
{\cal L} = -2\,R.
\ee

\noindent Ainsi, quand les \'equations du champ sont satisfaites, la
densit\'e lagrangienne est multiple du scalaire de courbure. Ceci
reste valable quel que soit la dimension de l'espace-temps. Pour $d
\geq 5$, on a

\be
{\cal L} = -\frac{2}{d-4}\,R\,.
\ee

\section{La loi du mouvement}
\setcounter{equation}{0}
Les \'equations (\ref{eq GB}) --ou l'action (\ref{eq
action})-- peuvent \^{e}tre g\'en\'eralis\'ees pour inclure un
terme de source $T_{AB}$ dans le second membre,

\be
\label{eq matiere}
G_{AB} + \gamma\,L_{AB} = \kappa\,T_{AB}\,;
\ee

\noindent puisque le premier membre de (\ref{eq matiere}) est sym\'etrique et
de divergence covariante nulle (par construction), le tenseur
de mati\`ere $T_{AB}$ doit l'\^{e}tre aussi:

\be
\label{eq T}
T^{AB}\,_{;\,A} = 0\,.
\ee

\noindent Or, la relation correspondante \`a quatre dimensions
$T^{\mu\nu}\,_{;\,\mu}=0$ conduit, pour $T^{\mu\nu}$ de la forme
$T^{\mu\nu} \equiv \rho\,u^{\mu}\,u^{\nu}$ qui correspond \`a une
poussi\`ere sans pression ($\rho \equiv$ densit\'e de masse et
$u^{\mu} \equiv dx^{\mu}/d\sigma$), \`a l'\'equation g\'eod\'esique pour
les particules libres:
$$
u^{\nu}\,u^{\mu}\,_{,\nu} +
{\Gamma}^{\mu}_{\nu\sigma}\,u^{\nu}\,u^{\sigma} = 0
$$
($u^{\nu}\,u^{\mu}\,_{,\nu}=d^{\,2}x^{\mu}/d\sigma^{\,2}$). De m\^{e}me \`a
cinq dimensions le choix $T^{AB}=\rho\,u^{A}\,u^{B}$ pour une poussi\`ere
(\'eventuellement charg\'ee) sans pression conduit \`a l'\'equation
g\'eod\'esique (pour les particules libres neutres ou charg\'ees):

\be
\label{eq geod}
\frac{d^{\,2}x^{A}}{d\sigma^{\,2}} +
{\Gamma}^{A}_{BC}\,\frac{dx^{B}}{d\sigma}\,\frac{dx^{C}}{d\sigma} = 0
\ee

\noindent o\`u les ${\Gamma}^{A}_{BC}$ ($\Gamma^{\mu}_{\nu\sigma}$) sont les
connexions (\ref{eq connexion}).

En d\'efinissant la 5-impulsion g\'en\'eralis\'ee

\be
p_{A} \equiv \mu\,u_{A}
\ee

\n (o\`u $\mu$ est proportionnelle \`a la masse de la particule), les
composantes d'espace-temps usuelles de (\ref{eq geod}) peuvent
s'\'ecrire

\be
\label{eq imp}
\frac{dp^{\lambda}}{d\sigma}=-\bar{\Gamma}^{\lambda}_{\mu\nu}\,p^{\mu}\,u^{\nu}
+\lambda\,p_{5}\,F^{\lambda}\,_{\nu}\,u^{\nu}+\frac{1}{2}\,
\frac{p_{5}^{2}}{\mu}\,\frac{\partial^{\lambda}g_{55}}{g_{55}^{2}}
\ee

\n o\`u les $\bar{\Gamma}^{\lambda}_{\mu\nu}$ sont les connexions
associ\'ees \`a la m\'etrique r\'eduite (\ref{eq projete}). Le premier
terme du membre de doite de (\ref{eq imp}) s'interpr\`ete comme une
``force" gravitationnelle, le deuxi\`eme comme une force
\'electromagn\'etique, et le troisi\`eme comme une force
scalaire. Cette interpr\'etation conduit \`a identifier
$$
q \equiv \lambda\,p_{5}
$$
(qui est une constante du mouvement d'apr\`es la cinqui\`eme
\'equation g\'eod\'esique) comme la charge \'electrique de la
particule. Cette charge est a priori quelconque en th\'eorie
classique, mais pas en th\'eorie quantique, o\`u elle est quantifi\'ee
(comme on peut le voir par exemple en utilisant la condition de
quantification de Bohr-Sommerfeld $\oint p_{5}dx^{5}=nh$, l'int\'egrale
portant sur le cercle de Klein):
$$
q=\frac{n\lambda \hbar}{a}\,.
$$
En supposant que $n=1$ pour l'\'electron, et en utilisant la valeur
(\ref{eq lambda}) de $\lambda$, on obtient ainsi la valeur (\ref{eq rayon})
du rayon de Klein $a$.

\section{Composantes du tenseur de Riemann dans la d\'ecomposition
        {\boldmath $n+p$}}
\setcounter{equation}{0}
Consid\'erons de fa\c{c}on g\'en\'erale un espace-temps \`a $d$
dimensions ($d = n+p$) de m\'etrique $g_{AB}$ ($A, B = 1,\ldots,d$).
Si la m\'etrique ne d\'epend explicitement que de $n$ coordonn\'ees
$x^{i}$ ($i=1,\ldots,n$), les autres $p$ coordonn\'ees $x^{a}$
($a = n+1,\ldots,d$) sont dites cycliques ($g_{AB,a} = 0$).
C'est le cas des th\'eories de
Kaluza-Klein o\`u toutes les dimensions suppl\'ementaires sont
suppos\'ees cycliques (voir section \ref{sec-Klein}). La
m\'etrique est dite alors {\em p-stationnaire} et admet $p$ vecteurs
de Killing donn\'es par

\be
\label{eq cycl}
{\xi}_{a}\,^{A} = {\delta}_{a}\,^{A} \;\;\;\;\;\forall\;a,
\ee

\noindent qui satisfont bien \`a l'\'equation g\'en\'erale de Killing

\be
\label{eq Killing}
{\xi}_{a\,A;B} + {\xi}_{a\,B;A} = 0.
\ee

\noindent Ces vecteurs de Killing engendrent un groupe d'isom\'etrie ab\'elien
\`a $p$ param\`etres dont l'action, qui peut \^{e}tre d\'efinie par
des translations infinit\'esimales parall\`element aux vecteurs de
Killing
$$
x^{\prime A} = x^{A} + \epsilon\,{\xi}^{A}, \quad \mbox{avec} \;\;
|\epsilon| \ll 1,
$$
assure d'apr\`es (\ref{eq Killing}) l'invariance de la m\'etrique

\be
\label{eq inv}
g^{\prime}_{AB}(x^{i}) = g_{AB}(x^{i})
\ee

\noindent Si de plus les \'el\'ements $g_{ai}(x^{j})$ sont
identiquement nuls, la m\'etrique $g_{AB}(x^{i})$ est dite alors {\em
p-statique} et peut \^{e}tre \'ecrite sous la forme

\be
\label{eq n+p}
ds^{\,2} = {\lambda}_{ab}(x^{k})\,dx^{a}\,dx^{b} +
{\tau}^{-1}\,h_{ij}(x^{k})\,dx^{i}\,dx^{j}
\ee

\noindent o\`u

\be
\label{eq dt}
{\tau}(x^{k}) \equiv |\mbox{det}{\lambda}_{ab}|.
\ee

Dans le cas de solutions statiques aux th\'eories de
Kaluza-Klein ou de Gauss-Bonnet, les $x^{i}$
sont les coordonn\'ees d'espace ordinaire ou certaines
d'entre elles, et le reste des coordonn\'ees y compris le temps sont
des coordonn\'ees cycliques. La m\'etrique $h_{ij}$ est donc celle de
l'espace (plus pr\'ecis\'ement c'est la m\'etrique ${\tau}^{-1}\,h_{ij}$
conforme \`a $h_{ij}$), ${\lambda}_{ab}$ est une matrice dont les
\'el\'ements sont des scalaires par rapport \`a la m\'etrique d'espace
$h_{ij}$.

On terminera ce chapitre par donner les composantes du tenseur de
Riemann dans la d\'ecomposition $n+p$ de la m\'etrique (\ref{eq n+p})

\bn
\label{abcd}
R_{abcd} &=&
\frac{1}{4}\,\tau\,\left({\lambda}_{ad,i}\,{\lambda}_{bc}\,^{,i}
- {\lambda}_{ac,i}\,{\lambda}_{bd}\,^{,i}\right) \\
\label{aijb}
R_{aijb} &=& \frac{1}{4}\left(2\,{\lambda}_{,j;i} -
{\lambda}_{,j}\,{\lambda}^{-1}{\lambda}_{,i} -
{\tau}\,({\tau}^{-1})_{,i}\,{\lambda}_{,j} -
{\tau}\,({\tau}^{-1})_{,j}\,{\lambda}_{,i} +
{\tau}\,h_{ij}\,({\tau}^{-1})_{,k}\,{\lambda}^{,k}\right)_{ab}\nonumber \\\\
\label{abij}
R_{abij} &=&
\frac{1}{4}\left({\lambda}_{,j}\,{\lambda}^{-1}{\lambda}_{,i} -
{\lambda}_{,i}\,{\lambda}^{-1}{\lambda}_{,j}\right)_{ab} \\
\label{ijkl}
R_{ijkl}&=& {\tau}^{-1}\;\;^{(h)}R_{ijkl} + \frac{1}{2}\,
\left(h_{il}\,({\tau}^{-1})_{,j;k} - h_{ik}\,({\tau}^{-1})_{,j;l} -
h_{jl}\,({\tau}^{-1})_{,i;k} + h_{jk}\,({\tau}^{-1})_{,i;l}\right)
\nonumber \\ & & \mbox{} +
\frac{1}{4}\,{\tau}\,\left[-3\,h_{il}\,({\tau}^{-1})_{,j}\,({\tau}^{-1})_{,k}
+ 3\,h_{jl}\,({\tau}^{-1})_{,i}\,({\tau}^{-1})_{,k} +
3\,h_{ik}\,({\tau}^{-1})_{,j}\,({\tau}^{-1})_{,l}\right. \nonumber \\
& & \left. \mbox{} - 3\,h_{jk}\,({\tau}^{-1})_{,i}\,({\tau}^{-1})_{,l}
+ (h_{il}\,h_{jk} - h_{ik}\,h_{jl})
\,({\tau}^{-1})_{,m}\,({\tau}^{-1})^{,m}\right]
\en

\noindent o\`u $^{(h)}R_{ijkl}$ et la d\'erivation covariante $(;)$
sont d\'efinis par rapport \`a la m\'etrique $h_{ij}$ dont l'inverse
$h^{ij}$ sert \`a \'elever les indices $i, j, k, \ldots$. Cette
d\'ecomposition ainsi que les \'equations (\ref{abcd} $\rightarrow$ \ref{ijkl})
nous seront tr\`es utiles dans les deux chapitres suivants, consacr\'es
respectivement \`a l'\'etude de la stabilit\'e des solutions de
Kaluza-Klein et \`a la recherche et l'\'etude des cordes cosmiques
dans la th\'eorie de Gauss-Bonnet.

\newpage

\chapter{\'Etude analytique de la stabilit\'e\\
         des solutions statiques\\
         \`a sym\'etrie sph\'erique}
\thispagestyle{plain}
\markboth{\'Etude analytique de la stabilit\'e...}{}
\label{chap stabilite}
Dans ce chapitre on \'etudiera la stabilit\'e, vis-\`a-vis des petites
excitations radiales, de toutes les solutions, {\em 2-statiques},
asymptotiquement plates, \`a sym\'etrie sph\'erique de la
th\'eorie de Kaluza-Klein. La m\'ethode que nous suivrons consiste \`a
consid\'erer un tenseur m\'etrique diff\'erant de la solution statique
par des petites perturbations qui ne d\'ependent que du temps $t$ et de la
variable radiale $r$, puis \`a \'etudier l'\'equation de Klein-Gordon,
qui d\'ecoule --apr\`es d\'ecouplage-- des \'equations d'Einstein
lin\'earis\'ees, pour une d\'ependance en $t$ de la forme
e$^{-i\,\omega\,t}$ et avec un syst\`eme de conditions de
r\'egularit\'e qui sera d\'efini dans la section \ref{sec-discussion}.
Si, pour $\omega=i\,k\;(k>0)$, toutes les fonctions de perturbations
introduites ont un comportement spatial physiquement acceptable, une
petite perturbation cro\^{\i}tra exponentiellement, d'o\`u l'instabilit\'e.
On concluera \`a la stabilit\'e dans le cas contraire.

Dans la section \ref{sec-solution}, on exposera bri\`evement la
m\'ethode de Maison \cite{Maison} qui conduit, dans la d\'ecomposition
3+2, aux solutions 2-statiques \`a sym\'etrie sph\'erique que nous
regrouperons en deux classes. Dans la section \ref{sec-mono}, on
introduira l'ansatz de travail et r\'eduira les \'equations
``lin\'eaires" d'Einstein (dans le vide) \`a une \'equation de type
de {\em Klein-Gordon dans un espace courbe}\,. L'\'etude de la
stabilit\'e est trait\'ee dans la section \ref{sec-discussion};
apr\`es des g\'en\'eralit\'es, l'\'etude analytique
est conduite cas par cas dans chaque classe.

\section{Solutions statiques}
\label{sec-solution}
\setcounter{equation}{0}
Consid\'erons d'abord en th\'eorie de Kaluza-Klein, le cas
des m\'etriques {\em 2-stationnai-}{\em res} qui ne d\'ependent pas du temps
$t$
et de la dimension suppl\'ementaire $x^{5}$, pour lesquelles le
carr\'e de l'intervalle est une g\'en\'eralisation \cite{ClementLett}
de (\ref{eq n+p}) (valable pour les m\'etriques {\em 2-statiques})

\be
\label{eq 3+2}
ds^{\,2} =
{\lambda}_{ab}(x^{k})\,(dx^{a}
+ \mbox{\LARGE $a$}^{a}_{i}\,dx^{i})
(dx^{b} + \mbox{\LARGE $a$}^{b}_{j}\,dx^{j}) +
{\tau}^{-1}\,h_{ij}(x^{k})\,dx^{i}\,dx^{j}
\ee

\noindent ($i, j, k,\ldots=1, 2, 3$\,; $a, b,\ldots=4, 5$) avec

\be
\label{eq A}
  \mbox{\LARGE $a$}^{a}_{i} \equiv
{\lambda}^{ab}\,g_{ib}
\;\;\;(={\lambda}^{ab}\,{\xi}_{b\,i})
\ee

\noindent (o\`u ${\lambda}^{ab}$ est la matrice inverse de ${\lambda}_{ab}$
et ${\xi}_{a\,i}$ sont les vecteurs de Killing (\ref{eq cycl})).
Dans (\ref{eq 3+2}), la m\'etrique $h_{ij}$ et les composantes de la matrice
${\lambda}_{ab}$ ont la m\^{e}me signification que dans (\ref{eq n+p}):
respectivement {\em m\'etrique d'espace et scalaires d'espace}.

Dans sa m\'ethode de r\'esolution des \'equations d'Einstein dans le
vide (\ref{eq E}), Maison \cite{Maison} cherche d'abord
\`a param\'etriser la m\'etrique $g_{AB}$ par des grandeurs {\em
3-tensorielles}. Or cette derni\`ere comporte 15 composantes dont 9
composantes sont d\'ej\`a 3-tensorielles \`a savoir les composantes
$h_{ij}$ et ${\lambda}_{ab}$; il introduit les 6 composantes
suivantes appel\'ees twists \footnote{En effet, les twists
de l'espace-temps peuvent \^{e}tre d\'efinis de fa\c{c}on covariante par
$$
{\omega}_{a\,A} \equiv \mbox{\Large $\varepsilon$}_{ABCDE}\,{\xi}_{4}\,^{B}\,
{\xi}_{5}\,^{C}\,{\xi}_{a}\,^{D;E}\,,
$$
({\Large $\varepsilon$}$_{ABCDE}$ \'etant le tenseur antisym\'etrique).
Or, par d\'efinition, la 2-surface engendr\'ee, en un point donn\'e,
par les 2 vecteurs de Killing ${\xi}_{a}\,^{A}$ est homog\`ene (voir
(\ref{eq inv})), cette derni\`ere sera d\'epourvue de twists qui ne
peuvent se manifester alors que dans l'espace ordinaire qui lui est
``normal". Ceci se traduit math\'ematiquement par
$$
{\xi}_{b}\,^{A}\,{\omega}_{a\,A}=0\;\;\;\;\;\forall a, b\;;
$$
${\omega}_{a\,i}$ sont les seules composantes non nulles des
${\omega}_{a\,A}$.}

\be
\label{eq torsion}
{\omega}_{ai} \equiv h^{-1/2}\,{\tau}\,{\lambda}_{ab}\,
h_{ij}\,\mbox{\Large $\eta$}^{jkl}\,\mbox{\LARGE $a$}^{b}_{j,k}\,,
\ee

\noindent ($h \equiv - |h_{ij}|$ et {\Large $\eta$}$^{jkl}$ est le
symbole antisym\'etrique) et montre, que pour $a$ fix\'e, les 3
composantes ${\omega}_{ai}$ sont bien celles d'un vecteur covariant
associ\'e \`a la m\'etrique d'espace
(voir aussi R\'ef. \cite{theseA}). En se servant des \'equations
$R_{AB} = 0$ et de la d\'efinition (\ref{eq torsion}), il d\'eduit l'\'equation
$$
{\omega_{ai,j}} - {\omega_{aj,i}} = 0
$$
conduisant \`a introduire les potentiels scalaires
${\omega}_{a}$ tels que

\be
{\omega}_{ai} = {\omega}_{a,i}.
\ee

Les \'equations d'Einstein $R_{AB}=0$ \'ecrites en fonction des grandeurs
3-tensorielles $h_{ij}$, ${\lambda}_{ab}$, ${\omega}_{a}$
(d\'ecomposition $3+2$ des \'equations d'Einstein) \cite{Maison} sont
invariantes sous certaines transformations qui constituent une
r\'ealisation non lin\'eaire du groupe SL(3, $\Re$) \cite{Maison}. En
r\'ealisant une transformation convenable sur les scalaires ${\lambda}_{ab}$
et ${\omega}_{a}$, Maison introduit la matrice $3 \times 3$ r\'eelle,
sym\'etrique et anti-unimodulaire $\chi$ de signature ($+ - +$)

\be
\label{eq X}
\chi = \left( \begin{array}{cc}
{\lambda}_{ab}+
{\tau}^{-1}\,{\omega}_{a}\,{\omega}_{b} & {\tau}^{-1}\,{\omega}_{a}\\
{\tau}^{-1}\,{\omega}_{b} &  {\tau}^{-1} \end{array} \right),
\ee

\noindent qui r\'ealise une repr\'esentation lin\'eaire du
groupe SL(3, $\Re$) \cite{Maison} c'est \`a dire que
$\chi_{uv}$ ($u, v=4, 5, 6$) se transforme comme un tenseur covariant
du second rang sous l'action du groupe SL(3, $\Re$).

\`A l'aide de cette param\'etrisation, les \'equations
$R_{AB} = 0$ s'\'ecrivent

\bn
\label{eq Maison1}
{^{(h)}R_{ij} \;\;\;\;\;\;} & = &
\frac{1}{4}\,\mbox{Tr}(\chi^{-1}\,\chi_{,i}\,
\chi^{-1}\,\chi_{,j})\\
\label{eq Maison2}
(\chi^{-1}\,\chi^{,i})_{;i} & = & 0
\en

\noindent ($^{(h)}R_{ij}$ et les d\'erivations se rapportent \`a
la m\'etrique $h_{ij}$). Ces \'equations sont bien invariantes sous
l'action du groupe SL(3, $\Re$):
$\chi^{\prime} \rightarrow P^{T}\,\chi\,P$,
$\chi^{\prime\,-1} \rightarrow P^{-1}\,\chi^{-1}\,P^{T\,-1}$ et
$\chi^{\prime}_{,i} \rightarrow P^{T}\,\chi_{,i}\,P$ avec $P \in$
SL(3, $\Re$). Consid\'erons le cas o\`u les diff\'erentes composantes
de la matrice $\chi$ d\'ependent d'une seule fonctions
$\sigma(x^{i})$, appel\'ee potentiel

\be
\chi = \mbox{\LARGE $\eta$}\,\,\mbox{\Large e}^{N\,\sigma}
\ee

\noindent o\`u {\LARGE $\eta$} et $N$ sont des matrices constantes
$3 \times 3$. Alors les \'equations d'Einstein \mbox{(\ref{eq Maison1},
\ref{eq Maison2})} se r\'eduisent au syst\`eme suivant

\bn
\label{eq sigma}
\triangle \sigma & = & 0\\
\label{eq Rij}
^{(h)}R_{ij} & = &
\frac{1}{4}\,\mbox{Tr}(N^{2})\,{\sigma}_{,i}\,{\sigma}_{,j}
\en

\noindent ($\triangle$ \'etant le laplacien d\'efini par rapport \`a
$h_{ij}$)).

Dans le cas particulier des solutions 2-statiques, ${\omega}_{a} \equiv 0$
et la matrice $N$ est de la forme

\be
\label{eq N}
N = \left( \begin{array}{cc}
            M & 0\\
            0 & -x \end{array} \right)
\ee

\noindent o\`u $x \equiv $\,Tr$(M)$ et $M$ est une matrice $2 \times 2$
param\'etris\'ee par

\be
\label{eq M}
M = \left( \begin{array}{cc}
            x-a & b\\
            -b & a \end{array} \right)
\ee

\noindent avec

\be
\label{eq b}
b^{2} \equiv a^{2} - x\,a + y
\ee

\noindent o\`u $y \equiv \mbox{det}M$. Pour un potentiel $\sigma$ s'annulant
\`a l'infini spatial, la m\'etrique de l'espace-temps est
asymptotiquement minkowskienne si

\be
\mbox{\LARGE $\eta$} \equiv \left( \begin{array}{ccc}
                            1 & 0 & 0\\
                            0 & -1 & 0\\
                            0 & 0 & 1 \end{array} \right)\,.
\ee

\noindent Alors la matrice $\lambda$ est donn\'ee par

\be
\label{eq matlamb}
{\lambda} =  \left( \begin{array}{cc}
                     1 & 0\\
                     0 & -1 \end{array} \right)\,
\mbox{\Large e}^{M\,\sigma}
\ee

\noindent puis son d\'eterminant $-{\tau}$ par

\be
\label{eq tau}
{\tau} = \mbox{e}^{x\,\sigma}\,.
\ee

\noindent D'autre part de (\ref{eq N}), (\ref{eq M}), (\ref{eq b}) on
obtient

\be
\label{eq N2}
\mbox{Tr}(N^{2}) = 2\,(x^{2} - y)\,.
\ee

Pour des solutions \`a sym\'etrie sph\'erique on param\'etrise la
m\'etrique d'espace par

\be
\label{eq spherique}
- h_{ij}(r)\,dx^{i}\,dx^{j} = dr^{\,2} + H(r)\,d{\Omega}^{2}\;\;\; ; \;\;\;
H(r) \geq 0
\ee

\noindent ($x^{1}=r,\; x^{2}=\theta,\; x^{3}=\varphi$ et $d{\Omega}^{2}=
d{\theta}^{2}+{\sin}^{2}\,\theta\,d{\varphi}^{2}$), ceci donne par
int\'egration de (\ref{eq sigma})

\be
\label{eq sig,r}
{\sigma}_{,r} = H^{-1}(r)
\ee

\noindent (\`a une constante multiplicative pr\`es). En reportant
les \'equations (\ref{eq N2} $\rightarrow$
\ref{eq sig,r}) dans les \'equations (\ref{eq Rij}), celles-ci
s'\'ecrivent respectivement pour $i=j=2$ et pour $i=j=1$ comme
$$
1 - \frac{1}{2}\,H_{,rr} = 0
$$
$$
- H^{-1}\,H_{,rr} + \frac{1}{2}\,H^{-2}\,(H_{,r})^{2} =
\frac{1}{2}\,(x^{2} - y)\,H^{-2}\,.
$$
La premi\`ere \'equation donne

\be
\label{eq H}
H(r) = r^{2} - {\nu}^{2}
\ee

\noindent o\`u ${\nu}^{2}$ est une constante r\'eelle, et la
deuxi\`eme \'equation d\'etermine ${\nu}^{2}$ par

\be
{\nu}^{2} = \frac{1}{4}\,(x^{2} - y)\,.
\ee

\noindent On obtient alors par int\'egration de (\ref{eq sig,r})

\be
\sigma(r) = - \int^{\infty}_{r}\,\frac{dr^{\prime}}{r^{\prime\,2} -
{\nu}^{2}}\,.
\ee

\noindent On peut distinguer essentiellement deux classes de solutions
suivant que ${\nu}^{2} < 0$ ou ${\nu}^{2} \geq 0$\,:\\

\noindent {\Large \boldmath $\alpha)$} Les solutions pour lesquelles
$y > x^{2}\;({\nu}^{2} < 0)$ sont r\'eguli\`eres ainsi que la
\mbox{5-m\'etrique} (\ref{eq 3+2}), pour tout $r$ r\'eel; la topologie
spatiale est donc du type wormhole, avec 2 points \`a l'infini
spatial, $r \rightarrow \pm\infty$. La fonction ${\sigma(r)}$ correspondante

\be
\label{eq alpha}
{\sigma}(r) = -\frac{1}{\mu}\,\left(\frac{\pi}{2} -
\arctan\left(\frac{r}{\mu}\right)\right)
\ee

\noindent (o\`u ${\mu}^{2}=-{\nu}^{2}$, donc $\mu=(1/2)\sqrt{y-x^{2}}$) est
d\'efinie dans l'intervalle $I_{\sigma}=]-\frac{\pi}{\mu}, 0[$ quand
$r$ varie de $-\infty$ \`a $+\infty$. La 5-g\'eom\'etrie
correspondant au cas $x=a=0$ (wormhole sym\'etrique sans masse) a
\'et\'e \'etudi\'ee en d\'etail dans \cite{theseA} et \cite{AAGC1}.\\

\noindent {\Large \boldmath $\beta)$} Les solutions pour lesquelles
$y \leq x^{2}\;({\nu}^{2} \geq 0)$ sont singuli\`eres:\\
{\Large \boldmath $-$} en $r=\nu > 0$ si $y<x^{2}\;({\nu}^{2} > 0)$,
la fonction ${\sigma(r)}$ donn\'ee par

\be
\label{eq beta1}
{\sigma}(r) = \frac{1}{2\,\nu}\,\ln\left(\frac{r-\nu}{r+\nu}\right)
\ee

\noindent est d\'efinie dans l'intervalle $I_{\sigma}=]-\infty, 0[$
quand $r$ varie de $\nu$ \`a $+\infty$. Le cas $y=a=0$, $x>0$ ($\nu
=x/2$) correspond \`a la solution de Schwarzschild;\\
{\Large \boldmath $-$} ou en $r=0$ si $y=x^{2}\;({\nu}^{2}=0)$, la fonction
${\sigma(r)}$ donn\'ee par

\be
\label{eq beta2}
{\sigma}(r) = - \frac{1}{r}
\ee

\noindent est d\'efinie dans l'intervalle $I_{\sigma}=]-\infty, 0[$
quand $r$ varie de $0$ \`a $+\infty$.\\

\noindent C'est le param\`etre $b$, d\'efini par (\ref{eq b}), qui
d\'etermine si le champ \'electrique (voir (\ref{eq potentiel}))

\be
\label{eq elec}
E \equiv -\frac{1}{\lambda}{\left(\frac{{\lambda_{45}}}{{\lambda_{55}}}
\right)}_{,r}
\ee

\noindent est pr\'esent ou absent. Quand $b=0$, le champ
\'electrique est nul, et la 5-m\'etrique est diagonale; on voit
de (\ref{eq b}) que ceci n'est possible que pour les solutions pour
lesquelles $y \leq x^{2}/4$. En particulier les solutions ``wormhole"
($y>x^{2}$) sont toujours charg\'ees \'electriquement. Le param\`etre
$a$ peut \^{e}tre associ\'e \`a la charge scalaire de la solution, le
champ scalaire ${\lambda}_{55}$ \'etant \`a courte port\'ee dans le
cas $a=0$. Enfin la masse inerte associ\'ee \`a la solution est
donn\'ee par \cite{theseA}

\be
\label{eq masse}
M = \frac{1}{2G}\,\left(x-\frac{a}{2}\right)\,.
\ee

\n Terminons cette section en donnant l'expression de ${\lambda}_{55}$
tir\'ee de (\ref{eq matlamb})

\be
\label{eq lambda55}
{\lambda}_{55} = \left\{ \begin{array}{ll}
-\left(\left(a - \frac{\mbox{\normalsize $x$}}{\mbox{\normalsize $2$}}\right)
\frac{\mbox{\normalsize $\sinh\,q\sigma$}}{\mbox{\normalsize $q$}} +
\cosh\,q\sigma\right)\,\mbox{e}^{x\sigma/2} &
\mbox{si $y<\frac{\mbox{\normalsize $x^{2}$}}{\mbox{\normalsize $4$}}$} \\\\
-\left(\left(a-\frac{\mbox{\normalsize $x$}}{\mbox{\normalsize $2$}}\right)\,
\sigma + 1\right)\,
\mbox{e}^{x\sigma/2} & \mbox{si $y=
\frac{\mbox{\normalsize $x^{2}$}}{\mbox{\normalsize $4$}}$}\\\\
-\left(\left(a-\frac{\mbox{\normalsize $x$}}{\mbox{\normalsize $2$}}\right)
\frac{\mbox{\normalsize $\sin\,p\sigma$}}{\mbox{\normalsize $p$}} +
\cos\,p\sigma\right)\,\mbox{e}^{x\sigma/2} &
\mbox{si $y>\frac{\mbox{\normalsize $x^{2}$}}{\mbox{\normalsize $4$}}$}
                            \end{array}  \right.
\ee

\noindent o\`u $q$ et $p$ sont d\'efinis par

\be
\label{eq qp}
q \equiv \sqrt{\frac{x^{2}}{4}-y}\;\;\;;\;\;\;
p \equiv \sqrt{y-\frac{x^{2}}{4}}\,.
\ee

\noindent Ces expressions nous serons utiles pour l'\'etude de la
stabilit\'e effectu\'ee dans la section \ref{sec-discussion}.

\section{Petites oscillations monopolaires}
\label{sec-mono}
\setcounter{equation}{0}
On suppose que la solution 2-statique est excit\'ee de fa\c{c}on que
les modifications d\'ependant du temps apport\'ees \`a la m\'etrique
conservent \`a celle-ci sa sym\'etrie sph\'erique (c'est ce qu'on
appelle excitation monopolaire).

La forme g\'en\'erale que peut prendre une m\'etrique \`a sym\'etrie
sph\'erique  d\'ependant du temps est \footnote{Toutes les grandeurs
{\em dynamiques} (d\'ependant du temps) et ayant le m\^{e}me
symbole math\'ematique que celles de la section pr\'ec\'edente
(section \ref{sec-solution}), seront d\'esign\'ees par la lettre $(d)$ en
haut \`a gauche.}

\be
\label{eq generale}
ds^{\,2} =^{(d)}{\hspace{-2.mm}}{\lambda}_{ab}(r, t)\,dx^{a}\,dx^{b} +
2\,{\mu}_{a}(r, t)\,dx^{a}\,dr - l^{2}(r, t)\,dr^{2} -
g^{2}(r, t)\,r^{2}\,d{\Omega}^{2}\,.
\ee

\noindent Elle d\'epend de 7 fonctions inconnues, n\'eanmoins
on peut r\'eduire ce nombre \`a 4 \cite{AAGC2} moyennant
des transformations de coordonn\'ees convenablement choisies laissant,
en particulier, invariante la condition de p\'eriodicit\'e de la
cinqui\`eme coordonn\'ee $x^{5}$. On peut ainsi r\'eduire la m\'etrique
(\ref{eq generale}) \`a une forme partiellement diagonalis\'ee o\`u
les termes mixtes en $dr\,dx^{a}$ ne figurent plus et o\`u l'un des
coefficients de $dr^{2}$ ou de $d{\Omega}^{2}$ est ind\'ependant du
temps. Dans \cite{AAGC2}, nous avons opt\'e pour la deuxi\`eme
possibilit\'e \footnote{En effectuant la transformation de coordonn\'ees
\be
\label{eq tr1}
g^{\prime}(r^{\prime})=g(r,t)\,r,\;
{\theta}^{\prime}=\theta,\;{\varphi}^{\prime}=\varphi,\;t^{\prime}=t,\;
x^{\prime\,5}=x^{5}
\ee
o\`u $g^{\prime}(r^{\prime})$ est choisi a priori, on ram\`ene la
m\'etrique (\ref{eq generale}) \`a une forme o\`u le coefficient de
$d{\Omega}^{\prime\,2}$ ne d\'epend que de $r^{\prime}$. Effectuons
ensuite la transformation suivante
\be
\label{eq tr2}
r^{\prime\prime}=r^{\prime},\;{\theta}^{\prime\prime}={\theta}^{\prime},\;
{\varphi}^{\prime\prime}={\varphi}^{\prime},\;
x^{\prime\prime\,a}={\psi}^{a}(r^{\prime},t^{\prime})+
{\delta}^{a}_{5}\,x^{\prime\,5}
\ee
qui conserve bien la sym\'etrie sph\'erique et la condition de
p\'eriodicit\'e de $x^{5}$. Si, les 2 fonctions ${\psi}^{a}$ sont
choisies de fa\c{c}on \`a satisfaire le syst\`eme lin\'eaire suivant
$$
\frac{\partial{\psi}^{a}}{\partial r^{\prime}}-({\lambda}^{\prime\,-1}\,
{\mu}^{\prime})^{4}\,\frac{\partial{\psi}^{a}}{\partial t^{\prime}}=
({\lambda}^{\prime\,-1}\,{\mu}^{\prime})^{5}\,{\delta}^{a}_{5}
$$
(o\`u ${\mu}^{\prime}$ est la matrice colonne des
${\mu}^{\prime}_{a}(r^{\prime},t^{\prime})$ qui sont les \'el\'ements
$g^{\prime}_{1a}(r^{\prime},t^{\prime})$ de la m\'etrique issue de la
transformation (\ref{eq tr1})), la m\'etrique prend alors la forme
(\`a comparer avec (\ref{eq reduite}))
$$
ds^{\,2} =^{(d)}{\hspace{-2.mm}}
{\lambda}_{ab}(r^{\prime\prime},t^{\prime\prime})\,
dx^{\prime\prime\,a}\,dx^{\prime\prime\,b} -
 l^{\prime\prime\,2}(r^{\prime\prime},t^{\prime\prime})\,dr^{\prime\prime\,2}
 - g^{\prime\,2}(r^{\prime\prime})\,d{\Omega}^{\prime\prime\,2}\,.
$$
}, et montr\'e que la m\'etrique excit\'ee peut \^{e}tre
param\'etris\'ee par

\be
\label{eq reduite}
ds^{\,2} =^{(d)}{\hspace{-2.mm}}{\lambda}_{ab}(r, t)\,dx^{a}\,dx^{b} -
{\tau}^{-1}(r)\,\left[L(r, t)\,dr^{2} + H(r)\,d{\Omega}^{2}\right]
\ee

\noindent o\`u ${\tau}(r)$ et $H(r)$ sont donn\'es respectivement par
(\ref{eq tau}), (\ref{eq H}).

Au voisinage de la m\'etrique statique ((\ref{eq 3+2}), avec
{\LARGE $a$}$^{a}_{i} \equiv 0$ et (\ref{eq spherique})) la matrice
$^{(d)}{\hspace{-1.mm}}{\lambda}(r, t)$ et la fonction $L(r, t)$ peuvent
\^{e}tre \'ecrites

\bn
\label{eq ordre1}
^{(d)}{\hspace{-1.mm}}{\lambda}(r, t) & = & {\lambda}(r) + P(r, t)\\
\label{eq ordre2}
L(r, t) & = & 1 - A(r, t)
\en

\noindent avec

\be
\label{eq P}
P(r, t) = \left( \begin{array}{ll}
        B(r, t) & C(r, t)\\
        C(r, t) & R(r, t)
       \end{array} \right)
\ee

\noindent o\`u les fonctions $A(r, t),\;B(r, t),\;C(r, t),\;R(r, t)$
sont consid\'er\'ees comme des petites perturbations de la m\'etrique.
La lin\'earisation des \'equations d'Einstein dans le vide conduit
alors au syst\`eme

\bn
\label{eq 14}
2\hat{R}_{14} & \equiv & - \left({\lambda}^{-1}\dot{P}_{,r}
\right)^{5}_{\,5} +
\frac{1}{2}\left({\lambda}^{-1}\dot{P}{\lambda}^{-1}{\lambda}_{,r}
\right)^{5}_{\,5} -
\frac{1}{2}\left({\lambda}^{-1}\dot{P}{\lambda}^{-1}{\lambda}_{,r}
\right)^{4}_{\,4} \nonumber \\
& & \mbox{} + \frac{1}{2}\left({\lambda}^{-1}{\lambda}_{,r}\right)^{4}_{\,4}
\,\mbox{Tr}({\lambda}^{-1}\dot{P})
- \left[{\tau}H^{-1}({\tau}^{-1}H)_{,r} +
\frac{1}{2}\left({\lambda}^{-1}{\lambda}_{,r}\right)^{5}_{\,5}\right]
\dot{A} = 0 \nonumber \\\\ \nonumber \\
\label{eq 15}
2\hat{R}_{15} & \equiv & - \left({\lambda}^{-1}\dot{P}_{,r}
\right)^{4}_{\,5}
- \left({\lambda}^{-1}\dot{P}{\lambda}^{-1}{\lambda}_{,r}\right)^{4}_{\,5}
+ \frac{1}{2}\left({\lambda}^{-1}{\lambda}_{,r}\right)^{4}_{\,5}\,
\left[\mbox{Tr}({\lambda}^{-1}\dot{P}) + \dot{A}\right] = 0 \nonumber \\\\
\nonumber \\
\label{eq 11}
2\hat{R}_{11} & \equiv & - {\tau}^{-1}{\lambda}^{44}\ddot{A}
- {\tau}^{1/2}H^{-1}({\tau}^{-1/2}H)_{,r}\,A_{,r} -
\frac{1}{2}\,\mbox{Tr}({\lambda}^{-1}{P})_{,rr} \nonumber \\
& & \mbox{} - \mbox{Tr}\left[{\lambda}^{-1}{\lambda}_{,r}\,
({\lambda}^{-1}P)_{,r}
\right] -
\frac{1}{2}\,{\tau}^{-1}\,{\tau}_{,r}\,\mbox{Tr}({\lambda}^{-1}P)_{,r} = 0 \\
\nonumber \\
\label{eq 22}
2\hat{R}_{22} & \equiv & - 2A - \frac{\tau}{2}
\,({\tau}^{-1}H)_{,r}\,
\left[\mbox{Tr}({\lambda}^{-1}P) + A\right]_{,r} = 0 \\ \nonumber \\
\label{eq ab}
2\hat{R}_{ab} & \equiv & {\delta}^{4}_{a}\,{\delta}^{4}_{b}
\,\ddot{A} + {\tau}^{-1}{\lambda}_{ab}\,\ddot{P}_{55} +
\tau{H}^{-1}(HP_{ab,r})_{,r} \nonumber \\
& & \mbox{} - {\tau}({\lambda}_{,r}\,{\lambda}^{-1}P_{,r} +
P_{,r}\,{\lambda}^{-1}{\lambda}_{,r})_{ab} + {\tau}
({\lambda}_{,r}\,{\lambda}^{-1}P{\lambda}^{-1}{\lambda}_{,r})_{ab} \nonumber \\
& & \mbox{} + \frac{1}{2}\,{\tau}\,{\lambda}_{ab,r}\,
\left[\mbox{Tr}({\lambda}^{-1}P) + A\right]_{,r} = 0
\en

\noindent o\`u $(\,\dot{}\,) \equiv (\partial/\partial t)$, et $\hat{R}_{AB}$
est la perturbation du tenseur de Ricci ($^{(d)}{\hspace{-1.mm}}R_{AB}=R_{AB}+
\hat{R}_{AB}+\cdots$). Le coefficient original --provenant du calcul
direct-- de $A$ dans (\ref{eq 22}), \'egal \`a
$-\left(\tau({\tau}^{-1}H)_{,r}\right)_{,r}\,$, a \'et\'e remplac\'e par
$-2$ en se servant de l'\'equation
$$
R_{22} \equiv -\frac{1}{2}\left(\tau({\tau}^{-1}H)_{,r}\right)_{,r}+1=0
$$
Les \'equations (\ref{eq 14} $\rightarrow$ \ref{eq ab}), au nombre de
sept, ne sont pas toutes ind\'ependantes; on peut montrer directement
deux relations entre elles en utilisant certaines des \'equations
$R_{AB}=0$ --\`a savoir les \'equations $R_{ab}=0$--. Ces relations
sont
\bnn
2\,{\tau}H^{-1}(H\hat{R}_{14})_{,r}&=&\left(\tau\hat{R}_{11}+
2\,{\tau}H^{-1}\hat{R}_{22}+{\lambda}^{44}\hat{R}_{44}-
{\lambda}^{55}\hat{R}_{55}\right)_{,t}\,,\\
{\tau}H^{-1}(H\hat{R}_{15})_{,r}&=&\left({\lambda}^{45}\hat{R}_{55}+
{\lambda}^{44}\hat{R}_{45}\right)_{,t}\,;
\enn
qui sont bien les identit\'es de Bianchi (\ref{eq Bianchi}) relatives
\`a $\hat{R}_{A}\,^{B}$:
$$
\hat{R}_{A}\,^{B}\,_{;B} - \frac{1}{2}\,\hat{R}_{;A} \equiv 0
$$
pour $A=4, 5$. La troisi\`eme relation est l'identit\'e de Bianchi
pour $A=1$.

Afin de d\'ecoupler les \'equations (\ref{eq 14} $\rightarrow$ \ref{eq ab}),
remarquons que les expressions de $\hat{R}_{14}$ et $\hat{R}_{15}$
sont des d\'eriv\'ees totales par rapport au temps $t$. Annulons les
primitives correspondantes

\bn
2\mbox{\Large $\hat{\cal R}$}_{14} & \equiv & -
\left({\lambda}^{-1}P_{,r}\right)^{5}_{\,5} +
\frac{1}{2}\left({\lambda}^{-1}P{\lambda}^{-1}{\lambda}_{,r}\right)^{5}_{\,5}
- \frac{1}{2}\left({\lambda}^{-1}P{\lambda}^{-1}{\lambda}_{,r}
\right)^{4}_{\,4} \nonumber \\
& & \mbox{} + \frac{1}{2}\left({\lambda}^{-1}{\lambda}_{,r}\right)^{4}_{\,4}
\,\mbox{Tr}({\lambda}^{-1}P)
- \left[{\tau}H^{-1}({\tau}^{-1}H)_{,r} +
\frac{1}{2}\left({\lambda}^{-1}{\lambda}_{,r}\right)^{5}_{\,5}\right]A = 0
\nonumber \\\\ \nonumber \\
2\mbox{\Large $\hat{\cal R}$}_{15} & \equiv & -
\left({\lambda}^{-1}P_{,r}\right)^{4}_{\,5}
- \left({\lambda}^{-1}P{\lambda}^{-1}{\lambda}_{,r}\right)^{4}_{\,5}
+ \frac{1}{2}\left({\lambda}^{-1}{\lambda}_{,r}\right)^{4}_{\,5}\,
\left[\mbox{Tr}({\lambda}^{-1}P) + A\right] = 0 \nonumber \\
\en

\noindent et r\'ealisons la combinaison suivante

\be
\label{eq comb1}
-{\lambda}_{55}\,\mbox{\Large $\hat{\cal R}$}_{14}+
{\lambda}_{54}\,\mbox{\Large $\hat{\cal R}$}_{15}=0
\ee

\noindent Le premier membre de (\ref{eq comb1}) comporte les fonctions
$R_{,r}$, $C$, $R$, $A$, factoris\'ees par des fonctions
d'\'el\'ements de la matrice $\lambda$. Or, le coefficient de $C$
\'egal \`a
$$
\frac{1}{4}\left[\mbox{det}({\lambda}^{-1})\,{\lambda}_{55}{\lambda}_{54\,,r}+
{\lambda}_{55\,,r}\,{\lambda}^{54}-
\left({\lambda}^{-1}{\lambda}_{,r}\right)^{4}_{\,5}\right]\,,
$$
s'annule identiquement. On obtient alors

\be
\label{eq AR}
A = -2\,\frac{{\tau}^{-3/2}\,H^{2}\,({\tau}^{-1/2}\,R)_{,r}}
{({\tau}^{-2}\,H^{2}\,{\lambda}_{55})_{,r}}
\ee

\noindent qui constitue une relation entre les fonctions $A$ et $R$.
R\'ealisons ensuite la combinaison suivante

\bn
\label{eq comb2}
\hspace{-7mm}2\left(-{\lambda}_{55\,,r}\mbox{\Large $\hat{\cal R}$}_{14}+
{\lambda}_{54\,,r}\mbox{\Large $\hat{\cal R}$}_{15}\right)\hspace{-2mm}
& \equiv &
\hspace{-2mm}\underbrace{({\lambda}_{,r}{\lambda}^{-1}P_{,r})_{55}-\frac{1}{2}
({\lambda}_{,r}{\lambda}^{-1}P{\lambda}^{-1}{\lambda}_{,r})_{55}}_{\Xi}
\nonumber\\
\hspace{-2mm}& &\hspace{-2mm}\underbrace{\mbox{}+
\frac{1}{2}\left\{-({\lambda}_{,r}
{\lambda}^{-1}P{\lambda}^{-1}{\lambda}_{,r})_{55}+{\lambda}_{55\,,r}
\mbox{Tr}({\lambda}^{-1}P{\lambda}^{-1}{\lambda}_{,r})\right.}_{\Pi}
\nonumber \\ \nonumber \\
\hspace{-2mm}& &\hspace{-2mm}\underbrace{\left.\mbox{} +
\left[{\lambda}_{54\,,r}({\lambda}^{-1}{\lambda}_{,r})^{4}_{\,5}
-{\lambda}_{55\,,r}({\lambda}^{-1}{\lambda}_{,r})^{4}_{\,4}
\right]\,\mbox{Tr}({\lambda}^{-1}P)\right\}}_{\Pi} \nonumber \\ \nonumber \\
\hspace{-2mm}& &\hspace{-2mm}\mbox{} +\left[{\lambda}_{55\,,r}{\tau}H^{-1}
({\tau}^{-1}H)_{,r}+
\frac{1}{2}({\lambda}_{,r}{\lambda}^{-1}{\lambda}_{,r})_{55}\right]A=0
\nonumber \\
\en

\noindent que l'on a \'ecrite de fa\c{c}on que les 2 premiers termes ($\Xi$),
contenant les fonctions $B, C$, puissent se compenser avec des termes
correspondants provenant de $\hat{R}_{55}=0$, \`a savoir le troisi\`eme
et le quatri\`eme termes; les 4 termes suivants ($\Pi$) se simplifient par
$$
\Pi = -\frac{1}{2}\,\mbox{det}({\lambda}^{-1}{\lambda}_{,r})\,R\,;
$$
et dans le dernier terme, $A$ s'\'elimine \`a l'aide de (\ref{eq AR}).
Or, en ins\'erant $\hat{R}_{55}=0$ dans (\ref{eq comb2}), pour \'eliminer
les termes ($\Xi$) (autrement dit, pour \'eliminer les fonctions $B, C$),
on y introduit le terme
$$
\left[\mbox{Tr}({\lambda}^{-1}P) + A\right]_{,r}
$$
qui, lui aussi, contient les fonctions $B, C$ et que l'on doit
\'eliminer --tout entier-- en y ins\'erant $\hat{R}_{22}=0$. Finalement,
ces deux derni\`eres op\'erations (\`a savoir l'insertion de
$\hat{R}_{55}=0$, $\hat{R}_{22}=0$ dans (\ref{eq comb2})) se
r\'ealisent par la combinaison suivante (avec l'insertion de (\ref{eq
  AR}) pour exprimer $A$ en fonction de $R$):

\be
\label{eq comb3}
({\tau}^{-1}H)_{,r}\,\hat{R}_{55}+2\,{\tau}\,({\tau}^{-1}H)_{,r}\,
\left(-{\lambda}_{55\,,r}\,\mbox{\Large $\hat{\cal R}$}_{14}+
{\lambda}_{54\,,r}\,\mbox{\Large $\hat{\cal R}$}_{15}\right)+
{\lambda}_{55\,,r}\,\hat{R}_{22} = 0
\ee

\noindent qui conduit (en utilisant \'egalement l'\'equation
$\left({\lambda}^{-1}{\lambda}^{,i}\right)_{;i}=0$ qui d\'ecoule de
l'\'equation statique matricielle $R_{ab}=0$) \`a l'\'equation d'onde
pour la fonction $R$

\be
\label{eq onde1}
{\tau}^{-1}\,{\lambda}^{44}\,\ddot{R}-H^{-1}(H\,R_{,r})_{,r}-
2f^{-1}\,f_{,r}\,R_{,r}+\left[\mbox{det}({\lambda}^{-1}{\lambda}_{,r})+
{\tau}^{-1}\,{\tau}_{,r}\,f^{-1}\,f_{,r}\right]R = 0
\ee

\noindent o\`u on a pos\'e

\be
\label{eq f}
f \equiv -\,\frac{({\tau}^{-2}\,H^{2})_{,r}}
{({\tau}^{-2}\,H^{2}\,{\lambda}_{55})_{,r}}.
\ee

\noindent L'\'equation (\ref{eq onde1}) se transforme, \`a l'aide des
changements de variables et de coordonn\'ees suivant

\be
\begin{array}{lll}
\Phi & \equiv & f\,R\,, \\
du & \equiv & {\tau}^{-1/2}\,dr, \end{array}
\ee

\noindent en une \'equation du type de Klein-Gordon pour la fonction
$\Phi$ dans l'espace 3-dimensi-\\onnel de m\'etrique
$d{\ell}^{\,2}=du^{\,2}+{\tau}^{-1/2}\,H\,d{\Omega}^{\,2}$:

\be
\label{eq onde2}
{\lambda}^{44}\,\ddot{\Phi}-
{\tau}^{1/2}\,H^{-1}\,({\tau}^{-1/2}\,H\,{\Phi}_{,u})_{,u}+
\left[\mbox{det}({\lambda}^{-1}{\lambda}_{,u})+{\tau}^{-1/2}\,H^{-1}\,f^{-1}\,
({\tau}^{1/2}\,H\,f_{,u})_{,u}\right]{\Phi} = 0
\ee

Les solutions stationnaires de cette \'equation sont de la forme
$\Phi(r,t)=\Phi(r)\mbox{e}^{-i\,\omega\,t}$, o\`u $\Phi(r)$ et
$\omega$ sont solutions du probl\`eme aux valeurs propres associ\'e
\`a (\ref{eq onde2}) (remplacer $\ddot{\Phi}$ par
$-{\omega}^{2}\Phi$). La solution 2-statique consid\'er\'ee est
instable si $\omega$ est imaginaire, $\omega=ik$ (avec $k^{2}>0$).

Pour des raisons de commodit\'e nous utiliserons aussi dans
la section suivante consacr\'ee \`a l'\'etude de la stabilit\'e, les
\'equations aux valeurs propres suivantes o\`u le potentiel $\sigma$
est utilis\'e comme coordonn\'ee radiale

\bn
\label{eq onde3}
R_{,\sigma\sigma}+2f^{-1}\,f_{,\sigma}\,R_{,\sigma}-
(k^{2}\,{\tau}^{-1}\,H^{2}\,{\lambda}^{44}+y+x\,f^{-1}\,f_{,\sigma})R&=&0\\
\label{eq onde4}
-{\Phi}_{,\sigma\sigma}+(k^{2}\,{\tau}^{-1}\,H^{2}\,{\lambda}^{44}+y+
f^{-1}\,f_{,\sigma\sigma}+x\,f^{-1}\,f_{,\sigma}){\Phi}=0
\en

\noindent ($x$ et $y$ sont d\'efinis par (\ref{eq N} $\rightarrow$
\ref{eq b})).

\section{Discussion de la stabilit\'e}
\label{sec-discussion}
\setcounter{equation}{0}
Notre but est de chercher si, {\em en partant d'une petite perturbation de
la 5-m\'etrique \`a l'instant $t=0$, celle-ci peut cro\^{\i}tre
ind\'efiniment dans le temps}. Pour des perturbations stationnaires de
la forme $R(\sigma,t)=R(\sigma)$e$^{-i\,\omega \,t}$, etc., ceci
revient \`a chercher des solutions {\em physiquement acceptables} des
\'equations d'Einstein lin\'earis\'ees (\ref{eq 14} $\rightarrow$
\ref{eq ab}) pour $\omega=ik$ ($k>0$).

En relativit\'e g\'en\'erale, une d\'efinition naturelle des
perturbations physiquement acceptables d'une m\'etrique r\'eguli\`ere
est d'imposer
que la m\'etrique perturb\'ee soit aussi r\'eguli\`ere. Une condition
suffisante pour ceci est que les petites perturbations $R$,
$A$, $B$, $C$ soient born\'ees \cite{AAGC2}. Mais cette condition semble
trop faible dans le cas d'une m\'etrique admettant une singularit\'e
(c'est le cas des solutions statiques avec $y\leq x^{2}$ qui sont
singuli\`eres en $r=\nu$). Dans ce cas nous supposerons seulement que
les perturbations ne modifient pas le caract\`ere de la singularit\'e,
c'est \`a dire que la perturbation $A$ (\'equation (\ref{eq ordre2}))
ainsi que les perturbations relatives $P_{ab}/{\lambda}_{ab}$ des
diff\'erentes composantes de la matrice $\lambda$ restent
born\'ees (c'est d'ailleurs une condition n\'ecessaire pour que la
lin\'earisation (\ref{eq ordre1}), (\ref{eq ordre2}) ait un sens).
Une approche plus rigoureuse pourrait consister \`a
consid\'erer au lieu de
perturbations stationnaires des paquets d'ondes dont le support
n'inclut pas la singularit\'e en $r=\nu$.

Les perturbations peuvent a priori diverger aux bornes de l'intervalle
de variation de $r$ (]$-\infty,+\infty$[ dans le cas $y>x^{2}$,
]$\nu,+\infty$[ dans le cas $y\leq x^{2}$), ainsi qu'aux \'eventuels
z\'eros $r_{0}$ de la fonction $f$ figurant dans (\ref{eq
  onde1}). Nous allons d'abord \'etudier le comportement des
pertubations \`a l'infini et au voisinage de $r=r_{0}$, et montrer
qu'il exite un (et un seul) z\'ero $r_{0}$ de $f$ pour un certain
domaine de valeurs des param\`etres ($x, y, a$). Les conclusions sur
la stabilit\'e d\'ependront du comportement au voisinage de la borne
inf\'erieure ($-\infty$ ou $\nu$) qui sera \'etudi\'e cas par cas.

\subsection{Comportement asymptotique et au voisinage de \mbox{{\boldmath
$r=r_{0}$}}}
D\'eterminons d'abord le comportement asymptotique ($r \rightarrow
+\infty$ ou $\sigma \rightarrow 0_{-}$) des perturbations. La forme
asymptotique de l'\'equation (\ref{eq onde2}) ($\tau \rightarrow
1,\;{\lambda}^{44} \rightarrow 1$, \mbox{$f \rightarrow 1$})

\be
\label{eq Schr}
k^{2}\,\Phi - r^{2}\,(r^{2}\,{\Phi}_{,r})_{,r} \simeq 0
\ee

\noindent est la m\^{e}me que celle de l'\'equation de Schr\"{o}dinger
habituelle, conduisant au comportement asymptotique de $\Phi$
($r \rightarrow +\infty$, $\sigma \rightarrow 0_{-}$)

\be
\label{eq comp1}
\begin{array}{lll}
\Phi & \simeq & r^{-1}\,\left(c_{1}\,\mbox{e}^{k\,r}+
c_{2}\,\mbox{e}^{-k\,r}\right) \\
     & \simeq & \mbox{}-\sigma\,\left(c_{1}\,\mbox{e}^{-k/\sigma}+
c_{2}\,\mbox{e}^{k/\sigma}\right) \end{array}
\ee

\noindent ($c_{1}$ et $c_{2}$ sont des constantes r\'eelles). Le
param\`etre  $k$ \'etant suppos\'e positif, on doit exclure le comportement
en e$^{k\,r}$, ce qui impose l'annulation de la constante $c_{1}$,

\be
\label{eq compPhi}
\Phi \simeq -\,c_{2}\,\sigma\,\mbox{e}^{k/\sigma}\,.
\ee

\noindent Ce comportement se transmet aux perturbations $R$, $A$, $B$,
$C$ de la 5-m\'etrique.

Montrons maintenant que la perturbation $R$ diverge
en un z\'ero de la fonction $f$ (\ref{eq f}). Au voisinage d'un z\'ero
--simple comme on le verra-- ${\sigma}_{0} \in I_{\sigma}$ (ouvert)
de $f$, celle-ci se comporte comme $(\sigma-{\sigma}_{0})$ et,
de ce fait, l'\'equation (\ref{eq onde1}) comme

\be
\label{eq comp2}
R_{,\sigma\sigma}+\frac{2}{\sigma-{\sigma}_{0}}\,R_{,\sigma}-
\frac{x}{\sigma-{\sigma}_{0}}\,R \simeq 0\,.
\ee

\noindent La recherche de $R(\sigma)$ sous la forme
$(\sigma-{\sigma}_{0})^{s}$
fournit les valeurs $s_{1}=0$ et $s_{2}=-1$ et donc la solution
g\'en\'erale \`a (\ref{eq comp2}) diverge en ${\sigma}_{0}$ comme
$(\sigma-{\sigma}_{0})^{-1}$, ainsi que le rapport $R/{\lambda}_{55}\,$.
 De plus, il r\'esulte de la relation
(\ref{eq AR}) \'ecrite sous la forme

\be
\label{eq ARsigma}
 A=\frac{{\tau}^{-2}\,H^{2}}{({\tau}^{-2}\,H^{2}\,{\lambda}_{55})_{,\sigma}}\,
(x\,R-2\,R_{,\sigma})
\ee

\noindent que la perturbation $A$ ne peut qu'y diverger (sachant
que le d\'enominateur de (\ref{eq ARsigma}) ne peut pas diverger en
${\sigma}_{0}$).

Cherchons maintenant dans quelle(s) condition(s) puisse exister un
z\'ero $r=r_{0}$ ou $\sigma={\sigma}_{0}$ de $f$;
celle-ci peut s'\'ecrire comme
$$
f=-\,\frac{({\tau}^{-2}\,H^{2})_{,r}}
{({\tau}^{-2}\,H^{2}\,{\lambda}_{55})_{,r}}=
-\,\frac{({\tau}^{-2}\,H^{2})_{,\sigma}}
{({\tau}^{-2}\,H^{2}\,{\lambda}_{55})_{,\sigma}}
$$
Le point $r_{0}$ est z\'ero de $f$ si

\be
\label{eq mini}
\begin{array}{ccc}
({\tau}^{-2}\,H^{2})_{,r}(r_{0})&=&0 \\
{\lambda}_{55\,,r}(r_{0})&\neq&0 \end{array}
\ee

\noindent D\'eterminons d'abord $r_{0}$ pour $y<x^{2}$. Dans ce cas,

\be
\label{eq Hr}
H = r^{2}-{\nu}^{2}\qquad ,\qquad
\tau = \left(\frac{r-\nu}{r+\nu}\right)^{x/(2\nu)}
\ee

\n d'o\`u

\be
\label{eq ()()}
{\tau}^{-1}\,H = (r+\nu)^{1+x/(2\nu)}\,(r-\nu)^{1-x/(2\nu)}\,.
\ee

\n En d\'erivant (\ref{eq ()()}) on obtient

\be
\frac{({\tau}^{-1}\,H)_{,r}}{{\tau}^{-1}\,H}=\frac{1+x/(2\nu)}{r+\nu}
+\frac{1-x/(2\nu)}{r-\nu}
\ee

\n qui s'annule pour

\be
\label{eq r0}
r = r_{0} \equiv \frac{x}{2}
\ee

\noindent (ce r\'esultat se g\'en\'eralise ais\'ement au cas $y \geq
x^{2}$). Il s'agit bien d'une racine \mbox{d'ordre $1$} (c'est \`a dire qu'au
voisinage de $r_{0}$: $({\tau}^{-2}\,H^{2})_{,r} \simeq
(r-r_{0})$). Or, la coordonn\'ee radiale $r$ est limit\'ee
inf\'erieurement, pour les solutions de la classe {\Large \boldmath
$\beta$}, $y\leq x^{2}$, par $\nu$. Donc la valeur $r=r_{0}$
appartient au domaine de variation de $r$ dans les cas

\bnn
& &y>x^{2} \\
& &y\leq x^{2}\,,\quad \frac{x}{2}>\nu
\enn

\n conditions qui sont \'equivalentes \`a

\be
\label{eq condition}
\begin{array}{l}
y>x^{2} \\ 0 < y \leq x^{2}\,,\quad x>0\,. \end{array}
\ee

Si maintenant

\be
\label{eq 55,s}
{\lambda}_{55\,,\sigma}({\sigma}_{0}) = 0
\ee

\noindent (avec les conditions (\ref{eq condition})), ${\sigma}_{0}$
n'est plus un z\'ero de $f$, car dans ce cas le num\'erateur et
le d\'enominateur de $f$ s'annulent simultan\'ement.  Mais, vu
la d\'ependance (\ref{eq lambda55}) de ${\lambda}_{55}$ en $\sigma$,
${\sigma}_{0}$ ne peut \^{e}tre qu'un z\'ero simple (d'ordre 1) de
${\lambda}_{55\,,\sigma}$ et par cons\'equent de tout le d\'enominateur
de la fonction $f$, laquelle s'\'ecrira au voisinage de ${\sigma}_{0}$ comme

\be
f \simeq C_{1}\,\left[1+C_{2}\,(\sigma-{\sigma}_{0})\right]
\ee

\noindent (o\`u $C_{1}$, $C_{2}$ sont des constantes). En reportant
dans (\ref{eq onde3}) on obtient pour $\sigma
\sim {\sigma}_{0}$

\be
R_{,\sigma\sigma}+2\,C_{2}\,R_{,\sigma}+C_{3}\,R \simeq 0\,.
\ee

\noindent Pour $R(\sigma)$ de la forme $(\sigma-{\sigma}_{0})^{s}$, on
obtient $s_{1}=0$ et $s_{2}=1$ et donc la perturbation $R(\sigma)$ est
finie en ${\sigma}_{0}$; par contre $A(\sigma)$ y diverge comme
$(\sigma-{\sigma}_{0})^{-1}$

\be
A(\sigma \sim {\sigma}_{0}) \propto \frac{x\,R-2\,R_{,\sigma}}
{\sigma-{\sigma}_{0}} \propto \frac{1}{\sigma-{\sigma}_{0}}\,.
\ee

\noindent On conclut donc que, si les conditions (\ref{eq condition})
sont satisfaites, il existe un point $\sigma={\sigma}_{0}$,
correspondant par (\ref{eq alpha}), (\ref{eq beta1}) ou (\ref{eq beta2})
\`a $r=r_{0} \equiv x/2$, o\`u l'une, au moins, des perturbations
$R(\sigma)$ ou $A(\sigma)$ diverge. Ceci va nous permettre d'aborder
l'\'etude de la stabilit\'e dans les paragraphes suivants.

\subsection{Classe des solutions du type wormhole {\boldmath $y > x^{2}$}}

\addcontentsline{toc}{subsection}{A) Cas g\'en\'erique}

\addcontentsline{toc}{subsection}{B) Cas {\boldmath $x=a=0$}}
\vspace{0.2cm}
\noindent {\bf A) Cas g\'en\'erique} \\

La forme asymptotique de l'\'equation (\ref{eq onde2}) pour
$r \rightarrow - \infty$, est identique \`a sa forme asymptotique
pour $r \rightarrow + \infty$, d'o\`u le comportement suivant
pour $\Phi$ ($r \rightarrow - \infty$)

\be
\label{eq compwormdissy}
\Phi \simeq r^{-1}\,\left(c_{3}\,\mbox{e}^{k\,r}+
c_{4}\,\mbox{e}^{-k\,r}\right)\,.
\ee

\noindent Ce comportement se transmet aux perturbations $R$, $A$, $B$,
$C$. Le terme en e$^{-k\,r}$ doit \^{e}tre exclu afin que les rapports
$R/{\lambda}_{55}, \cdots$, restent finis quand $r \rightarrow -
\infty$; or, ayant d\'ej\`a annul\'e la constante $c_{1}$ dans
(\ref{eq comp1}), l'annulation de $c_{4}$ n'est possible que pour
certaines valeurs $k_{n}$ de $k$ qui constituent le spectre du
probl\`eme aux valeurs propres. Mais, d'autre part,  comme nous
l'avons vu, la fonction $f$ s'annule en $r=x/2$, point o\`u l'une au
moins des perturbations $R$ ou $A$ diverge. Cette divergence ne
pouvant pas \^{e}tre \'evit\'ee (nous ne disposons plus de param\`etre
libre \`a ajuster de fa\c{c}on \`a la supprimer), le probl\`eme aux
valeurs propres n'admet pas de solution. Il en r\'esulte que les
solutions statiques de cette classe sont stables.  \\

\noindent {\bf B) Cas {\boldmath $x=a=0$}} \\

Dans ce cas la 5-m\'etrique (\ref{eq 3+2}) prend la forme

\be
ds^{\,2} = \frac{r^{2}+{\mu}^{2}}{r^{2}-{\mu}^{2}}\,dt^{\,2}-dr^{\,2}-
(r^{2}+{\mu}^{2})\,(d{\theta}^{\,2}+\sin^{2}\theta\,d{\varphi}^{\,2})-
\frac{r^{2}-{\mu}^{2}}{r^{2}+{\mu}^{2}}\left(dx^{5}+\frac{2\,\mu\,r}
{r^{2}-{\mu}^{2}}\,dt\right)^{2}\,.
\ee

\n La 4-m\'etrique projet\'ee (\ref{eq projete}) est sym\'etrique dans
l'\'echange $r \leftrightarrow -r$, et
la masse associ\'ee (\ref{eq masse}) est nulle, d'o\`u le nom de
``wormhole sym\'etrique" donn\'e \`a cette m\'etrique statique. Il
r\'esulte de cette sym\'etrie que l'\'equation d'onde (\ref{eq onde2})
pour les petites perturbations, qui s'\'ecrit ici
$$
-[(r^{2}+\mu^{2})\,{\Phi}_{,r}]_{,r}+
\left[k^{2}\,(r^{2}-\mu^{2})+\frac{6\,\mu^{2}}{r^{2}}\right]
\,\Phi=0\,,
$$
est invariante aussi par la r\'eflexion $r \rightarrow -r$.  Les constantes
$c_{1}$, $c_{2}$ dans (\ref{eq comp1}) doivent donc se confondre
respectivement avec les constantes $c_{4}$, $c_{3}$ dans
(\ref{eq compwormdissy}) d'o\`u une seule condition de r\'egularit\'e
\`a l'infini spatial, qui entra\^{\i}ne l'annulation de la constante
$c_{1}$ dans (\ref{eq comp1}).

Dans ce cas $\tau = 1$ et $H^{2}{\lambda}_{55}=-r^{4}+\mu^{4}$, d'o\`u
(d'apr\`es (\ref{eq f}))
$$
f = \frac{r^{2}+\mu^{2}}{r^{2}}\,;
$$
donc contrairement au cas g\'en\'erique, la fonction $f$ n'a plus de
z\'ero mais un p\^{o}le double en $r=r_{0} \equiv 0$, au voisinage
duquel la fonction $\Phi$ se d\'eveloppe par

\be
\Phi = \frac{\bar{c}_{1}}{r^{2}}\left[1+
\frac{1}{6}\left(k^{2}+\frac{2}{\mu^{2}}\right)r^{2}+\cdots\right]+
\bar{c}_{2}\,r^{3}\left[1-\frac{1}{14}
\left(k^{2}-\frac{34}{3\mu^{2}}\right)r^{2}+\cdots\right]
\ee

\n o\`u $\bar{c}_{1}$ et $\bar{c}_{2}$ sont des constantes r\'eelles
dont le rapport ($\bar{c}_{1}/\bar{c}_{2}$) a \'et\'e choisi
de fa\c{c}on \`a annuler ${c}_{1}$ dans (\ref{eq comp1}). La fonction
$\Phi$ n'\'etant pas born\'ee en $r=0$, mais la fonction $R=f^{-1}\Phi$,
quand \`a elle, est bien born\'ee en $r=0$

\be
R = \frac{\bar{c}_{1}}{\mu^{2}}\left[1+
\frac{1}{6}\left(k^{2}-\frac{4}{\mu^{2}}\right)r^{2}+\cdots\right]+
\frac{\bar{c}_{2}\,r^{5}}{\mu^{2}}\left[1-\frac{1}{14}
\left(k^{2}+\frac{8}{3\mu^{2}}\right)r^{2}+\cdots\right]
\ee

\n et donc le rapport $R/{\lambda}_{55}$ aussi, \'etant donn\'e que
${\lambda}_{55}(r=0)=1$. De (\ref{eq AR}) on d\'eduit l'expression de $A$

\be
A = \frac{\bar{c}_{1}\mu^{2}}{6r^{2}}
\left(k^{2}-\frac{4}{\mu^{2}}\right)+\cdots
\ee

\n qui est singuli\`ere en $r=0$, sauf si on choisit

\be
\label{eq propre}
k^{2} = \frac{4}{\mu^{2}}
\ee

\n qui constitue donc une solution du probl\`eme aux valeurs propres
d\'efini dans l'introduction de cette section. On v\'erifie que pour
$k^{2}=4/\mu^{2}$ et ${c}_{1}=0$, les rapports $B(r)/{\lambda}_{44}$
et $C(r)/{\lambda}_{45}$ sont aussi born\'es en $r=0$. Les rapports
$P_{ab}/{\lambda}_{ab}$ ainsi que la fonction $A$ sont donc, pour
$k^{2}=4/\mu^{2}$ et ${c}_{1}=0$ ($\bar{c}_{1}/\bar{c}_{2}$
convenablement choisi), born\'es partout. On conclut que le wormhole
sans masse ($x=a=0$) a un mode propre qui cro\^{\i}t dans le temps
comme e$^{2\,t/\mu}$, et est ainsi instable.

Pour un wormhole de charge \'egale \`a celle de l'\'electron, la
constante de temps est donn\'ee par \footnote{Dans le syst\`eme
$c={\epsilon}_{0}=1$, on obtient en partant de (\ref{eq elec}) le
champ \'electrique asymptotique
$$
E \simeq \frac{2\,\mu}{\lambda}\,\frac{1}{r^{2}}\quad r \rightarrow \infty
$$
o\`u le coefficient de $1/r^{2}$ doit \^{e}tre \'egal \`a $e/4\pi$
($e$ \'etant la charge du wormhole suppos\'ee \'el\'ementaire), d'o\`u
en se servant de (\ref{eq lambda})
$$
{\mu}^{2}=\frac{e^{2}\,G}{4\,\pi}={\alpha}\,t_{P}^{2}\,.
$$
o\`u $t_{P}$ est le ``temps de Planck".}

\be
\frac{\mu}{2} = \frac{1}{2}\,\sqrt{\alpha}\,t_{P}
\ee

\n L'accroissement de la ``petite" perturbation
initiale est donc extr\^{e}mement
rapide. Ceci ne sanctionne pas l'int\'er\^{e}t de l'\'etude faite dans
\cite{theseA}, \cite{AAGC1} qui, rappelons le, n'est qu'une
approximation du cas o\`u la masse du wormhole --non nulle-- est
faible devant sa charge (pour une particule \'el\'ementaire telle que
l'\'electron, $\lambda M/Q \simeq 10^{-21}$).

\subsection{Classe des solutions pour lesquelles {\boldmath $y \leq x^{2}$}}
\label{subsec-Tom}
\vspace{0.2cm}
Vu l'expression de ${\lambda}_{55}$, \'equation (\ref{eq lambda55}),
qui diff\`ere suivant la relation d'ordre entre $y$ et $x^{2}/4$,
cette classe sera scind\'ee en deux sous-classes: $x^{2}/4<y \leq x^{2}$
et $y \leq x^{2}/4$. La classification dans chaque sous-classe se
fait de fa\c{c}on \`a faire appara\^{\i}tre les conditions (\ref{eq
  condition}).

\addcontentsline{toc}{subsection}{A) Sous-classe {\boldmath $x^{2}/4 < y
\leq x^{2}\;\;(x \neq 0)$}}
\vspace{0.5cm}

\noindent {\bf A) Sous-classe
{\boldmath $x^{2}/4 < y \leq x^{2}\;\;(x \neq 0)$}}\\

Dans ce cas la connaissance de $\tau = \mbox{e}^{x\,\sigma}$,
$H={\nu}^{2}/\sinh^{2}\nu\sigma$,
${\lambda}^{44}=-{\tau}^{-1}{\lambda}_{55}$ o\`u ${\lambda}_{55}$ est
donn\'ee par (\ref{eq lambda55}), permet de d\'eterminer les comportements
asymptotiques ($\sigma \rightarrow -\infty$) de la fonction $f$ et
des autres fonctions figurant dans l'\'equation (\ref{eq onde4})

\be
f \propto \frac{\mbox{e}^{-\,x\,\sigma/2}}{\sin(p\,\sigma+{\phi}_{1}
+{\phi}_{2})}
\ee
\be
\label{eq tph}
k^{2}\,{\tau}^{-1}\,H^{2}\,{\lambda}^{44} \simeq
\frac{k^{2}\,b}{p}\,(x^{2}-y)^{2}\sin(p\,\sigma+{\phi}_{1})\,
\mbox{e}^{(8\,\nu-3\,x)\,\sigma/2}
\ee
\be
\label{eq fph}
y+\frac{f_{,\sigma\sigma}}{f}+x\,\frac{f_{,\sigma}}{f} \simeq
\frac{2\,p^{2}}{\sin^{2}(p\,\sigma+{\phi}_{1}+{\phi}_{2})}
\ee

\noindent o\`u $0<{\phi}_{1}<\pi$, $0<{\phi}_{2}<\pi/2$ sont tels que

\be
\tan{\phi}_{1} \equiv \frac{2\,p}{2\,a-x}\;\;\;\;\;;\;\;\;\;\;
\tan{\phi}_{2} \equiv \frac{2\,p}{8\,\nu-3\,x}\,,
\ee

\noindent et on a suppos\'e $y<x^{2}$ (le cas $y=x^{2}$ sera trait\'e
plus loin). On peut distinguer deux cas suivant le signe de $8\,\nu-3\,x$
(\'equation (\ref{eq tph})).\\

\noindent \underline{Cas ($x<0$) ou ($(x^{2}/4)<y<7\,x^{2}/16$)}\\

On v\'erifie que dans ce cas

\be
8\,\nu-3\,x>0
\ee

\noindent le terme (\ref{eq tph}) est alors asymptotiquement n\'egligeable
devant le terme (\ref{eq fph}), dans ce cas l'\'equation (\ref{eq onde4})
prend la forme asymptotique suivante ($\sigma \rightarrow -\infty$)

\be
\label{eq p/sin}
{\Phi}_{,\sigma\sigma}-\frac{2\,p^{2}}
{\sin^{2}(p\,\sigma+{\phi}_{1}+{\phi}_{2})}\,\Phi \simeq 0
\ee

\noindent dont on connait la solution g\'en\'erale

\be
\label{eq p/sin2}
\Phi \simeq \frac{\bar{c}_{3}}{\tan(p\,\sigma+{\phi}_{1}+{\phi}_{2})}+
\bar{c}_{4}\,\left(\frac{1}{p}-\frac{\sigma}
{\tan(p\,\sigma+{\phi}_{1}+{\phi}_{2})}\right)\,.
\ee

\noindent On en tire l'expression asymptotique de $R=\Phi/f$

\be
\label{eq R4}
R \simeq
c_{3}\,\cos(p\,\sigma+{\phi}_{1}+{\phi}_{2})\,\mbox{e}^{x\,\sigma/2}+
c_{4}\,\left(\frac{\sin(p\,\sigma+{\phi}_{1}+{\phi}_{2})}{p}-
\sigma\,\cos(p\,\sigma+{\phi}_{1}+{\phi}_{2})\right)\,\mbox{e}^{x\,\sigma/2}\,.
\ee

\noindent Si $c_{4} \neq 0$, le rapport $R/{\lambda}_{55}$ n'est pas
born\'e, \`a cause de la pr\'esence du facteur $\sigma$ dans le second
terme de (\ref{eq R4}). Avec $c_{4}=0$, on obtient
($\sigma \rightarrow -\infty$)

\be
\label{eq cos/sin}
\frac{R}{{\lambda}_{55}} \propto \frac{\cos(p\,\sigma+{\phi}_{1}+{\phi}_{2})}
{\sin(p\,\sigma+{\phi}_{1})}\;\;\;\;\;\;\;\;\;;\;\;\;\;\;\;\;\;\;A \propto 1\,.
\ee

\noindent Le rapport $R/{\lambda}_{55}$ diverge donc p\'eriodiquement
quand $\sigma \rightarrow -\infty$. Mais ces divergences \'etant dues
au d\'ephasage constant ${\phi}_{2}$ entre le champ scalaire statique
${\lambda}_{55}$ et sa perturbation, le champ scalaire perturb\'e
$^{(d)}{\hspace{-1.5mm}}{\lambda}_{55}(r, 0)={\lambda}_{55}(r)+R(r)$, qui
peut \^{e}tre mis sous la forme

\be
^{(d)}{\hspace{-1.5mm}}{\lambda}_{55}=-\frac{b}{p}\,\sin(p\sigma+{\phi}_{1}
+\alpha)\,\mbox{e}^{(x\,\sigma/2)+\beta}
\ee

\n ($\alpha\,,\beta$ constantes) a le m\^{e}me comportement asymptotique
que ${\lambda}_{55}$, et est donc physiquement
acceptable. La contante $c_{1}$ dans (\ref{eq comp1}) \'etant d\'ej\`a
prise \'egale \`a z\'ero, l'annulation de la constante $c_{4}$ n'est
possible que pour certaines valeurs discr\`etes du param\`etre $k$,
qui constituent le spectre de valeurs propres du probl\`eme. Les
solutions statiques de la th\'eorie de Kaluza-Klein correspondant
au cas $x<0$ sont donc probablement instables (elles ne pourraient
\^{e}tre stables que si le spectre de valeurs propres \'etait vide),
tandis que les solutions correspondant au cas $x>0$ et
$(x^{2}/4)<y<7x^{2}/16$ sont stables car dans ce cas au probl\`eme
aux valeurs propres s'ajoute la divergence
appara\^{\i}ssant au point ${\sigma}_{0}$. \\

\noindent \underline{Cas $x>0$ et $7\,x^{2}/16<y(<x^{2})$}\\

Ce cas correspond \`a $8\nu-3x<0$; le terme (\ref{eq tph}) est
pr\'epond\'erant par rapport au terme (\ref{eq fph}) sauf au voisinage
des p\^{o}les de ce dernier, o\`u on retrouve le comportement fini
(\ref{eq R4}). Sinon, l'\'equation (\ref{eq onde4})
prend la forme asymptotique suivante ($\sigma \rightarrow -\infty$)

\be
\label{eq Phisin}
{\Phi}_{,\sigma\sigma}-\frac{k^{2}\,b}{p}\,(x^{2}-y)^{2}
\sin(p\,\sigma+{\phi}_{1})\,\mbox{e}^{(8\,\nu-3\,x)\,\sigma/2}\,\Phi
\simeq 0
\ee

\noindent o\`u le potentiel asymptotique (au sens de l'\'equation de
Schr\"{o}dinger) est une exponentielle modul\'ee par une fonction
sinuso\"{\i}dale. Consid\'erons
d'abord l'\'equation voisine, \`a potentiel purement exponentiel,
$$
{\Phi}_{,\sigma\sigma}-d^{2}\,\mbox{e}^{2\,c\,\sigma}\,\Phi=0
$$
($c<0$). La solution asymptotique ($\sigma \rightarrow -\infty$) g\'en\'erale
de cette \'equation est une combinaison lin\'eaire d'une biexponentielle
croissante et d'une biexponentielle d\'ecroissante
$$
\mbox{exp}\left[\pm\frac{d}{c}\,\mbox{e}^{c\,\sigma}\right]\,.
$$
Ceci nous conduit \`a penser (nous n'avons pu le prouver) que la
solution g\'en\'erale de (\ref{eq Phisin}) est combinaison lin\'eaire
d'une biexponentielle croissante et d'une biexponentielle
d\'ecroissante modul\'ees p\'eriodiquement. Les perturbations
relatives pourront alors \^{e}tre born\'ees, pour $\sigma \rightarrow
-\infty$, pour les valeurs discr\`etes de $k$ pour lesquelles la
biexponentielle croissante est absente, mais divergeront toujours au
voisinage du point ${\sigma}_{0}$ (les conditions (\ref{eq condition}) \'etant
remplies), d'o\`u la conclusion \`a la stabilit\'e. \\

\noindent \underline{Cas $x>0$ et $y=7\,x^{2}/16$}\\

Dans ce cas --avec $8\nu-3x=0$-- les deux termes (\ref{eq tph}),
(\ref{eq fph}) sont pr\'esents dans l'\'equation (\ref{eq onde4}),
dont la forme est voisine de (\ref{eq p/sin}), conduisant au
comportement (\ref{eq p/sin2}) au voisinage des p\^{o}les du
potentiel, et \`a une conclusion analogue \`a celle du cas $x>0$ et
$x^{2}/4<y<7x^{2}/16$, \`a savoir la stabilit\'e.\\

\n \underline{Cas $y=x^{2}$ $(x \neq 0)$}\\

Dans ce cas l'\'equation (\ref{eq tph}) est remplac\'ee par

\be
\label{eq tphs}
k^{2}\,{\tau}^{-1}\,H^{2}\,{\lambda}^{44} \simeq
\frac{k^{2}\,b}{p}\,\frac{\sin(p\,\sigma+{\phi}_{1})}{{\sigma}^{4}}\,
\mbox{e}^{-3\,x\,\sigma/2}\,;
\ee

\noindent l'\'equation (\ref{eq fph}) restant valable. Si $x<0$, le terme
(\ref{eq tphs}) est n\'egligeable devant le terme (\ref{eq fph});
pour $x>0$, il se produit le contraire. On a alors les conclusions
suivantes: pour $x<0$, les \'equations (\ref{eq p/sin} $\rightarrow$
\ref{eq cos/sin}) restent valables ainsi que la discussion qui suit,
ce cas est un probl\`eme aux valeurs propres et la conclusion est qu'il
est probablement instable; pour $x>0$, l'\'equation d'onde (\ref{eq onde4})
s'\'ecrit

\be
{\Phi}_{,\sigma\sigma}-\frac{k^{2}\,b}{p}\,\frac{\sin(p\,\sigma+{\phi}_{1})}
{{\sigma}^{4}}\,\mbox{e}^{-3\,x\,\sigma/2}\,\Phi \simeq 0\,,
\ee

\noindent du m\^{e}me type que (\ref{eq Phisin}) (le facteur
$1/{\sigma}^{4}$ variant lentement devant l'exponentielle). La discussion
qui a suivi l'\'equation (\ref{eq Phisin}) reste donc valable, d'o\`u la
conclusion \`a la stabilit\'e.\\

En r\'esum\'e, la sous-classe A): $x^{2}/4<y\leq x^{2}$ ($x \neq 0$) est
caract\'eris\'ee par la stabilit\'e de toutes les solutions pour
lesquelles $x>0$; les solutions pour lesquelles $x<0$ sont toutes
probablement instables.

\addcontentsline{toc}{subsection}{B) Sous-classe {\boldmath $y \leq x^{2}/4$}}

\addcontentsline{toc}{subsection}{B.1) {\boldmath $y \neq 0$}}
\vspace{0.5cm}

\noindent {\bf B) Sous-classe {\boldmath $y \leq x^{2}/4$}}\\

On doit distinguer respectivement les cas $y \neq 0$, $y=0$. \\

\noindent {\bf B.1)} {\boldmath $y \neq 0$} \\

\noindent \underline{Cas ($y<x^{2}/4$ ($a$ quelconque)) ou
($y=x^{2}/4$ et $a=x/2$)}\\

Dans ce cas, la fonction ${\lambda}_{55}$ se comporte pour $\sigma
\rightarrow -\infty$ comme

\be
{\lambda}_{55} \propto \mbox{e}^{(x \pm 2\,q)\,\sigma/2}
\ee

\noindent avec le signe inf\'erieur si $a\neq (x/2)+q$, et le signe
sup\'erieur si $a=(x/2)+q$. On obtient ensuite

\be
f \propto \mbox{e}^{-\,(x \pm 2\,q)\,\sigma/2}
\ee

\noindent d'o\`u

\bn
\label{eq fq}
\frac{f_{,\sigma}}{f}& \simeq &-\,\left(\frac{x}{2} \pm q\right)\,, \\
\label{eq tq}
k^{2}\,{\tau}^{-1}\,H^{2}\,{\lambda}^{44}& \simeq &
k^{2}\,\mbox{e}^{(8\,\nu-3\,x \pm 2\,q)\,\sigma/2}\,.
\en

\noindent Or, on v\'erifie que, pour $y \neq 0$ ($\forall x$)

\be
\label{eq nq}
8\,\nu-3\,x \pm 2\,q > 0
\ee

\noindent ceci permet de n\'egliger le terme (\ref{eq tq}) devant le
terme (\ref{eq fq}) dans l'\'equation (\ref{eq onde3}) qui prend la
forme asymptotique ($\sigma \rightarrow -\infty$)

\be
R_{,\sigma\sigma}-2\,\left(\frac{x}{2} \pm q\right)\,R_{,\sigma}+
\left(\frac{x}{2} \pm q\right)^{2}\,R \simeq 0
\ee

\noindent et admet pour solution

\be
R(\sigma) \simeq (c_{3}+c_{4}\,\sigma)\,
\mbox{e}^{-\,(x \pm 2\,q)\,\sigma/2}
\ee

\noindent et donc

\be
\frac{R}{{\lambda}_{55}} \simeq c_{3}+c_{4}\,\sigma
\ee

\noindent qui diverge sauf si $c_{4}=0$. Cette condition conduisant
\`a un probl\`eme aux valeurs propres, on en conclut que:
les solutions statiques, correspondant \`a ce cas, pour lesquelles
$y<0$ ou $x<0$, sont probablement instables, tandis que les solutions pour
lesquelles $y>0$ et $x>0$ sont stables, les perturbations relatives
divergent toujours au point ${\sigma}_{0}$.\\

\noindent \underline{Cas $y=x^{2}/4$ et $a \neq x/2$}\\

Ce cas ($q=0$) se distingue du cas pr\'ec\'edent par le comportement de
${\lambda}_{55}$ pour $\sigma \rightarrow -\infty$

\be
{\lambda}_{55} \simeq \left(\frac{x}{2}-a\right)\,\sigma\,
\mbox{e}^{x\,\sigma/2}
\ee

\noindent et donc de $f$ aussi, toutefois la d\'eriv\'ee logarithmique
de $f$ reste inchang\'ee

\be
\frac{f_{,\sigma}}{f} \simeq -\,\frac{x}{2}\,.
\ee

\noindent L'expression de $k^{2}{\tau}^{-1}H^{2}{\lambda}^{44}$ est la
m\^{e}me --avec $q=0$-- que dans (\ref{eq tq}) multipli\'ee par
$\sigma$. Or, l'in\'egalit\'e (\ref{eq nq}) \'etant toujours
satisfaite, ce terme reste toujours n\'egligeable devant
$f_{,\sigma}/f$, d'o\`u

\be
R_{,\sigma\sigma}-x\,R_{,\sigma}+\frac{x^{2}}{4}\,R \simeq 0
\ee

\noindent et

\be
\label{eq Rx}
R(\sigma) \simeq (c_{3}+c_{4}\,\sigma)\,\mbox{e}^{x\,\sigma/2}
\ee

\noindent avec maintenant un rapport

\be
\frac{R}{{\lambda}_{55}} \simeq
\frac{2\,c_{4}}{x-2\,a}
\ee

\noindent fini ($\forall\;c_{3}$ et $c_{4}$); l'annulation de $c_{4}$
n'est donc plus n\'ecessaire (on v\'erifie que la perturbation $A$
tend vers z\'ero pour $\sigma \rightarrow -\infty$). Il en r\'esulte
que si $x<0$, il existe pour tout $k^{2}<0$ une perturbation
physiquement acceptable (on peut toujours ajuster le rapport
$c_{4}/c_{3}$ de fa\c{c}on \`a annuler la constante $c_{1}$ dans
l'\'equation (\ref{eq comp1})); les solutions statiques de ce cas sont
donc instables. Si $x>0$, il appara\^{\i}t une divergence en
${\sigma}_{0}$ qui ne pourra \^{e}tre \'elimin\'ee que pour certaines
valeurs discr\`etes $k_{n}$ de $k$. Les solutions correspondant \`a
$x>0$ sont donc probablement instables.

En r\'esum\'e, le cas $y \neq 0$ de la sous-classe B): $y \leq
x^{2}/4$ est caract\'eris\'e
par la stabilit\'e de toutes les solutions pour lesquelles $y>0$ et $x>0$
sauf le cas particulier $y= x^{2}/4$ et $a \neq x/2$ ($x>0$) lequel
est probablement instable. Les solutions pour lesquelles $y<0$, $x<0$ sont
toutes probablement instables, \`a part le cas particulier $y= x^{2}/4$ et
$a \neq x/2$ ($x<0$), donc on sait qu'il est instable. \\

\addcontentsline{toc}{subsection}{B.2) {\boldmath $y=0$}}

\noindent {\bf B.2)} {\boldmath $y=0$}\\

\noindent \underline{Cas $y=a=0$ ($x \neq 0$)}\\

Dans ce cas la solution statique (\ref{eq 3+2})

\be
ds^{\,2}=\left(1-\frac{x}{\bar{r}}\right)\,dt^{\,2}-
\left(1-\frac{x}{\bar{r}}\right)^{-1}d{\bar{r}}^{\,2}-{\bar{r}}^{\,2}\,
d{\Omega}^{\,2}-(dx^{5})^{2}
\ee

\n (o\`u $\bar{r}\equiv r+(x/2)$) est simplement le produit tensoriel
de la m\'etrique de Schwarzschild (pour $x>0$) par le cercle de Klein.

Avec ${\lambda}_{55}=-1$ et $f=1$, l'\'equation (\ref{eq onde3}) prend
la forme

\be
\label{eq Schw}
R_{,\sigma\sigma}-k^{2}\,x^{2}\,\frac{\mbox{e}^{-\,4\,x\,\sigma}}
{\left(1-\mbox{e}^{-\,4\,x\,\sigma}\right)^{4}}\,R = 0\,.
\ee

\noindent Cette \'equation entra\^{\i}ne que $(R^{2})_{,\sigma\sigma}$ est
positif dans tout l'intervalle de variation de $\sigma\in
]\infty,0[$. Il en r\'esulte que la perturbation $R(\sigma)$ ne peut
pas \^{e}tre born\'ee simultan\'ement aux deux extr\'emit\'es de cet
intervalle. Ce cas est donc stable. En particulier la solution de
Schwarzschild ($x>0$) est stable vis-\`a-vis des perturbations
monopolaires de la 5-m\'etrique, comme d\'ej\`a observ\'e par
Tomimatsu \cite{Tom}. \\

\noindent \underline{Cas $y=0$, $a \neq 0$, $x<0$}\\

Quand $\sigma \rightarrow - \infty$, la fonction ${\lambda}_{55}$ se
comporte comme

\be
{\lambda}_{55} \propto \mbox{e}^{x\,\sigma}
\ee

\noindent d'o\`u l'on tire

\be
f \propto \mbox{e}^{-\,x\,\sigma}
\ee

\noindent et
\be
\label{eq enk2}
k^{2}\,{\tau}^{-1}\,H^{2}\,{\lambda}^{44} \propto k^{2}\,
\mbox{e}^{-\,3\,x\,\sigma}\,.
\ee

\noindent On peut n\'egliger le terme (\ref{eq enk2}) devant le terme
en $f$ dans (\ref{eq onde3}) qui s'\'ecrit pour $\sigma \rightarrow - \infty$

\be
R_{,\sigma\sigma}-2\,x\,R_{,\sigma}+x^{2}\,R \simeq 0\,.
\ee

\noindent La solution g\'en\'erale de cette \'equation est donn\'ee par

\be
R \simeq (c_{3}+c_{4}\,\sigma)\,\mbox{e}^{x\,\sigma}
\ee

\noindent d'o\`u l'expression du rapport

\be
\label{eq apSchw}
\frac{R}{{\lambda}_{55}} \simeq c_{3}+c_{4}\,\sigma
\ee

\noindent qui diverge sauf si $c_{4}=0$. Cette condition conduisant
\`a un probl\`eme aux valeurs propres, les solutions de ce cas sont
probablement instables.\\

\noindent \underline{Cas $y=0$, $a \neq 0$, $x>0$}\\

Dans ce cas l'expression de ${\lambda}_{55}$ se r\'eduit \`a

\be
{\lambda}_{55}=\frac{a-x}{x}-\frac{a}{x}\,\mbox{e}^{x\,\sigma}\,.
\ee

\n Pour \underline{$x=a$}, on trouve ($\sigma \rightarrow -\infty$)

\be
f \simeq 4\,\left(1-3\,\mbox{e}^{x\,\sigma}\right)
\ee

\noindent en substituant dans (\ref{eq onde3}), on obtient finalement
l'\'equation asymptotique

\be
R_{,\sigma\sigma}-6\,x\,\mbox{e}^{x\,\sigma}\,R_{,\sigma}-(k^{2}\,x^{2}-3)\,
x^{2}\,\mbox{e}^{x\,\sigma}\,R \simeq 0
\ee

\noindent qui admet

\be
R \simeq c_{3}+c_{4}\,\sigma
\ee

\noindent comme solution g\'en\'erale. D'o\`u l'on d\'eduit les
expressions du rapport $R/{\lambda}_{55}$

\be
\frac{R}{{\lambda}_{55}} \simeq (c_{3}+c_{4}\,\sigma)\,\mbox{e}^{-\,x\,\sigma}
\ee

\noindent qui diverge quelles que soient les valeurs des constantes $c_{3}$,
$c_{4}$, d'o\`u l'on conclut \`a la stabilit\'e.\\
Pour \underline{$x \neq a$}, on obtient

\be
f \propto
\frac{1}{4\,x-3\,a+3\,a\,\mbox{e}^{x\,\sigma}}\;\;\;\;\;(\forall \sigma)\,.
\ee

\noindent Commen\c{c}ons d'abord par traiter le cas
\underline{$x>a$}. Pour $\sigma \rightarrow - \infty$ on peut \'ecrire

\be
\label{eq fx>a}
f \propto 1-\frac{3\,a}{4\,x-3\,a}\,\mbox{e}^{x\,\sigma}
\ee

\noindent d'o\`u

\be
\frac{f_{,\sigma}}{f} \simeq
\frac{-\,3\,a\,x}{4\,x-3\,a}\,\mbox{e}^{x\,\sigma}.
\ee

\noindent On obtient de m\^{e}me ($\sigma \rightarrow - \infty$)

\be
\label{eq ktHl}
k^{2}\,{\tau}^{-1}\,H^{2}\,{\lambda}^{44} \simeq k^{2}\,x^{3}\,(x-a)\,.
\ee

\noindent En reportant ces deux derni\`eres \'equations dans (\ref{eq
  onde3}), celle-ci s'\'ecrit pour \mbox{$\sigma \rightarrow - \infty$} comme

\be
\label{eq Rx>a}
R_{,\sigma\sigma}+k^{2}\,x^{3}\,(a-x)\,R \simeq 0
\ee

\noindent et admet pour solution g\'en\'erale l'expression

\be
R(\sigma) \simeq c_{3}\,\mbox{e}^{m\,\sigma}+
c_{4}\,\mbox{e}^{-\,m\,\sigma}\;\;\;\;\;\;(x>a)
\ee

\noindent (avec $m \equiv kx\sqrt{x(x-a)}$) o\`u le terme en $c_{4}$ doit
\^{e}tre exclu pour que le rapport $R/{\lambda}_{55}$ soit fini quand
$\sigma \rightarrow - \infty$, ceci suppose que le probl\`eme aux
valeurs propres admet des solutions $k_{n}$; dans ce cas la fonction $A$ se
comporte comme

\be
A \simeq \frac{c_{3}}{3\,a-4\,x}\,(x-2\,m)\,\mbox{e}^{(m-x)\,\sigma}\,.
\ee

\noindent On v\'erifie ensuite que

\be
\frac{B}{{\lambda}_{44}} \sim \frac{C}{{\lambda}_{45}} \sim
c_{3}\,\mbox{e}^{(m-x)\,\sigma}\,.
\ee

\noindent La condition de convergence commune de $A$ et des rapports
$B/{\lambda}_{44}$, $C/{\lambda}_{45}$ n'est donc remplie que si

\be
\label{eq k2>}
k_{n}^{2} \geq \frac{1}{x\,(x-a)}\,.
\ee

\noindent En l'absence d'information sur les valeurs num\'eriques
des $k_{n}$, on peut conjecturer que ce cas est instable.

\noindent Traitons maintenant le cas \underline{$x<a$}. Si $4x-3a \neq
0$ ou

\be
x \neq \frac{3\,a}{4}
\ee

\noindent les \'equations (\ref{eq fx>a} $\rightarrow$ \ref{eq Rx>a})
restent valables, et on obtient pour $R$ l'expression ($\sigma
\rightarrow - \infty$)

\be
\label{eq Rmbar}
R \simeq c_{3}\,\cos\,\bar{m}\sigma+c_{4}\,\sin\,\bar{m}\sigma\;\;\;\;\;\;(x<a)
\ee

\noindent (avec $\bar{m} \equiv kx\sqrt{x(a-x)}$) qui est bien
born\'ee. N\'eanmoins, en poussant un peu les calculs, on montre que ($\sigma
\rightarrow - \infty$)

\bn
\label{eq tHs}
{\tau}^{-2}\,H^{2}& \simeq &x^{4}\,\left(1+4\,\mbox{e}^{x\,\sigma}\right)\,,\\
\label{eq tHls}
{\tau}^{-2}\,H^{2}\,{\lambda}_{55}& \simeq &x^{3}\,(a-x)\,
\left(1+\frac{4\,x-3\,a}{x-a}\,\mbox{e}^{x\,\sigma}\right)\,.
\en

\noindent En reportant les \'equations (\ref{eq Rmbar} $\rightarrow$
\ref{eq tHls}) dans (\ref{eq ARsigma}) on obtient pour $A(\sigma)$ l'expression

\be
A \simeq (\bar{c}_{3}\,\cos\,\bar{m}\sigma+\bar{c}_{4}\,\sin\,\bar{m}\sigma)\,
\mbox{e}^{-\,x\,\sigma}\;\;\;\;\;\;(x<a)
\ee

\noindent qui n'est pas born\'ee quand $\sigma$ tend vers $-\infty$.
Cette situation est inacceptable physiquement, et on conclut donc
\`a la stabilit\'e de ce cas de solutions.\\
Si maintenant

\be
x = \frac{3\,a}{4}
\ee

\noindent on aura ($\sigma \rightarrow - \infty$)

\be
f \propto \mbox{e}^{-\,x\,\sigma}
\ee

\noindent et

\be
\label{eq f,s}
\frac{f_{,\sigma}}{f} \simeq -\,x
\ee

\noindent En reportant (\ref{eq ktHl}), (\ref{eq f,s}) dans (\ref{eq
  onde3}) on obtient

\be
R_{,\sigma\sigma}-2\,x\,R_{,\sigma}+x^{2}\,[k^{2}\,x\,(a-x)+1]\,R
\simeq 0
\ee

\noindent puis, par int\'egration, on d\'eduit l'expression de $R$
($\sigma \rightarrow - \infty$)

\be
\label{eq Rms}
R \simeq (c_{3}\,\cos\,m^{\prime}\sigma+c_{4}\,\sin\,m^{\prime}\sigma)\,
\mbox{e}^{x\,\sigma}
\ee

\noindent (avec $m^{\prime} \equiv kx^{2}/\sqrt{3}$) qui est bien
born\'ee quelles que soient les constantes $c_{3}$, $c_{4}$. La
pr\'esence du facteur e$^{x\,\sigma}$ dans $R$ fait que $A$
est maintenant aussi born\'ee

\be
\label{eq Ams}
A \simeq \bar{c}_{3}\,\cos\,m^{\prime}\sigma+
\bar{c}_{4}\,\sin\,m^{\prime}\sigma
\ee

\noindent On v\'erifie que les perturbations $B$ et $C$ sont aussi born\'ees
pour $\sigma \rightarrow - \infty$. Quel que soit $k>0$, on peut donc
toujours choisir une constante d'int\'egration de fa\c{c}on \`a
annuler ${c}_{1}$ dans l'\'equation (\ref{eq comp1}) et \`a rendre les
perturbations finies. Ce sous-cas particulier est donc instable.

Le cas ($y=0$, $a \neq 0$, $x>0$) est caract\'eris\'e par la
stabilit\'e de toutes les solutions pour lesquelles $x \leq a$, sauf la
solution correspondant \`a $x=3a/4$ laquelle est instable; les
solutions correspondant \`a $x>a$ sont probablement instables.\\

\noindent \underline{Cas $y=x=0$}\\

Dans ce cas $b=\pm a$, $\sigma=-1/r$, et la 5-m\'etrique statique est
de la forme

\be
ds^{\,2}=\left(1-\frac{a}{r}\right)^{-1}dt^{\,2}-dr^{\,2}-r^{2}\,
d{\Omega}^{\,2}-\left(1-\frac{a}{r}\right)\left(dx^{5}\pm
\frac{a}{r-a}\,dt\right)^{2}\,.
\ee

\n On obtient

\be
f = \frac{4}{4+3\,a\,\sigma}\,,
\ee

\n et l'\'equation d'onde (\ref{eq onde4}) prend la forme

\be
\label{eq Fyx0}
-\,{\Phi}_{,\sigma\sigma}+\left(\frac{k^{2}\,(1+a\,\sigma)}{{\sigma}^{4}}+
\frac{18\,a^{2}}{(4+3\,a\,\sigma)^{2}}\right)\,\Phi = 0\,.
\ee

\noindent On peut  distinguer deux cas. Pour \underline{$a<0$}, la
fonction
$$
\Phi=-4\,\frac{1+a\,\sigma}{4+3\,a\,\sigma}\,\frac{R}{{\lambda}_{55}}
$$
doit \^{e}tre born\'ee comme le rapport $R/{\lambda}_{55}$, ce qui est
impossible vu que $({\Phi}^{2})_{,\sigma\sigma}$ est positif dans tout
l'intervalle de variation de $\sigma$ (comparer avec le cas $y=a=0$ $(x
\neq 0)$, \'equation (\ref{eq Schw})); ce cas est donc stable. Pour
\underline{$a>0$} la fonction $f$ a un p\^{o}le en $\sigma=-4/(3a)$,
et il est pr\'ef\'erable de raisonner sur l'\'equation (\ref{eq
  onde3}) qui prend la forme

\be
R_{,\sigma\sigma}-\frac{6\,a}{4+3\,a\,\sigma}\,R_{,\sigma}-
\frac{k^{2}\,(1+a\,\sigma)}{{\sigma}^{4}}\,R = 0\;\;\;\;\;(\forall \sigma)
\ee

\noindent dont la solution g\'en\'erale est donn\'ee par

\be
\label{eq Ryx0}
R(\sigma) = c_{3}\,S_{1}({\sigma}^{-1})+c_{4}\,{\sigma}^{3}\,
S_{2}({\sigma}^{-1})
\ee

\noindent o\`u $S_{1}$ et $S_{2}$ sont des s\'eries enti\`eres en
${\sigma}^{-1}$ dont les termes ind\'ependants sont non nuls. Le
rapport $R/{\lambda}_{55}$ est born\'e pour les valeurs $k_{n}$ de $k$
telles que $c_{4}=0$, et ce cas est probablement instable.

\section{Conclusion}
\setcounter{equation}{0}
R\'esumons les r\'esultats de cette \'etude rendue complexe par la
diversit\'e des cas de figure. Nous avons montr\'e la stabilit\'e,
vis-\`a-vis des excitations monopolaires, de toutes les solutions
r\'eguli\`eres \`a
sym\'etrie sph\'erique (classe des wormholes $y>x^{2}$) de la
th\'eorie de Kaluza-Klein, \`a l'exception du wormhole sym\'etrique
charg\'e sans masse ($x=a=0$, ($y>0$)), lequel est instable.
Parmi les solutions singuli\`eres ($y \leq x^{2}$), la sous-classe
$y>0,\;x>0$ est stable, \`a l'exception du cas $y=x^{2}/4,\;a\neq x/2
>0$ (probablement instable), les autres solutions de cette classe \'etant
probablement instable, \`a l'exception des cas $y=x^{2}/4,\;a\neq x/2
<0$ (certainement instable); $y=0,\;0<x\leq a$ (stable si $a\neq
4x/3$, instable si $a=4x/3$); $y=a=0$ (solution de Schwarzschild,
stable); $y=x=0,\;a<0$ (stable). Le tableau suivant rassemble ces r\'esultats:

\newpage
\n {\Large $\alpha)$} \underline{$y>x^{2}$} $\longrightarrow$
{\em stable}\\
\mbox{}\hspace{3.cm}sauf $x=a=0$ $\longrightarrow$ {\em instable} \\\\\\

\n {\Large $\beta)$} \underline{$y\leq x^{2}$} \\\\

\n $\beta1)$ $y>0,\;x>0$ $\longrightarrow$ {\em stable}\\
\mbox{}\hspace{4.cm}$\left(\mbox{{\em probablement instable}}\; \mbox{pour}\;
  y=x^{2}/4\; \mbox{et}\; a\neq x/2\right)$\\

\n $\beta2)$ $y>0,\;x<0$ $\longrightarrow$ {\em probablement instable}\\
\mbox{}\hspace{4.cm}$\left(\mbox{{\em instable}}\; \mbox{pour}\;
  y=x^{2}/4\; \mbox{et}\; a\neq x/2\right)$\\

\n $\beta3)$ $y<0$ $\longrightarrow$ {\em probablement instable}\\

\n $\beta4)$ $y=0$ \hspace{1.cm}----- $\;\;0<x\leq a$ $\longrightarrow$
{\em stable}\\
\mbox{}\hspace{4.cm}sauf $a=4x/3$ $\longrightarrow$ {\em instable}\\

\mbox{}\hspace{2.2cm}----- $\;\;0<x,\;a<x$ $\longrightarrow$
{\em probablement instable}\\
\mbox{}\hspace{4.cm}sauf $a=0$ $\longrightarrow$ {\em stable}\\

\mbox{}\hspace{2.2cm}----- $\;\;x\leq 0$ $\longrightarrow$
{\em probablement instable}\\
\mbox{}\hspace{4.cm}sauf $a=0$ $\longrightarrow$ {\em stable}\\
\mbox{}\hspace{4.cm}sauf $x=0,\;a<0$ $\longrightarrow$ {\em stable}

\newpage

Abstraction faite des cas: $y=x^{2}/4,\;x>0\;(a\neq x/2)$ et $y=0$, on
remarque que le domaine de stabilit\'e \`a 3 dimensions dans l'espace
des param\`etres ($x, y, a$) co\"{\i}ncide avec le domaine ($y>x^{2}$
ou $0<y\leq x^{2},\;x>0$) d'existence d'un z\'ero de la fonction $f$
(\'equation (\ref{eq f})). Or annuler $f$ revient, d'apr\`es (\ref{eq
  mini}), \`a extr\'emiser $4\,\pi\,{\tau}^{-1}\,H$ qui est, d'apr\`es
(\ref{eq 3+2}) et (\ref{eq H}), l'aire d'une sph\`ere de rayon
$r$. Donc les solutions pour lesquelles il existe une sph\`ere d'aire
minimale (situ\'ee \`a l'ext\'erieur de l'horizon $r=\nu$ dans le cas
$0<y\leq x^{2},\;x>0$) sont stables (\`a l'exception du wormhole
sym\'etrique sans masse et du cas $y=x^{2}/4,\;x>0\;(a\neq x/2)$).

Comme il ressort de cette \'etude, la m\'ethode analytique ainsi
appliqu\'ee pr\'esente de s\'erieuses limites. Remarquons que
les cas qui ont \'et\'e consid\'er\'es comme \'etant probablement
instables sont des probl\`emes aux valeurs propres, auxquels, seule
l'analyse num\'erique peut {\em trancher}. Vu les diff\'erences appr\'eciables
entre ces cas, et la complexit\'e de certains d'entre
eux, en particulier de la sous-classe $x^{2}/4<y\leq x^{2}$ ($x\neq 0$),
nous n'avons pas pu \'elaborer un programme de calcul num\'erique
unique et uniforme.

Le cas des excitations non radiales comporte plus de fonctions --huit
dans le cas des excitations dipolaires-- ind\'ependantes, contre
quatre dans notre cas, ce qui donnerait au probl\`eme de {\em d\'ecouplage}
des \'equations d'Einstein lin\'earis\'ees, ainsi qu'au probl\`eme de
{\em recherche de points singuliers} o\`u peuvent appara\^{\i}tre des
divergences des unes ou des autres fonctions ind\'ependantes, une
allure plus complexe. L'\'etude de la stabilit\'e, g\'en\'eralis\'ee
aux cas des excitations non radiales, ne peut \^{e}tre envisag\'e que
num\'eriquement; nous conjecturons que certaines solutions, comme
les wormholes avec masse ($y>x^{2}$, $a\neq 0$)
\cite{AAGC2} continueront, en vertu de leur topologie sp\'eciale, \`a
\^{e}tre stables vis-\`a-vis des petites excitations, m\^{e}me multipolaires.

Comparons maintenant les r\'esultats de notre \'etude avec ceux de Tomimatsu
\cite{Tom}. Il se restreint dans son \'etude au cas o\`u le champ
\'electrique $E$ est nul, il prend donc (voir (\ref{eq b})) $b\equiv
0$; autrement dit, il se restreint au cas des solutions singuli\`eres
pour lesquelles $y\leq x^{2}/4$. En fait, on peut montrer qu'il ne
consid\`ere que le sous-cas $y<x^{2}/4$ pour lequel

\be
\label{eq conditionT}
\begin{array}{l} y\leq 0\quad,\quad x>0\quad,\quad
a=\frac{\mbox{\normalsize $x$}}{\mbox{\normalsize $2$}}-q\\
                 y>0\quad,\quad x<0\quad,\quad
a=\frac{\mbox{\normalsize $x$}}{\mbox{\normalsize $2$}}+q\,. \end{array}
\ee

En effet, Tomimatsu param\'etrise sa m\'etrique par

\be
\label{eq Tom}
ds^{\,2}=B_{T}(r_{_{T}})\,dt^{\,2}-A_{T}(r_{_{T}})\,{dr_{_{T}}}^{2}
-{r_{_{T}}}^{2}\,d{\Omega}^{\,2}-R^{2}_{T}(r_{_{T}})\,(dx^{5})^{2}
\ee

\n o\`u

\be
{\lambda}_{44}\equiv B_{T}={\psi}^{p_{\mbox{}+}} \quad,\quad
{\lambda}_{55}\equiv -R^{2}_{T}=-{\psi}^{p_{\mbox{}-}}
\ee
\be
\label{eq l/r}
A_{T}=\left(2\,q_{_{T}}\,{\lambda}_{_{T}}^{-1}\,{r_{_{T}}}^{2}\,
{\psi}_{,\,r_{_{T}}}\right)^{2}\quad,\quad
\frac{{\lambda}_{_{T}}}{r_{_{T}}}=\psi\,
\left({\psi}^{-q_{_{T}}}-{\psi}^{q_{_{T}}}\right)
\ee

\n avec
\bnn
& &p_{\mbox{}\pm}\equiv 1\pm \sqrt{1+{\kappa}_{_{T}}}\,,\quad
({\kappa}_{_{T}}>-1)\\
& &q_{_{T}}\equiv \sqrt{1+\frac{{\kappa}_{_{T}}}{4}}\,.
\enn

\n (${\lambda}_{_{T}}$ et ${\kappa}_{_{T}}$ sont suppos\'es de
m\^{e}me signe). Or, pour $b=0$, on a
${\lambda}_{44}=$ e$^{(x-a)\,\sigma}$, \mbox{${\lambda}_{55}=-$e$^{a\,\sigma}$}
avec $a=(x/2)\pm q$ d'o\`u

\be
\label{eq psi2}
{\psi}^{p_{\mbox{}+}}\,{\psi}^{p_{\mbox{}-}}={\psi}^{2}=\mbox{e}^{x\,\sigma}
\,.
\ee

\n On calcule ensuite ${\psi}^{p_{\mbox{}+}}$ par
$$
{\psi}^{p_{\mbox{}+}}=\mbox{e}^{x\,(p_{\mbox{}+})\,\sigma/2}\;
(=\mbox{e}^{(x-a)\,\sigma})
$$
d'o\`u

\be
\sqrt{1+{\kappa}_{_{T}}}=\mp \frac{2q}{x}\quad (x=\mp |x|)
\ee
puis
\be
{\kappa}_{_{T}}=-\frac{4y}{x^{2}}\quad,\quad q_{_{T}}=\mp
\frac{2\nu}{x}\,.
\ee

\n En substituant $\psi =$e$^{x\,\sigma/2}$ et $q_{_{T}}=\mp 2\nu/x$
dans la deuxi\`eme relation de (\ref{eq l/r}) on calcule $r_{_{T}}$ par

\be
\label{eq rt}
r_{_{T}}=\mp \frac{{\lambda}_{_{T}}}{2\nu}\,\left(r+\nu\right)^{(2\nu +x)/4\nu}
\left(r-\nu\right)^{(2\nu -x)/4\nu}
\ee

\n qui doit \^{e}tre \'egal \`a $({\tau}^{-1}H)^{1/2}$ (\'equation
(\ref{eq ()()})), d'o\`u

\be
{\lambda}_{_{T}}=\mp 2\,\nu\,.
\ee

\n Dans le cas $\kappa_{_{T}}\neq 0$, Tomimatsu suppose que la borne
inf\'erieure de sa variable $r_{_{T}}$ est \'egale \`a z\'ero; l'on
doit donc imposer la condition $x<2\nu$ pour assurer que le second
membre de (\ref{eq rt}) soit monotone et croissant et que, pour
$r=\nu$, $r_{_{T}}=0$. D'o\`u les conditions

\be
\label{eq T2}
\begin{array}{cll}
\kappa_{_{T}}>0&\Rightarrow &y<0\quad,\quad x>0\quad,\quad
a=\frac{\mbox{\normalsize $x$}}{\mbox{\normalsize $2$}}-q\\
-1<\kappa_{_{T}}<0&\Rightarrow &0<y<x^{2}/4\quad,\quad x<0\quad,\quad
a=\frac{\mbox{\normalsize $x$}}{\mbox{\normalsize $2$}}+q\,. \end{array}
\ee

\n Dans le cas $\kappa_{_{T}}=0$, il prend $r_{_{T\,(min)}}=
{\lambda}_{_{T}}$; dans ce cas l'on doit poser $x=2\nu$. On obtient ensuite
$r_{_{T\,(min)}}=2\nu$, d'o\`u ${\lambda}_{_{T}}=+2\nu$ et donc $x>0$
et $a=0$ qui est bien la solution de Schwarzschild. Or, pour
$\kappa_{_{T}}=0$, on a

\be
\label{eq T3}
y = 0 \qquad (x>0)\,.
\ee

\n Les conditions (\ref{eq T2}), (\ref{eq T3}) rassembl\'ees redonnent
l'ensemble des sous-cas (\ref{eq conditionT}) trait\'es par Tomimatsu.
 Dans l'espace des param\`etres
($x, y, a$), l'\'etude de Tomimatsu n'a concern\'e que les portions de
surfaces $a=(x/2)-q$ et $a=(x/2)+q$ ($y<x^{2}/4$) pour lesquelles les
conditions (\ref{eq conditionT}) sont remplies.

Tomimatsu d\'efinit la fonction de perturbation ${\phi}_{_{T}}$ de
${\lambda}_{55}$ par

\be
\label{eq pphi}
^{(d)}{\hspace{-1.5mm}}{\lambda}_{55}=-R_{T}^{2}-2\,R_{T}\,{\phi}_{_{T}}\,.
\ee

\n En posant ${\cal R}\equiv R/{\lambda}_{55}$, on obtient

\be
\label{eq pphi1}
{\phi}_{_{T}}=\frac{1}{2}\,\sqrt{|{\lambda}_{55}|}\,{\cal R}\,.
\ee

\n Il introduit ensuite la fonction d'onde ${\Phi}_{T}$ reli\'ee \`a
${\phi}_{_{T}}$ par

\bn
\label{eq Pphi}
{\Phi}_{T}&\equiv &\frac{r_{_{T}}\,R_{T}^{1/2}}{(R_{T}+r_{_{T}}\,
R_{T,\,r_{_{T}}}/2)}\,{\phi}_{_{T}}\\
\label{eq Pphi1}
 &=&\frac{1}{2}\,\left({\tau}^{-2}\,H^{2}\,|{\lambda}_{55}|\right)^{1/4}\,
(f\,{\lambda}_{55})\,{\cal R}\,.
\en

Dans les cas o\`u nous avons observ\'e un d\'esaccord avec Tomimatsu,
\`a savoir les cas ($\kappa_{_{T}}\neq 0$):

\bn
& &y<0\,,\;x\geq 0>a\\
& &y>0\,,\;x<0\,,\;a<0
\en

\n qui sont certainement instables pour lui (spectre continu) et
seulement probablement instables pour nous (spectre discret), les
fonctions ${\cal R}$, $f{\lambda}_{55}$,
${\tau}^{-2}\,H^{2}\,|{\lambda}_{55}|$ se comportent pour \mbox{$\sigma
\rightarrow -\infty$} comme
$$
{\cal R} \simeq c_{3}+c_{4}\,\sigma \quad,\quad f\,{\lambda}_{55}
\propto 1
$$
$$
{\tau}^{-2}\,H^{2}\,|{\lambda}_{55}| \propto \mbox{e}^{(8\nu-3x-2q)\sigma/2}
$$
(avec $8\nu-3x-2q>0$ pour $y\neq 0$). On obtient

\be
\label{eq compPhiT}
{\Phi}_{T} \simeq \mbox{e}^{(8\nu-3x-2q)\sigma/8}\,(\bar{c}_{3}+
\bar{c}_{4}\,\sigma)\,.
\ee

\n Le comportement (\ref{eq compPhiT}) est bien celui qu'a obtenu Tomimatsu:
$$
{\Phi}_{T} \simeq {c}_{1}\,(r_{_{T}}^{*})^{1/2}\,(1+{c}_{2}\,\ln r_{_{T}}^{*})
$$
avec, quand $\sigma \rightarrow -\infty$ :

\be
r_{_{T}}^{*} \propto \mbox{e}^{(8\nu-3x-2q)\sigma/4}
\ee

\n o\`u $r_{_{T}}^{*}$ est la coordonn\'ee radiale de l'\'equation aux
valeurs propres de Tomimatsu. On voit ainsi que quand
$\sigma \rightarrow -\infty$, ${\cal R}$ diverge, mais la solution
g\'en\'erale de ${\Phi}_{T}$ est bien born\'ee \footnote{La fonction
de perturbation ${\phi}_{_{T}}$ est non
  born\'ee.}, d'o\`u la conclusion \`a l'instabilit\'e de Tomimatsu
car il suffit de choisir convenablement le rapport
$\bar{c}_{3}/\bar{c}_{4}$ pour rendre le comportement \`a l'infini
spatial de sa fonction d'onde ${\Phi}_{T}$ born\'e; Tomimatsu a donc
bien \'etudi\'e le probl\`eme aux valeurs propres de Schr\"{o}dinger
pour la fonction ${\Phi}_{T}$, et non celui de la relativit\'e
g\'en\'erale, tel que nous l'avons formul\'e, pour le rapport ${\cal R}$.

\newpage

\chapter{Cordes cosmiques en th\'eorie de Kaluza-Klein et de Gauss-Bonnet}
\thispagestyle{plain}
\markboth{Cordes cosmiques en th\'eorie de K-K et de G-B}{}
\label{ch corde}
Ce chapitre est consacr\'e \`a la recherche analytique de solutions
dites {\em 4-statiques} de la th\'eorie de Kaluza-Klein d\'ependant
d'une seule variable $\rho$. En imposant la sym\'etrie cylindrique,
cette variable sera choisie telle qu'elle mesure la distance (propre)
\`a l'axe de sym\'etrie et les 4 vecteurs de Killing associ\'es \`a la
g\'eom\'etrie seront li\'es aux 4 coordonn\'ees cycliques (section
3.1). La m\'ethode de r\'esolution consiste \`a introduire une matrice
$X$ r\'eelle $4\times4$, dont la d\'efinition \cite{Clement3dim}
rappelle celle des connexions affines qui figurent explicitement dans
les tenseurs d'Einstein et de Lanczos, ce qui permet d'\'ecrire les
\'equations de la th\'eorie en fonction de $X$ et des traces de ses
puissances ($X, X^{2}, X^{3}$) (section 3.1); la m\'ethode est
quasiment matricielle. Mais le but principal {\em consiste \`a en extraire
les solutions qui peuvent \^{e}tre interpr\'et\'ees comme des cordes
cosmiques et \`a \'etudier leurs g\'eom\'etries}. Dans la section 3.2
nous \'epuisons toutes les solutions 4-statiques \`a  sym\'etrie
cylindrique de la th\'eorie de Kaluza-Klein sans toutefois approfondir
l'\'etude de leurs propri\'et\'es g\'eom\'etriques. Dans la section 3.3
nous g\'en\'eralisons l'\'etude de la section 3.2 en tenant compte du
terme de Gauss-Bonnet mais trouvons des solutions ind\'ependantes de
$\gamma$. C'est dans la section 3.4 que nous nous int\'eressons \`a la
recherche de solutions d\'ependant explicitement de $\gamma$ et \`a en
construire une famille qui s'interpr\`ete comme une corde cosmique
supraconductrice dont nous \'etudions les propri\'et\'es physiques et
\`a laquelle nous avons consacr\'e la section 3.5 exposant qualitativement
l'\'etude des g\'eod\'esiques de sa g\'eom\'etrie.

\section{Sym\'etrie cylindrique}
\setcounter{equation}{0}
Dans ce chapitre nous nous int\'eressons au cas o\`u la 5-m\'etrique admet
4 vecteurs de Killing; dans un syst\`eme de coordonn\'ees particulier,
ces derniers prennent la forme ${\xi}_{a}\,^{A} = {\delta}_{a}\,^{A}$
(\'equation (\ref{eq cycl})) o\`u $a$ prend $p=4$ valeurs, donc $n=1$. Les
coordonn\'ees associ\'ees $x^{2}=\varphi$, $x^{3}=z$, $x^{4}=t$, $x^{5}$, y
sont cycliques, et la m\'etrique ne d\'epend que de la coordonn\'ee
$x^{1}=\rho$ prise comme coordonn\'ee radiale. Nous param\'etriserons
les m\'etriques 4-statiques par

\be
\label{eq met}
ds^{\,2} = {\Lambda}_{ab}(\rho)\,dx^{a}\,dx^{b}-d{\rho}^{2}
\ee

\n ($a,b=2,3,4,5$), qui sont bien de la forme (\ref{eq n+p})
avec

\be
h_{11}=-\tau(\rho)
\ee

\n o\`u $\tau \equiv \mbox{det}{\Lambda}_{ab}$ (\'equation (\ref{eq dt})).

Proc\'edons maintenant \`a la r\'eduction des \'equations (\ref{eq GB})
de la th\'eorie de Kaluza-Klein avec terme de Gauss-Bonnet en une
\'equation scalaire: $G_{11}+\gamma\,L_{11}=0$, et une \'equation matricielle
$4 \times 4$: $G^{a}_{\,b}+\gamma\,L^{a}_{\,b}=0$. Ces \'equations
s'\'ecrivent, apr\`es insertion de (\ref{eq 6R}) dans (\ref{eq L})
puis de (\ref{eq L}) dans (\ref{eq GB}), comme

\be
\label{eq champ}
R_{AB}+g_{AB}\,R+\gamma\left(R_{A}\,^{CDE}\,R_{BCDE}-2\,R^{CD}\,R_{ACBD}-
2\,R_{AC}\,R_{B}\,^{C}+R\,R_{AB}\right)=0
\ee

\n Dans le cas de la m\'etrique 4-statique (\ref{eq met}),
elles prennent la forme simplifi\'ee suivante

\bn
\label{eq scal1}
& &R_{11}-R+\gamma\left(-2\,R_{1}\,^{ab}\,_{1}\,R_{1ab1}+2\,R^{ab}\,R_{1ab1}+
2\,(R_{11})^{2}+R\,R_{11}\right)=0 \\ \nonumber \\
\label{eq matr1}
& &R^{a}\,_{b}+{\delta}^{a}\,_{b}\,R+\gamma\left(2\,R^{a11c}\,R_{b11c}+
R^{acde}\,R_{bcde}-2\,R^{11}\,R^{a}\,_{1b1} \right. \nonumber \\
& &\left.-2\,R^{cd}\,R^{a}\,_{cbd}-2\,R^{a}\,_{c}\,R_{b}\,^{c}+R\,R^{a}\,_{b}
\right)=0\,.
\en

\n Comme dans d'autres probl\`emes de relativit\'e g\'en\'erale o\`u la
m\'etrique d\'epend d'une seule variable \cite{Clement3dim}, introduisons
la matrice $4 \times 4$ $X_{ab}$ d\'efinie par

\be
\label{eq chii}
X \equiv {\Lambda}^{-1}\,{\Lambda}_{,\rho}
\ee

\n d'o\`u la relation

\be
\label{eq Tchii}
\mbox{Tr}\,X={\tau}^{-1}\,{\tau}_{,\rho}\,.
\ee

\n En se servant des \'equations (\ref{abcd}
$\rightarrow$ \ref{ijkl}), les termes des \'equations (\ref{eq scal1}),
(\ref{eq matr1}) se d\'eveloppent en fonction de $X$ par

\bnn
& &R_{1ab1}=\frac{1}{2}\,(\Lambda\,B)_{ab} \\
& &R_{11}=R^{a}\,_{1a1}=-\frac{1}{2}\,\mbox{Tr}B \\
& &R_{abcd}=-\frac{1}{4}\,[(\Lambda\,X)_{bc}\,(\Lambda\,X)_{ad}-
(\Lambda\,X)_{ac}\,(\Lambda\,X)_{bd}] \\
& &R_{ab}={\Lambda}^{cd}\,R_{cadb}-R_{1a1b}=\frac{1}{2}\,\left[\Lambda
\left(B-\frac{1}{2}\,X^{2}+\frac{1}{2}\,(\mbox{Tr}X)\,X\right)\right]_{ab} \\
& &R={\Lambda}^{ab}\,R_{ab}-R_{11}=\mbox{Tr}B+\frac{1}{4}\,
(\mbox{Tr}X)^{2}-\frac{1}{4}\,\mbox{Tr}X^{2}
\enn

\n (o\`u on a utilis\'e la relation ${\Lambda}_{,\rho}=\Lambda\,X$ qui
d\'ecoule de la d\'efinition (\ref{eq chii}), ainsi que la convention
(\ref{eq Ricci}) pour calculer $R_{11}$ et $R_{ab}$). $B$ est une matrice $4
\times 4$ d\'efinie par

\be
B \equiv X_{,\rho} + \frac{1}{2}\,X^{2}\,.
\ee

\n Les autres termes s'obtiennent en multipliant, une \`a
quatre fois, par la matrice inverse ${\Lambda}^{-1}$, par exemple
$$
R^{abcd}=-\frac{1}{4}\,[(X\,{\Lambda}^{-1})^{ad}\,(X\,{\Lambda}^{-1})
^{bc}-(X\,{\Lambda}^{-1})^{bd}\,(X\,{\Lambda}^{-1})^{ac}]
$$
d'o\`u
$$
R^{acde}\,R_{bcde}=\frac{1}{8}\,[(\mbox{Tr}X^{2})\,X^{2}-X^{4}]
^{a}\,_{b}\,.
$$
Finalement, les \'equations (\ref{eq scal1}), (\ref{eq matr1})
prennent respectivement les formes suivantes

\bn
\label{eq scal2}
& &3\,\mbox{Tr}B+\frac{1}{2}\,[(\mbox{Tr}X)^{2}-\mbox{Tr}X^{2}]+\gamma\,
\left\{\frac{1}{2}\,[\mbox{Tr}(B\,X^{2})-\mbox{Tr}(B\,X)\,\mbox{Tr}X]
\right. \nonumber \\
& &\left.\mbox{}+\frac{1}{4}\,\mbox{Tr}B\,[(\mbox{Tr}X)^{2}-\mbox{Tr}X^{2}]
\right\}=0 \\ \nonumber \\ \nonumber \\
\label{eq matr2}
& &X_{,\rho}+2\,\mbox{Tr}X_{,\rho}+\frac{1}{2}\,[(\mbox{Tr}X)\,X+
\mbox{Tr}X^{2}+(\mbox{Tr}X)^{2}]+\gamma\,\left\{\frac{1}{2}\,(X^{3})_{,\rho}
-\frac{1}{2}\,(\mbox{Tr}X)(X^{2})_{,\rho}\right. \nonumber \\
& &\mbox{}+\frac{1}{2}\,[(\mbox{Tr}X)^{2}-\mbox{Tr}X^{2}]\,X_{,\rho}
-\frac{1}{2}\,(\mbox{Tr}X_{,\rho})[X^{2}-(\mbox{Tr}X)\,X]-\frac{1}{4}\,
(\mbox{Tr}X^{2})_{,\rho}\,X \nonumber \\
& &\left.\mbox{}+\frac{1}{4}\,[(\mbox{Tr}X)\,X^{3}-(\mbox{Tr}X^{2})\,X^{2}
-(\mbox{Tr}X)(\mbox{Tr}X^{2})\,X+(\mbox{Tr}X)^{3}\,X]\right\}=0
\en

\n Nous pouvons rajouter \`a ces \'equations, l'\'equation
caract\'eristique pour les matrices $4 \times 4$

\be
\label{eq carac}
X^{4}-f\,X^{3}+\frac{1}{2}\,(f^{2}-g)\,X^{2}+\left(-\frac{1}{3}\,
h+\frac{1}{2}\,f\,g-\frac{1}{6}\,f^{3}\right)\,X+k \equiv 0
\ee

\n et sa trace

\be
\label{eq Tcarac}
\mbox{Tr}X^{4} \equiv\frac{4}{3}\,f\,h-f^{2}\,g+\frac{1}{2}\,g^{2}+
\frac{1}{6}\,f^{4}-4\,k
\ee

\n avec{,} par d\'efinition

\be
\label{eq def}
\begin{array}{lll}
f \equiv \mbox{Tr}X &;& g \equiv \mbox{Tr}X^{2} \\
h \equiv \mbox{Tr}X^{3} &;& k \equiv \mbox{det}X \end{array}
\ee

\section{R\'esolution des \'equations pour {\boldmath $\gamma=0$}}
\setcounter{equation}{0}
Dans cette section, nous recherchons les solutions 4-statiques de la
th\'eorie de Kaluza-Klein \footnote
{J. A. Ferrari \cite{Fer} fut le premier, \`a
notre connaissance, qui s'est int\'eress\'e \`a \'etudier une classe de
solutions \`a sym\'etrie cylindrique de Kaluza-Klein, plus
pr\'ecis\'ement les solutions \`a l'int\'erieur d'un sol\'eno\"{\i}de
infini, en l'absence d'un champ \'electrique.}. Pour
$\gamma=0$, les \'equations (\ref{eq scal2}), (\ref{eq matr2})
deviennent respectivement

\bn
\label{eq KK1}
& &3\,f_{,\rho}+\frac{1}{2}\,f^{2}+g=0 \\
\label{eq KK2}
& &X_{,\rho}+\frac{1}{2}\,f\,X +2\,f_{,\rho}+\frac{1}{2}\,f^{2}+
\frac{1}{2}\,g=0\,.
\en

\n En prenant la trace de (\ref{eq KK2}), on a

\be
\label{eq KK3}
9\,f_{,\rho}+\frac{5}{2}\,f^{2}+2\,g=0
\ee

\n en \'eliminant ensuite $g$ entre (\ref{eq KK1}) et (\ref{eq KK3}),
on trouve l'\'equation

\be
\label{eq f2}
f_{,\rho}+\frac{1}{2}\,f^{2}=0
\ee

\n qui s'int\`egre par

\be
\label{eq f=0}
f = 0
\ee

\n ou par

\be
\label{eq fneq0}
f = \frac{2}{\rho}
\ee

\subsection{Cas {\boldmath $f = 0$}}
\label{sec-f=0}
Dans ce cas on obtient, en substituant $f=0$ dans (\ref{eq KK1}) ou
(\ref{eq KK3}), $g=0$ puis, en substituant dans (\ref{eq KK2}),
$X_{,\rho}=0$, d'o\`u

\be
\label{eq X=A}
X = A
\ee

\n o\`u $A$ est une matrice $4 \times 4$ constante (r\'eelle). La
matrice $\Lambda$ associ\'ee est alors donn\'ee par

\be
\Lambda = C\,\mbox{\large e}^{A\,\rho}
\ee

\noindent o\`u $C$ est une matrice $4 \times 4$ r\'eelle et constante telle que

\be
\label{eq Crelation}
\mbox{det}C < 0\;\;\;\;\;;\;\;\;\;C = C^{T}\;\;\mbox{et}\;\;C\,A = (C\,A)^{T}
\ee

\n pour assurer, d'une part, que det$\Lambda=\tau$ ait le signe
n\'egatif, et d'autre part, que la matrice $\Lambda$ soit sym\'etrique
($A^{T}$ et $C^{T}$ sont les matrices transpos\'ees).

Les valeurs propres $p_{i}$ de $A$ satisfont sa propre \'equation
caract\'eristique (\ref{eq carac}) en rempla\c{c}ant $A$ par $p$:

\be
p^{4}-\frac{h}{3}\,p+k = 0
\ee

\n avec les conditions $f=0$ et $g=0$, qui impliquent

\be
\label{eq petp2}
\sum_{i=1}^{4}\,p_{i} = 0\qquad; \qquad  \sum_{i=1}^{4}\,p_{i}^{2} = 0\,.
\ee

Commen\c{c}ons par traiter le cas g\'en\'erique o\`u l'un au moins de $h$
et $k$ est non nul. Les valeurs propres de $A$ peuvent s'\'ecrire

\bn
& &p_{1}=-x+i\,\sqrt{2x^{2}-y^{2}} \nonumber \\
& &p_{2}=-x-i\,\sqrt{2x^{2}-y^{2}} \nonumber \\
& &p_{3}=x+i\,y \nonumber \\
& &p_{4}=x-i\,y \nonumber
\en

\n (o\`u $x$ et $y$ sont suppos\'es r\'eels), avec
$x^{2}-y^{2}=h/(12x)$ et $(x^{2}+y^{2})(3x^{2}-y^{2})=k$. Nous distinguons:\\

\n 1) Les 4 valeurs propres sont complexes ($3x^{4}<k\leq 4x^{4}$).
La formule (\ref{eq Lagrange}) d'interpolation de Lagrange \cite{L}
donne dans ce cas

\bn
\label{eq L1}
\Lambda&=&\mbox{e}^{-x\,\rho}\,C\,M_{1}\,[(A+x)\,\sin(\sqrt{2x^{2}-y^{2}}\,
\rho+\theta_{1})+\sqrt{2x^{2}-y^{2}}\,\cos(\sqrt{2x^{2}-y^{2}}\,\rho+
\theta_{1})] \nonumber \\
 & &\mbox{}+\mbox{e}^{x\,\rho}\,C\,M_{2}\,[(A-x)\,\sin(y\,\rho+\theta_{2})+
y\,\cos(y\,\rho+\theta_{2})]
\en

\n o\`u $M_{1}$, $M_{2}$ sont des matrices quadratiques en $A$ \`a
coefficients r\'eels et $\theta_{1}$, $\theta_{2}$ sont des constantes
r\'eelles. \\

\n 2) Deux valeurs propres sont r\'eelles et distinctes $p_{1}$,
$p_{2}$ ($k<3x^{4}$). On a alors

\bn
\label{eq L2}
\Lambda&=&\mbox{e}^{-x\,\rho}\,C\,M_{1}\,[(A+x)\,\sinh(\sqrt{y^{2}-2x^{2}}\,
\rho+\theta_{3})+\sqrt{y^{2}-2x^{2}}\,\cosh(\sqrt{y^{2}-2x^{2}}\,\rho+
\theta_{3})] \nonumber \\
 & &\mbox{}+\mbox{e}^{x\,\rho}\,C\,M_{2}\,[(A-x)\,\sin(y\,\rho+\theta_{2})+
y\,\cos(y\,\rho+\theta_{2})]
\en

\n o\`u $\theta_{3}$ est une constante r\'eelle. Dans le cas particulier
$k=0$ ($h\neq0$), les valeurs propres sont
$0,\,p,\,jp,\,j^{2}p$ avec $p\equiv(h/3)^{1/3}$ et
$j=$e$^{2i\pi/3}$. Comme exemple de ce cas, on choisit

\be
\label{eq k=0}
\label{eq p^3}
A = p\,\left(\begin{array}{cccc} 1&0&0&0 \\ 0&0&0&0\\
                    0&0&0&-1 \\ 0&0&1&-1 \end{array} \right)\;\;\;;\;\;\;
C = \left(\begin{array}{cccc} -1&0&0&0 \\ 0&-1&0&0\\
                    0&0&1&-2 \\ 0&0&-2&1 \end{array} \right)
\ee

\n d'o\`u, apr\`es un r\'earrangement des coordonn\'ees

\bn
ds^{\,2}&=&-d{\rho}^{\,2}-\mbox{e}^{p\,\rho}\,dz^{\,2}+
2\,\cos\left(\frac{\sqrt{3}p\rho}{2}+\frac{\pi}{3}\right)\,\mbox{e}^{-p\,\rho
 /2}\,dt^{\,2}-(dx^{5})^{\,2} \nonumber \\
 & &\mbox{}+2\,\cos\left(\frac{\sqrt{3}p\rho}{2}-\frac{\pi}{3}\right)\,
\mbox{e}^{-p\,\rho/2}\,{d\varphi}^{\,2}-4\,\cos\frac{\sqrt{3}p\rho}{2}\,
\mbox{e}^{-p\,\rho/2}\,dt\,d\varphi\,.\nonumber \\
\en

\n Cet espace-temps est le produit cart\'esien d'une solution \`a
sym\'etrie cylindrique des \'equations d'Einstein \`a 4 dimensions
(voir aussi \cite{Zouzou}) par le cercle de Klein.\\

Consid\'erons maintenant le cas $h=k=0$. L'\'equation caract\'eristique de la
matrice $A$
se r\'eduit alors \`a

\be
\label{eq 37}
A^{4} = 0
\ee

\n d'o\`u la solution g\'en\'erale

\be
\label{eq h=0}
\Lambda = C\,\left(I+\rho\,A+\frac{{\rho}^{2}}{2}\,A^{2}+
\frac{{\rho}^{3}}{6}\,A^{3}\right)\,.
\ee

\n Les solutions peuvent se classer suivant le rang de $A$.\\

\n a) r$(A)=3$ ou $A^{3}\neq0$. La forme normale de Jordan de la matrice
$A$ est

\be
A =\left( \begin{array}{cccc}
          0&1&0&0\\0&0&1&0\\0&0&0&1\\0&0&0&0 \end{array} \right)
\ee

\n pour laquelle la matrice $C$ la plus g\'en\'erale r\'ealisant
$C=C^{T}$ et $CA=(CA)^{T}$ est donn\'ee par

\be
C =\left( \begin{array}{cccc}
          0&0&0&a\\0&0&a&b\\0&a&b&c\\a&b&c&d \end{array} \right)
\ee

\n o\`u $a, b, c, d$ sont des constantes r\'eelles. Or

\be
\mbox{det}C = a^{4} \geq 0
\ee

\n ne satisfaisant pas la condition (\ref{eq Crelation}), la solution
correspondante $\Lambda$ n'a pas la \mbox{signature} lorentzienne.\\

\n b) r$(A)=2$ ou $A^{3}=0$. Dans le cas o\`u $A^{2}\neq0$,
on choisit $A$ sous la forme

\be
A = \left( \begin{array}{cccc}
0&0&0&0\\0&0&1&0\\0&0&0&1\\0&0&0&0 \end{array} \right)\;\;\;,\;\;\;
C = \left( \begin{array}{cccc}
-1&0&0&0\\0&0&0&-1\\0&0&-1&0\\0&-1&0&0 \end{array} \right)
\ee

\n d'o\`u, apr\`es un r\'earrangement des coordonn\'ees

\be
\label{eq esp1}
ds^{\,2} =-d{\rho}^{\,2}-\frac{{\rho}^{2}}{2}\,d{\varphi}^{\,2}-dt^{\,2}
-(dx^{5})^{\,2}-2\,dz\,d\varphi -2\,\rho\,dt\,d\varphi \,.
\ee

\n Dans le cas o\`u $A^{2}=0$, la forme normale de Jordan de la
matrice $A$ est

\be
A = \left( \begin{array}{cccc}
0&1&0&0\\0&0&0&0\\0&0&0&1\\0&0&0&0 \end{array} \right)
\ee

\n et la matrice $C$ la plus g\'en\'erale associ\'ee est de la forme

\be
C = \left( \begin{array}{cccc}
0&a&0&b\\a&c&b&d\\0&b&0&e\\b&d&e&m \end{array} \right)
\ee

\n o\`u $a, b, c, d, e, m$ sont des constantes r\'eelles. D'o\`u

\be
\mbox{det}C = (a\,e-b^{2})^{2} \geq 0\,;
\ee

\n ce cas est donc non lorentzien.\\

\n c) r$(A)=1$ ou $A^{2}=0$ et $A\neq0$. On choisit $A$ sous la forme
normale et $C$ sous une forme particuli\`ere

\be
A = \left( \begin{array}{cccc}
0&1&0&0\\0&0&0&0\\0&0&0&0\\0&0&0&0 \end{array} \right)\;,\;
C = \left( \begin{array}{cccc}
0&1&0&0\\1&0&0&0\\0&0&-1&0\\0&0&0&-1 \end{array} \right)
\ee

\n d'o\`u, apr\`es un r\'earrangement ad\'equat des coordonn\'ees

\be
\label{eq esp2}
ds^{\,2} = -d{\rho}^{\,2}-dz^{\,2}+\rho\,dt^{\,2}-(dx^{5})^{\,2}+
2\,d\varphi\,dt\,.
\ee

\n Cet espace-temps est plat \cite{Zouzou}.\\

\n d) r$(A)=0$ ou $A=0$. La matrice $\Lambda$ est dans ce cas
constante, conduisant \`a la m\'etrique

\be
\label{eq esp3}
ds^{\,2} =
-d{\rho}^{\,2}-{d\varphi}^{\,2}-dz^{\,2}+dt^{\,2}-(dx^{5})^{\,2}\,.
\ee

\n qui est un espace-temps de Minkowski \`a 5 dimensions avec 2
dimensions compactifi\'ees ($\varphi,\,x^{5}$).\\

Remarquons que les espaces-temps (\ref{eq esp1}), (\ref{eq esp2}),
(\ref{eq esp3}) sont \`a nouveau produits de solutions \`a sym\'etrie
cylindrique des \'equations d'Einstein \`a 4 dimensions \cite{Zouzou}
par le cercle de Klein.

Les diff\'erentes solutions qui pr\'ec\`edent ($f=0$) sont telles que
det$\Lambda=$ det$C=const.$, donc que la m\'etrique est r\'eguli\`ere
pour $\rho$ variant de $-\infty$ \`a $+\infty$: la topologie spatiale
n'est pas euclidienne, mais plut\^{o}t du type wormhole. D'autre part,
ces m\'etriques ne sont pas asymptotiquement voisines de la m\'etrique
de Minkowski, ce qui limite leur int\'er\^{e}t physique. Dans la
sous-section suivante, nous pr\'esenterons d'autres solutions
($f=2/\rho$) ayant plus d'int\'er\^{e}t physique.

\subsection{Cas {\boldmath $f=2/\rho$}}
\label{sec-f=2/r}
Dans ce cas le d\'eterminant de la 5-m\'etrique, \'egal \`a $-\tau$,
donn\'e par
$$
\label{eq tr2}
-\tau = const.\,{\rho}^{2}
$$
est proportionnel \`a celui de la m\'etrique de Minkowski (ce qui ne
suffit \'evidemment pas pour que la 5-m\'etrique soit asymptotiquement
voisine de celle de Minkowski). D'autre part, cette 5-m\'etrique sera
singuli\`ere en $\rho=0$.

En reportant la relation (\ref{eq f2}) dans (\ref{eq KK1}) ou
(\ref{eq KK3}) on obtient

\be
g = f^{2} = \frac{4}{{\rho}^{2}}\,.
\ee

\n Ceci transforme (\ref{eq KK2}) en

\be
X_{,\rho}+\frac{1}{\rho}\,X = 0
\ee

\n qui s'int\`egre par

\be
X = \frac{2}{\rho}\,A
\ee

\n o\`u $A$ est une matrice $4 \times 4$ constante (r\'eelle) avec

\be
\label{eq A=1}
\mbox{Tr}A^{2} = \mbox{Tr}A = 1\,.
\ee

\noindent La matrice $\Lambda$, reli\'ee \`a $A$ par (\ref{eq chii}):
${\Lambda}^{-1}{\Lambda}_{,\rho}=(2/\rho)A$, s'\'ecrit

\be
\label{eq UA1}
\Lambda = C\,\mbox{\large e}^{U(\rho)\,A}
\ee

\noindent o\`u $C$ est une matrice constante satisfaisant les
relations (\ref{eq Crelation}) et $U(\rho)$ est d\'efini par
$$
U_{,\rho} \equiv \frac{2}{\rho}
$$
d'o\`u, en omettant une constante d'int\'egration \footnote{Cette
constante produirait un facteur {\normalsize e}$^{Const\,A}$ qui serait
absorb\'e dans $C$.}

\be
U(\rho) = \ln {\rho}^{2}\,.
\ee

La matrice $A$ admet 4 valeurs propres, solutions de l'\'equation

\be
p^{4}-p^{3}-\frac{1}{3}\,(c-1)\,p+d = 0
\ee

\n o\`u

\be
c \equiv \mbox{Tr}A^{3} \qquad; \qquad d \equiv \mbox{det}A\,,
\ee

\n avec les conditions (\ref{eq A=1}):

\be
\label{eq sum}
\sum_{i=1}^{4}p_{i} = 1 \qquad, \qquad \sum_{i=1}^{4}p_{i}^{2} = 1\,.
\ee

Les valeurs propres $p_{i}$ peuvent \^{e}tre param\'etris\'es par

\be
\label{eq propres}
p_{1}=x+i\,y\;\;;\;\;p_{2}=x-i\,y\;\;;\;\;p_{3}=\frac{1}{2}-x+i\,q\;\;;\;\;
p_{4}=\frac{1}{2}-x-i\,q
\ee

\n o\`u $x, y^{2}, q^{2}$ sont suppos\'es r\'eels. On calcule ensuite
$y^{2}, q^{2}, d$ par

\bnn
y^{2}+q^{2}&=&2\,x^{2}-x-\frac{1}{4}\\
y^{2}&=&\frac{3\,x^{2}\,(4\,x-3)+1-c}{3\,(4\,x-1)}\\
d&=&(x^{2}+y^{2})\left[\left(\frac{1}{2}-x\right)^{2}+q^{2}\right]\,.
\enn

\n Nous distinguons:\\

\n 1) Les 4 valeurs propres sont complexes. La matrice $\Lambda$
s'obtient de (\ref{eq L1}) par reparam\'etrisation

\bn
\label{eq LL1}
\Lambda&=&{\rho}^{2\,x}C\,M_{3}\,\left[(A-x)\,\sin(2y\ln\rho+{\delta}_{1})+
y\,\cos(2y\ln\rho+{\delta}_{1})\right] \nonumber \\
 & &\mbox{}+{\rho}^{1-2\,x}\,C\,M_{4}\,\left[\left(A+x-\frac{1}{2}\right)
\sin(2q\ln\rho+{\delta}_{2})+q\,\cos(2q\ln\rho+{\delta}_{2})\right]\nonumber \\
\en

\n o\`u $M_{3}$, $M_{4}$ sont des matrices quadratiques en $A$, et
${\delta}_{1}, {\delta}_{2}$ sont des constantes r\'eelles.\\

\n 2) Deux valeurs propres sont complexes.\\
Si les 2 valeurs propres r\'eelles sont distinctes, la matrice
$\Lambda$ s'obtient de (\ref{eq LL1}) en rempla\c{c}ant, dans la
premi\`ere ligne, $y$ par $\bar{y}=iy$ et les fonctions trigonom\'etriques
par des fonctions
hyperboliques. Si
les 2 valeurs propres r\'eelles sont \'egales la formule de Lagrange
(\ref{eq Lagrange}) appliqu\'ee aux 3 valeurs propres $x,\,1/2-x\pm iq$
conduit alors \`a une matrice $\Lambda$ du type de (\ref{eq LL1}) avec $y=0$.\\

\n 3) Les 4 valeurs propres sont r\'eelles et distinctes. La matrice $A$
peut \^{e}tre diagonalis\'ee. Si on choisit $C$ sous la forme
$C=$ diag$(-1, -1, +1, -1)$, on obtient une m\'etrique de Kasner \cite{Kasner}

\be
\label{eq 4reelle}
ds^{\,2} =
-d{\rho}^{2}-{\rho}^{2\,p_{1}}\,d{\varphi}^{\,2}-{\rho}^{2\,p_{2}}\,
dz^{\,2}+{\rho}^{2\,p_{3}}\,dt^{\,2}-{\rho}^{2\,p_{4}}\,
(dx^{5})^{\,2}\,.
\ee

\n o\`u les $p_{i}$ sont reli\'es par les relations (\ref{eq sum}).\\

Si deux ou plusieurs des valeurs propres co\"{\i}ncident, on obtient
des cas d\'eg\'en\'er\'es que nous allons \'etudier successivement.\\

\n 4) Deux des valeurs propres r\'eelles sont \'egales,
$p_{1}=p_{2}$. La forme normale de Jordan de la matrice $A$ est

\be
\label{eq JA}
A=\left(\begin{array}{cccc} p_{1}&{\epsilon}_{1}&0&0\\0&p_{1}&0&0\\
0&0&p_{3}&0\\0&0&0&p_{4} \end{array} \right)
\ee

\n avec ${\epsilon}_{1}=0$ ou 1 (${\epsilon}_{1}^{2}={\epsilon}_{1}$).
On d\'ecompose $A$ sous la forme

\be
A = D + H
\ee

\n avec $D=\mbox{diag}(p_{1}, p_{1}, p_{3}, p_{4})$, et $H_{12}=
{\epsilon}_{1}$ est le seul \'el\'ement
non nul de $H$. Remarquons que $D$ et $H$ commutent
($\forall\;{\epsilon}_{1}$) et que $H^{2}=0$, d'o\`u
$$ \mbox{\large e}^{\ln{\rho}^{2}\,A}=\mbox{\large
  e}^{\ln{\rho}^{2}\,D}\, \mbox{\large e}^{\ln{\rho}^{2}\,H}=
\mbox{diag}({\rho}^{2\,p_{1}}, {\rho}^{2\,p_{1}}, {\rho}^{2\,p_{3}},
{\rho}^{2\,p_{4}})\,(1+\ln{\rho}^{2}\,H)\,.
$$
Et finalement

\be
\label{eq 2egale}
\Lambda = C\,\mbox{diag}({\rho}^{2\,p_{1}}, {\rho}^{2\,p_{1}},
{\rho}^{2\,p_{3}},
{\rho}^{2\,p_{4}})+{\epsilon}_{1}\,{\rho}^{2\,p_{1}}\,\ln{\rho}^{2}\,C\,H
\ee

\n avec seul l'\'el\'ement $(CH)_{22}={\epsilon}_{1}b_{1}$
de $CH$ non identiquement nul \footnote{La matrice $C$ la plus
  g\'en\'erale associ\'ee \`a $A$ (\ref{eq JA}) est dans ce cas
  donn\'ee par (\ref{eq JC}) avec ${\epsilon}_{2}=0$ en prenant
  $b_{2}\equiv 0$.}. Si ${\epsilon}_{1}=0$, on retrouve
une m\'etrique de Kasner particuli\`ere.\\

\n 5) Les 4 valeurs propres r\'eelles sont \'egales deux \`a deux
($8c=5,\,64d=1$):

\be
p_{1}=p_{2}=\frac{1-\sqrt{3}}{4} \qquad; \qquad
p_{3}=p_{4}=\frac{1+\sqrt{3}}{4}.
\ee

\n La forme normale de Jordan de $A$ est

\be
A = \left( \begin{array}{cccc}
    p_{1}&{\epsilon}_{1}&0&0\\0&p_{1}&0&0\\
    0&0&p_{3}&{\epsilon}_{2}\\0&0&0&p_{3} \end{array} \right)
\ee

\n d'o\`u la matrice $C$ la plus g\'en\'erale r\'ealisant $C=C^{T}$ et
$CA=(CA)^{T}$

\be
\label{eq JC}
C = \left( \begin{array}{cccc}
    (1-{\epsilon}_{1})a_{1}&b_{1}&0&0\\b_{1}&c_{1}&0&0\\
    0&0&(1-{\epsilon}_{2})a_{2}&b_{2}\\0&0&b_{2}&c_{2}
\end{array} \right)\,,
\ee

\n et son d\'eterminant

\be
\mbox{det}C = [(1-{\epsilon}_{1})\,a_{1}\,c_{1}-{b_{1}}^{2}]\,
[(1-{\epsilon}_{2})\,a_{2}\,c_{2}-{b_{2}}^{2}]\,.
\ee

\n On distingue ainsi deux cas:\\

\n a) ${\epsilon}_{1}{\epsilon}_{2}=1$. On a alors det$C>0$, ce qui
constitue une solution non lorentzienne.\\

\n b) ${\epsilon}_{1}{\epsilon}_{2}=0$. Dans ce cas, si
${\epsilon}_{2}=0$, on retrouve un cas particulier du cas 4)
(m\'etrique \ref{eq 2egale}) avec $p_{4}=p_{3}$; le cas
${\epsilon}_{1}=0$ s'en d\'eduit en \'echangeant $p_{1}$ et $p_{3}$.\\

\n 6) Trois valeurs propres r\'eelles sont \'egales et non nulles
($4c=-16d=1$):

\be
p_{1} = p_{2} = p_{3} = -p_{4} = \frac{1}{2}\,.
\ee

\n Dans ce cas on a
$$
\left(A-\frac{1}{2}\right)^{4} = - \left(A-\frac{1}{2}\right)^{3}\,.
$$

\n En posant $B=A-(1/2)$, on trouve
\bnn
\mbox{\large e}^{\ln{\rho}^{2}\,A}&=&\rho\,\mbox{\large e}^{\ln{\rho}^{2}\,B}\\
 &=&\rho\left[1+\ln{\rho}^{2}\,B+\frac{(\ln{\rho}^{2})^{2}}{2}\,B^{2}
+\left(1-\ln{\rho}^{2}+\frac{(\ln{\rho}^{2})^{2}}{2}-
\mbox{e}^{-\ln{\rho}^{2}}\right)B^{3}\right]
\enn
et enfin

\be
\label{eq LL2}
\Lambda = \rho\,C\,(1+B^{3})+2\,\rho\,\ln\rho\,C\,(B-B^{3})+
2\,\rho\,(\ln\rho)^{2}\,C\,(B^{2}+B^{3})-\frac{1}{\rho}\,C\,B^{3}\,.
\ee

\n La forme normale de Jordan de la matrice $B$ est

\be
B = \left( \begin{array}{cccc}
    0&{\epsilon}_{1}&0&0\\0&0&{\epsilon}_{2}&0\\0&0&0&0\\0&0&0&-1
  \end{array} \right)\,.
\ee

\n D'o\`u, si \\

\n a) ${\epsilon}_{1}={\epsilon}_{2}=1$ ou r$(B)=3$, il n'y a pas de
simplification dans (\ref{eq LL2}).\\

\n b) ${\epsilon}_{1}{\epsilon}_{2}=0$ et ${\epsilon}_{1}+{\epsilon}_{2}\neq
0$, ou r$(B)=2$, les termes en $(\ln\rho)^{2}$ dans (\ref{eq LL2})
s'\'eliminent.\\

\n c) ${\epsilon}_{1}={\epsilon}_{2}=0$ ou r$(B)=1$, tous
les termes logarithmiques dans (\ref{eq LL2}) s'\'eliminent.\\

\n 7) Trois valeurs propres sont nulles ($c=1,\,d=0$). L'\'equation
caract\'eristique de $A$ se r\'eduit alors \`a

\be
\label{eq A4A3}
A^{4} = A^{3}
\ee

\n avec 1, 0, 0, 0 comme valeurs propres. Le calcul direct --semblable
\`a celui qui a men\'e \`a (\ref{eq LL2})-- conduit \`a

\be
\label{eq 77}
\Lambda = C\,(1-A^{3})+2\,\ln{\rho}\,C\,(A-A^{3})+
2\,(\ln{\rho})^{2}\,C\,(A^{2}-A^{3})+{\rho}^{2}\,C\,A^{3}\,.
\ee

\n La forme normale de Jordan de la matrice $A$ est

\be
\label{eq A77}
A = \left( \begin{array}{cccc}
           0&{\epsilon}_{1}&0&0\\0&0&{\epsilon}_{2}&0\\
           0&0&0&0\\0&0&0&1 \end{array} \right)
\ee

\n d'o\`u l'on d\'eduit; si:\\

\n a) ${\epsilon}_{1}{\epsilon}_{2}=1$ ou r$(A)=3$, on a alors
$A^{3} \neq A^{2}$ (voir ci-dessous).\\

\n b) ${\epsilon}_{1}{\epsilon}_{2}=0$ et ${\epsilon}_{1}+{\epsilon}_{2}=1$
ou r$(A)=2$, on a $A^{3}=A^{2}$ mais $A^{2}\neq A$. Dans ce cas les
termes en $(\ln\rho)^{2}$ disparaissent de l'expression de $\Lambda$
(voir sous-section \ref{sec-XaA}).\\

\n c) ${\epsilon}_{1}={\epsilon}_{2}=0$ ou r$(A)=1$, on a $A^{2}=A$.
L'expression de $\Lambda$ ne comporte plus de termes logarithmiques;
la 5-m\'etrique est le produit de la m\'etrique d'une corde cosmique
\cite{Zouzou} par le cercle de Klein:

\be
ds^{\,2}=dt^{\,2}-d{\rho}^{\,2}-{\alpha}^{2}\,{\rho}^{2}d{\varphi}^{\,2}-
dz^{\,2}-(dx^{5})^{2}
\ee

\n (si $\alpha =1$ la 5-m\'etrique est minkowskienne).\\

Dans le cas g\'en\'eral, la 5-m\'etrique associ\'ee est, comme on le
verra dans l'exemple suivant, minkowskienne \`a des logarithmes pr\`es.
Consid\'erons le cas r$(A)=3$ \mbox{(${\epsilon}_{1}{\epsilon}_{2}=1$),} et
choisissons pour $C$ la forme particuli\`ere

\be
C = \left( \begin{array}{cccc}
           0&0&-1&0\\0&-1&0&0\\
           -1&0&0&0\\0&0&0&-1 \end{array} \right)\,,
\ee

\n et l'on a

\be
\Lambda = \left( \begin{array}{cccc}
0&0&-1&0\\0&-1&-2\ln\rho&0\\
-1&-2\ln\rho&-2(\ln\rho)^{2}&0\\0&0&0&-{\rho}^{2} \end{array} \right)\,,
\ee

\n qui, \`a la suite d'un r\'earrangement ad\'equat des coordonn\'ees
$\varphi$, $z$, $t$ $x^{5}$, donne

\be
ds^{\,2} = -d{\rho}^{\,2}-{\rho}^{2}\,d{\varphi}^{\,2}-dt^{\,2}-
2\,(\ln\rho)^{2}\,(dx^{5})^{\,2}-2\,dz\,dx^{5}-4\,\ln\rho\,dt\,dx^{5}\,.
\ee

\n Cette m\'etrique se diagonalise par

\be
-1\,,\;\;-{\rho}^{2}\,,\;\;\,-1-(\ln\rho)^{2}+|\ln\rho|\sqrt{2+(\ln\rho)^{2}}
\,,\;\;\,+1\,,\;\;\,-1-(\ln\rho)^{2}-|\ln\rho|\sqrt{2+(\ln\rho)^{2}}\,.
\ee

\section{Une classe de solution pour {\boldmath $\gamma \neq0$}}
\label{sec-terme}
\setcounter{equation}{0}
Nous consid\'erons maintenant les \'equations (\ref{eq scal2}),
(\ref{eq matr2}) dans le cas g\'en\'eral o\`u $\gamma\neq0$. Ce
syst\`eme d'\'equations diff\'erentielles \'etant assez compliqu\'e,
nous n'aborderons pas le probl\`eme de sa solution g\'en\'erale, mais
nous contenterons de chercher, dans cette section et la suivante, deux
classes de solutions. Dans cette section, nous cherchons des solutions
de la forme

\be
\label{eq forme}
X(\rho) = \alpha(\rho)\,A
\ee

\n o\`u $\alpha(\rho)$ est une fonction (r\'eelle) de $\rho$ \`a
d\'eterminer, et $A$ une matrice $4\times4$ constante (r\'eelle). Par analogie
avec la section pr\'ec\'edente --consacr\'ee \`a Kaluza-Klein-- nous
scindons cette section en deux sous-sections, dans l'une nous int\`egrons
les \'equations (\ref{eq scal2}), (\ref{eq matr2}) pour $\alpha$
constante, donc $X=A$ et dans l'autre nous nous int\'eresserons au
cas des solutions avec $\alpha(\rho)$ non constante.

\subsection{Cas {\boldmath $X=A$}}
Dans ce cas les \'equations (\ref{eq scal2}), (\ref{eq
  matr2}) s'\'ecrivent respectivement

\bn
\label{eq scal3}
& &12(a^{2}+2b)+\gamma(a^{4}-3a^{2}b+2ac-24d)=0 \\
\label{eq matr3}
& &a\gamma A^{3}-b\gamma A^{2}+a[(a^{2}-b)\gamma+2] A+2(a^{2}+b)=0
\en

\n o\`u

\be
\label{eq defA}
\begin{array}{lcl}
a \equiv \mbox{Tr}A&,&b \equiv \mbox{Tr}A^{2}\\
c \equiv \mbox{Tr}A^{3}&,&d \equiv \mbox{det}A \,. \end{array}
\ee

\n La trace de (\ref{eq matr3}) permet d'exprimer $c$ en fonction de
$a,\;b$ par

\be
\label{eq cab}
\gamma ac=\gamma(-a^{4}+a^{2}b+b^{2})-10a^{2}-8b\,.
\ee

\n Combin\'ees ensemble, les \'equations (\ref{eq scal3}), (\ref{eq cab})
permettent d'exprimer $d$ en fonction de $a,\;b$ par

\be
\label{eq dab}
24\gamma d=\gamma(-a^{4}-a^{2}b+2b^{2})-8a^{2}+8b\,.
\ee

Remarquons que pour $a=0$ et $b=0$, $d$ s'annule et les \'equations
(\ref{eq scal3}), (\ref{eq matr3}) sont identiquement satisfaites alors que
$\,c\,$ reste ind\'etermin\'e. Ce cas pour lequel l'\'equation
caract\'eristique de $A$ s'\'ecrit

\be
A^{4}=\frac{1}{3}\,c\,A
\ee

\n correspond au cas $k=0$ de la sous-section \ref{sec-f=0}. Si
$c\neq 0\,(h\neq 0)$, la solution est localement de la forme (\ref{eq k=0}).
Si $\,c=0\,(h=0)$, les formes locales des solutions sont
(\ref{eq esp1}), (\ref{eq esp2}) ou (\ref{eq esp3}). Nous
montrerons dans l'Annexe 2 que le cas $a=0,\;b=0$
--avec par cons\'equent $d=0$-- repr\'esente l'unique solution des
\'equations de Gauss-Bonnet sous la forme $X=A$. Nous nous contentons
de montrer dans le reste de cette sous-section que pour les cas
($a=0,\;b\neq0$) ou ($a\neq0,\;b=0$) ne correspondent pas des
solutions de la forme $X=A$; le cas g\'en\'erique ($a\neq0,\;b\neq0$)
sera trait\'e dans l'Annexe 2.\\

\n \underline{Cas $a=0,\;b\neq0$}

Dans ce cas (\ref{eq scal3}) donne
$$
d=\frac{b}{\gamma}
$$
et (\ref{eq cab}) donne
$$
b=\frac{8}{\gamma}
$$
et permet de transformer (\ref{eq matr3}) en
$$
A^{2}=\frac{2}{\gamma}
$$
d'o\`u, en prenant le d\'eterminant des 2 membres
$$
d^{2}=\frac{16}{{\gamma}^{4}}
$$
qui est en d\'esaccord avec la valeur
$d^{2}=b^{2}/{\gamma}^{2}=64/{\gamma}^{4}$; ce cas est
donc impossible.\\

\n \underline{Cas $a\neq0,\;b=0$}

En substituant dans (\ref{eq cab}), (\ref{eq dab}), (\ref{eq matr3})
on obtient respectivement

\bn
\label{eq ca}
& &\gamma c=-\gamma a^{3}-10a\\
\label{eq da}
& &24\gamma d=-\gamma a^{4}-8a^{2}\\
\label{eq matr3a}
& &\gamma A^{3}+(\gamma a^{2}+2)A+2a=0
\en

\n En se servant de (\ref{eq ca}), (\ref{eq da}), l'\'equation
caract\'eristique (\ref{eq carac}) multipli\'ee par $\gamma$ devient

\be
\label{eq caracb=0}
\gamma(A^{4}-aA^{3})+\frac{1}{2}\,a^{2}A^{2}+\left(\frac{\gamma}{6}\,a^{3}
+\frac{10}{3}\,a\right)A-\frac{\gamma}{24}\,a^{4}-\frac{1}{3}\,a^{2}=0\,.
\ee

\n En \'eliminant $A^{3}$ et $A^{4}$ \`a l'aide de (\ref{eq matr3a})
et (\ref{eq matr3a})$\times A$, on obtient

\be
\label{eq resb=0}
-\left(\frac{\gamma}{2}\,a^{2}+2\right)A^{2}+\left(\frac{7\gamma}{6}\,a^{3}+
\frac{10}{3}\,a\right)A-\frac{\gamma}{24}\,a^{4}+\frac{5}{3}\,a^{2}=0\,;
\ee

\n puis, en prenant la trace, on obtient

\be
\label{eq 10}
\gamma a^{2}=-10
\ee

\n en substituant ensuite dans (\ref{eq ca}) on a
$$
c=0\,.
$$
La trace de (\ref{eq resb=0})$\times A$ fournit en tenant compte de
(\ref{eq 10})
$$
a^{3}=-\frac{36}{25}\,c\,.
$$
Avec $\,c=0\,$, on retrouve donc le cas $a=b=0$ d\'ej\`a trait\'e.

\subsection{Cas {\boldmath $X(\rho)=\alpha(\rho)A$}}
\label{sec-XaA}
Dans la sous-section pr\'ec\'edente nous avons montr\'e que certaines
solutions de la forme $X=A$ (avec Tr$A=$Tr$A^{2}=0$) de la
th\'eorie de Kaluza-Klein ($\gamma = 0$) sont aussi solutions de la
th\'eorie avec terme de Gauss-Bonnet ($\gamma \neq 0$). Nous allons
obtenir maintenant une propri\'et\'e analogue pour les solutions
$X=(2/\rho)A$ (avec Tr$A=$Tr$A^{2}=1$). On montre dans l'Annexe 3
que ces deux possibilit\'es \'epuisent la classe
des solutions $X(\rho)=\alpha(\rho)A$ de la th\'eorie de Kaluza-Klein
avec terme de Gauss-Bonnet.

Notre point de d\'epart est donc

\bn
\label{eq forme1}
& &X=\frac{2}{\rho}\,A \\
\label{eq depart}
& &a=1\quad,\quad b=1
\en

\n (o\`u $a, b, c, d$ sont d\'efinis par (\ref{eq defA})). Ce cas
\'etant d\'ej\`a solution des \'equations (\ref{eq scal2}), (\ref{eq matr2})
avec $\gamma =0$; il suffit donc de chercher s'il annule la somme des
termes \`a l'int\'erieur de chaque accolade dans les deux
\'equations. Commen\c{c}ons par l'\'equation (\ref{eq
  matr2}). Regroupons les termes \`a l'int\'erieur de l'accolade
suivant les puissances de $A$. On voit facilement que le terme en $A$
s'annule identiquement, tandis que la combinaison des termes en
$A^{2}$ et en $A^{3}$ donne

\bnn
& &\frac{1}{2}\left(\frac{8}{{\rho}^{3}}\right)_{,\,\rho}A^{3}-\frac{1}{2}
\frac{2}{\rho}\left(\frac{4}{{\rho}^{2}}\right)_{,\,\rho}A^{2}-\frac{1}{2}
\left(\frac{2}{\rho}\right)_{,\,\rho}\frac{4}{{\rho}^{2}}A^{2}+\frac{1}{4}
\frac{2}{\rho}\frac{8}{{\rho}^{3}}A^{3}-\frac{1}{4}\frac{4}{{\rho}^{2}}
\frac{4}{{\rho}^{2}}A^{2}=\\
& &-\frac{8}{{\rho}^{4}}(A^{3}-A^{2})\,.
\enn

\n L'\'equation (\ref{eq matr2}) est donc satisfaite si

\be
\label{eq supp1}
A^{3}-A^{2}=0
\ee

\n d'o\`u en prenant la trace

\be
\label{eq supp2}
c=1\,.
\ee

\n De (\ref{eq supp1}) on d\'eduit aussi ($A^{4}-A^{3}=0$)
$$
\mbox{Tr}A^{4}=1\,;
$$
or, Tr$A^{4}$ \'etant donn\'e par (\ref{eq Tcarac})

\bn
\label{eq A4d}
\mbox{Tr}A^{4}&\equiv&\frac{4}{3}\,a\,c-a^{2}\,b+\frac{1}{2}\,b^{2}+
\frac{1}{6}\,a^{4}-4\,d\\
 &=&1-4\,d\,; \nonumber
\en

\n ces deux valeurs ne sont compatibles que si

\be
\label{eq supp3}
d=0\,.
\ee

\n \`A l'aide de (\ref{eq forme1}), (\ref{eq depart}), (\ref{eq
  supp1}) on calcule la matrice $B$
$$
B=\frac{2}{{\rho}^{2}}\,(A^{2}-A)
$$
d'o\`u
$$
\mbox{Tr}B=\mbox{Tr}(B\,X)=\mbox{Tr}(B\,X^{2})=0\,;
$$
l'\'equation (\ref{eq scal2}) est donc bien satisfaite.

L'expression (\ref{eq forme1}) est donc solution des \'equations
  de Kaluza-Klein avec terme de Gauss-Bonnet avec les conditions
  $a=b=c=1,\;d=0$ et $A^{3}-A^{2}=0$. Cette solution correspond au
  sous-cas ${\epsilon}_{1}{\epsilon}_{2}=0$ du cas 7) de la
  sous-section \ref{sec-f=2/r}. Si la matrice $A$ est de rang 2
  ($A^{3}=A^{2},\,A^{2}\neq A$), on peut choisir

\be
A=\left(\begin{array}{cccc} 1&0&0&0\\0&0&1&0\\0&0&0&0\\0&0&0&0
\end{array} \right)\;,\;
C=\left(\begin{array}{cccc} -{\alpha}^{2}&0&0&0\\0&0&-1&0\\0&-1&0&0\\0&0&0&-1
\end{array} \right)
\ee

\n d'o\`u

\be
ds^{\,2}=-d{\rho}^{\,2}-{\alpha}^{2}\,{\rho}^{2}\,d{\varphi}^{\,2}-dz^{\,2}-
2\,\ln\rho\,(dx^{5})^{2}-2\,dt\,dx^{5}\,.
\ee

\n Cette solution, qui r\'epond au crit\`ere (\ref{eq tr2})
$\tau=\mbox{det}C{\rho}^{2}$, repr\'esente une corde cosmique
charg\'ee avec des potentiels \'electrique $A_{4}$ et gravitationnel
$\bar{g}_{44}$ (voir \'equations (\ref{eq potentiel}), (\ref{eq projete}))
proportionnels \`a $1/\ln\rho$. Si $A$ est de rang 1 ($A^{2}=A$),
la 5-m\'etrique

\be
ds^{\,2}=dt^{\,2}-d{\rho}^{\,2}-{\alpha}^{2}\,{\rho}^{2}\,d{\varphi}^{\,2}-
dz^{\,2}-(dx^{5})^{2}
\ee

\n est celle d'une corde cosmique neutre.\\

Remarquons que les espaces-temps consid\'er\'es dans cette section
sont de la forme $V_{4}\times\Re$ ou $V_{4}\times S^{1}$, o\`u $V_{4}$
est un espace-temps statique \`a sym\'etrie cylindrique solution des
\'equations d'Einstein \`a 4 dimensions \cite{Zouzou}, pour lequel le
terme de Gauss-Bonnet s'annule identiquement \cite{Madore}. Il n'est
donc pas \'etonnant que ces espaces-temps soient --trivialement--
solutions quel que soit $\gamma$ de la th\'eorie de Kaluza-Klein avec
terme de Gauss-Bonnet. Notre but dans la section suivante consiste
\`a chercher s'il peut exister des solutions repr\'esentant
des cordes cosmiques et d\'ependant explicitement de $\gamma$.

\section{Une solution exacte avec source de Gauss-Bonnet}
\setcounter{equation}{0}
Un observateur situ\'e \`a grande distance de la source du champ de
Kaluza-Klein peut d\'evelopper la matrice $X$ sous la forme

\be
\label{eq devel}
X=\frac{2}{\rho}\,A+\frac{2\gamma}{{\rho}^{2}}\,D+\frac{2\gamma}{{\rho}^{3}}
\,E+\cdots
\ee

\n o\`u $(2/\rho)A$ est une solution exacte de Kaluza-Klein, donc
telle que (voir sous-section \ref{sec-f=2/r}):

\be
\label{eq cond}
a=b=1,
\ee

\n et $\,c, d\,$ sont quelconques ($a, b, c, d$ sont d\'efinis par
(\ref{eq defA})). Les termes
$(2\gamma/{\rho}^{2})D+(2\gamma/{\rho}^{3})E$ sont consid\'er\'es
comme des perturbations dues \`a l'introduction du terme de
Gauss-Bonnet, consid\'er\'e comme source, dans les \'equations de la
th\'eorie de Kaluza-Klein, $D,\;E$ \'etant deux matrices
r\'eelles $4\times 4$ constantes \`a d\'eterminer.

\subsection{Solution perturbative}
En omettant les termes d'ordre sup\'erieur \`a $1/{\rho}^{4}$, les
\'equations (\ref{eq scal2}), (\ref{eq matr2}) se scindent chacune en 2
\'equations qui sont les coefficients de $1/{\rho}^{3}$ et de
$1/{\rho}^{4}$. On obtient en multipliant par des constantes
convenables, pour (\ref{eq scal2})

\be
\label{eq s3}
\mbox{Tr}(AD)=\mbox{Tr}D \qquad (\mbox{coef. de}\;1/{\rho}^{3})
\ee

\be
\label{eq s4}
-21\mbox{Tr}E+12\mbox{Tr}(AE)-2[2(c-1)+12d]+6\gamma
\mbox{Tr}D^{2}+3\gamma (\mbox{Tr}D)^{2}=0 \quad (\mbox{coef. de}\;
1/{\rho}^{4})
\ee

\n puis, pour (\ref{eq matr2})

\be
\label{eq m3}
D= (\mbox{Tr}D)A \qquad (\mbox{coef. de}\;1/{\rho}^{3})
\ee

\bn
& &2E-\gamma (\mbox{Tr}D)D-(\mbox{Tr}E)A+4(A^{3}-A^{2})+4\mbox{Tr}E
-2\mbox{Tr}(AE)\nonumber \\
\label{eq m4}
& &\mbox{}-\gamma (\mbox{Tr}D)^{2}-\gamma \mbox{Tr}D^{2}=0
\qquad (\mbox{coef. de}\;1/{\rho}^{4})
\en

\n o\`u l'\'equation (\ref{eq s3}) a \'et\'e utilis\'ee dans (\ref{eq m3})
et (\ref{eq A4d}) dans (\ref{eq s4}). En \'elevant (\ref{eq m3}) au
carr\'e et en prenant ensuite la trace, on obtient

\be
\label{eq int}
\mbox{Tr}D^{2}=(\mbox{Tr}D)^{2}\,.
\ee
En tenant compte de cette relation (\ref{eq int}), on obtient
en calculant la trace de (\ref{eq m4})

\be
\label{eq int1}
17\mbox{Tr}E-8\mbox{Tr}(AE)+4(c-1)-9\gamma (\mbox{Tr}D)^{2}=0\,.
\ee

\n En r\'ealisant la somme membre \`a membre des \'equations (\ref{eq s4}),
(\ref{eq int1}), on obtient

\be
\mbox{Tr}E-\mbox{Tr}(AE)=-6\,d\,.
\ee

\n En utilisant ces deux derni\`eres \'equations pour calculer Tr$E$
et Tr$(AE)$, on obtient \`a partir de (\ref{eq m4})

\be
\label{eq int4}
E=-2(A^{3}-A^{2})+\left[\gamma (\mbox{Tr}D)^{2}+\frac{2}{9}\,(1-c+12d)\right]A
+\frac{2}{9}\,[2(c-1)+3d]\,.
\ee

\n On a ainsi d\'etermin\'e, respectivement par les \'equations
(\ref{eq m3}), (\ref{eq int4}), les matrices $D,\,E$ en fonction de la
matrice $A$ et de ses param\`etres --restant-- libres $c,\,d$ \`a
Tr$D$ pr\`es.

\subsection{Solution exacte}
Imposons maintenant la condition suppl\'ementaire que la solution
cherch\'ee puisse \^{e}tre interpr\'et\'ee comme une corde cosmique
supraconductrice. La 5-m\'etrique d'une corde cosmique rectiligne
neutre doit avoir le comportement asymptotique
\mbox{\cite{Hiscock, Gott, Linet}}

\be
\begin{array}{ccc}
ds^{\,2} &\sim &dt^{\,2}-d{\rho}^{\,2}-{\alpha}^{2}{\rho}^{2}d{\varphi}^{\,2}
-dz^{\,2}-(dx^{5})^{2}\,.\\
 &(\rho \rightarrow \infty)&
\end{array}
\ee

\n Dans le cas d'une corde cosmique rectiligne supraconductrice,
l'existence d'une distribution de courant allant jusqu'\`a l'infini
conduit, comme en \'electrodynamique de Maxwell, \`a des contributions
logarithmiques au comportement asymptotique (\cite{Witten}: B. Linet).
Or, parmi toutes les solutions \`a sym\'etrie cylindrique de
Kaluza-Klein \'etudi\'ees dans la sous-section \ref{sec-f=2/r}, seules
les solutions pour lesquelles

\be
\label{eq cond1}
c=1 \quad, \quad d=0
\ee

\n ont un {\em comportement asymptotiquement minkowskien} \`a des
fonctions logarithmiques pr\`es \footnote{La pr\'esence de fonctions
  logarithmiques est donc due \`a l'hypoth\`ese simplificatrice, mais
  non physique, de sym\'etrie de r\'evolution qui implique
  une corde de longueur infinie.}. En reportant
$c=1,\;d=0$ dans (\ref{eq int4}) puis dans (\ref{eq devel}) on obtient

\be
\label{eq exacte}
X=\frac{2}{\rho}\,A+\frac{2\gamma \mbox{Tr}D}{{\rho}^{2}}\,A+
\frac{\gamma}{{\rho}^{3}}\,[-4(A^{3}-A^{2})+2\gamma (\mbox{Tr}D)^{2}\,A]
+\cdots\,.
\ee

\n L'ensemble des conditions (\ref{eq cond}), (\ref{eq cond1})
d\'efinissent une alg\`ebre particuli\`ere de la matrice $A$ qui se
r\'esume par

\be
\label{eq alg}
A^{4}-A^{3}=0\quad \Rightarrow A^{n+1}-A^{n}=0 \quad \mbox{pour}\;
n \geq 3
\ee

\n (voir l'\'equation (\ref{eq A4A3})). Cette alg\`ebre
particuli\`ere fait que maintenant le second membre de (\ref{eq
  exacte}) r\'esout exactement les \'equations (\ref{eq scal2}),
(\ref{eq matr2}) pour Tr$D=0$. Pour le v\'erifier, cherchons les termes d'ordre
sup\'erieur \`a $1/{\rho}^{4}$ --qui ont \'et\'e n\'eglig\'es lors du
d\'eveloppement (\ref{eq s3}) $\rightarrow$ (\ref{eq m4})-- dans les
\'equations (\ref{eq scal2}), (\ref{eq matr2}). Il r\'esulte de (\ref{eq alg})
que
$X^{2}=(4/{\rho}^{2})A^{2},\,X^{3}=(8/{\rho}^{3})A^{3},\,B=(2/{\rho}^{2})
(A^{2}-A)$, et que les diff\'erentes traces qui figurent dans (\ref{eq scal2}),
 (\ref{eq matr2}) ont la m\^{e}me valeur que dans le cas $\gamma =0$.
La seule contribution possible d'ordre sup\'erieur \`a
 $1/{\rho}^{4}$ proviendrait des termes lin\'eaires en $X$ dans
 (\ref{eq matr2}):
$$
\frac{1}{4}\,[2(\mbox{Tr}X_{,\rho})(\mbox{Tr}X)-(\mbox{Tr}X^{2})_{,\rho}-
(\mbox{Tr}X)(\mbox{Tr}X^{2})+(\mbox{Tr}X)^{3}]X
$$
mais on v\'erifie que le crochet ci-dessus est identiquement nul.

Finalement, la solution \footnote{L'\'equation (\ref{eq exacte1}) est sous la
  forme g\'en\'erale d'une solution d\'ependant d'une seule matrice $A$.
En effet, \'etant
donn\'e que toutes les puissances $A^{n}$ de $A$ pour lesquelles
$n\geq4$ peuvent \^{e}tre exprim\'ees en fonction de
$A^{3},\;A^{2},\;A$ en se servant plusieurs fois de l'\'equation
caract\'eristique, l'expression g\'en\'erale de $X$ d\'ependant d'une
seule matrice $A$ est donc de la forme
\be
\label{eq sugg}
X=\epsilon(\rho)+\alpha(\rho)\,A+\beta(\rho)\,A^{2}+\delta(\rho)\,A^{3}
\ee
\n o\`u
$\epsilon(\rho),\;\alpha(\rho),\;\beta(\rho),\;\delta(\rho)$ sont des
fonctions r\'eelles de $\rho$ \`a d\'eterminer. La m\'ethode g\'en\'erale de
r\'esolution est donc la suivante: en reportant (\ref{eq sugg}) dans
(\ref{eq scal2}), (\ref{eq matr2}) on obtient respectivement une
relation scalaire --non lin\'eaire-- entre les fonctions
$\epsilon(\rho),\;\alpha(\rho),\;\beta(\rho),\;\delta(\rho)$ et leurs
d\'eriv\'ees premi\`eres (et les param\`etres de $A$: $a, b,  c, d$),
et un polyn\^{o}me $P(A)$ d'ordre 3 en $A$, annulateur de celle-ci, et dont
les coefficients sont des fonctions de
$\epsilon(\rho),\;\alpha(\rho),\;\beta(\rho),\;\delta(\rho)$ et de
leurs d\'eriv\'ees premi\`eres. Pour qu'il soit annulateur de $A$ pour
toute valeur de $\rho$ (voir aussi Annexe 3), il faut que ses coefficients
soient ou bien identiquement nuls, ou bien proportionnels; dans le
premier cas on obtient 4 relations plus la relation scalaire et dans
le second cas 3 relations plus la relation scalaire. Dans les 2 cas on
doit pouvoir d\'eterminer les fonctions
$\epsilon(\rho),\;\alpha(\rho),\;\beta(\rho),\;\delta(\rho)$; seuls
les param\`etres de $A$ ou certains d'entre eux restent ajustables. La
solution (\ref{eq exacte1}) fait part du cas o\`u les coefficients de
$P(A)$ sont identiquement nuls.}

\bn
\label{eq exacte1}
& &X=\frac{2}{\rho}\,A-\frac{4\,\gamma}{{\rho}^{3}}\,(A^{3}-A^{2})\\
& &A^{4}-A^{3}=0\qquad (a=b=c=1,\;d=0)
\en

\n est solution exacte de la th\'eorie de Kaluza-Klein avec terme de
Gauss-Bonnet. La m\'etrique associ\'ee est construite dans la section suivante.

\subsection{Construction d'une m\'etrique corde cosmique supraconductrice}
\label{sec-sup}
La matrice $\Lambda(\rho)$ associ\'ee \`a $X$ se calcule de la
fa\c{c}on suivante. \'Ecrivons $\Lambda(\rho)$ sous la forme

\be
\label{eq met1}
\Lambda(\rho) = C\,[1+\gamma\,{\Lambda}_{\cal GB}(\rho)]\,
\mbox{\large e}^{\ln{\rho}^{2}\,A}
\ee

\n o\`u ${\Lambda}_{\cal GB}(\rho)$
est une matrice r\'eelle $4\times4$ \`a d\'eterminer. En d\'erivant
(\ref{eq met1}) on obtient

\be
\label{eq met2}
{\Lambda}_{,\,\rho}=C\left[\gamma\,
{\Lambda}_{{\cal GB},\,\rho}+\frac{2}{\rho}\,[1+\gamma\,{\Lambda}
_{\cal GB}]\,A\right]\,\mbox{\large e}^{\ln{\rho}^{2}\,A}\,.
\ee

\n Or, ${\Lambda}_{,\,\rho}=\Lambda X$ (voir (\ref{eq chii})), d'o\`u
en effectuant le produit de $\Lambda$ par $X$

\be
\label{eq met3}
{\Lambda}_{,\,\rho}=C\,(1+\gamma\,{\Lambda}_{\cal GB})\left[\frac{2}{\rho}\,A
-\frac{4\gamma}{{\rho}^{3}}\,(A^{3}-A^{2})\right]
\mbox{\large e}^{\ln{\rho}^{2}\,A}\,.
\ee

\n En comparant (\ref{eq met2}), (\ref{eq met3}) on obtient apr\`es
multiplication par $C^{-1}$ \`a gauche et par {\large e}$^{-\ln{\rho}^{2}\,A}$
\`a droite

\be
\label{eq met4}
(1+\gamma\,{\Lambda}_{\cal GB})_{,\,\rho}=-\frac{4\gamma}{{\rho}^{3}}\,
(1+\gamma\,{\Lambda}_{\cal GB})\,(A^{3}-A^{2})\,.
\ee

\n Ceci conduit,  en multipliant par $A$ \`a droite, \`a
$$
(1+{\gamma}\,{\Lambda}_{\cal GB})_{,\,\rho}\,A = 0
$$
d'o\`u

\be
\label{eq met5}
(1+{\gamma}\,{\Lambda}_{\cal GB})\,A = K_{1}
\ee

\n o\`u $K_{1}$ est une matrice r\'eelle $4\times4$ constante. En
reportant dans (\ref{eq met4}), puis en int\'egrant, on obtient

\be
\label{eq met6}
1+{\gamma}\,{\Lambda}_{\cal GB} = \frac{2\gamma}{{\rho}^{2}}\,K_{1}\,
(A^{2}-A)+K_{2}
\ee

\n o\`u $K_{2}$ est une matrice r\'eelle $4\times4$ constante.
La multiplication de (\ref{eq met6}) par $A$
fournit en comparant avec (\ref{eq met5})

\be
K_{1}=K_{2}\,A\,.
\ee

\n Finalement, la matrice (\ref{eq met1}) s'\'ecrit

\be
\Lambda =C\,K_{2}\left[1+\frac{2\gamma}{{\rho}^{2}}\,
(A^{3}-A^{2})\right]\mbox{\large e}^{\ln{\rho}^{2}\,A}\,.
\ee

\n On peut toujours, moyennant une red\'efinition de la matrice
constante $C$, se ramener \`a $K_{2}=1$, d'o\`u

\be
\label{eq met9}
\Lambda =C\left[1+\frac{2\gamma}{{\rho}^{2}}\,
(A^{3}-A^{2})\right]\mbox{\large e}^{\ln{\rho}^{2}\,A}\,.
\ee

\n Remarquons que
$$
1+\frac{2\gamma}{{\rho}^{2}}\,(A^{3}-A^{2})= \mbox{\large e}^
{2\gamma (A^{3}-A^{2})/{\rho}^{2}}
$$
d'o\`u

\be
\label{eq met11}
\tau \equiv \mbox{det}\Lambda = \mbox{det}C\,{\rho}^{2}\,;
\ee

\n cette solution est donc aussi singuli\`ere le long de l'axe $\rho =0$. En
d\'eveloppant (\ref{eq met9}) on obtient

\bn
\label{eq met12}
\Lambda &=&C\,\mbox{\large e}^{\ln{\rho}^{2}\,A}+\frac{2\gamma}{{\rho}^{2}}
\,C\,(A^{3}-A^{2})\\
\label{eq met13}
 &=&C\left[1-A^{3}+2\,(\ln\rho)\,(A-A^{3})+2\left((\ln\rho)^{2}-
\frac{\gamma}{{\rho}^{2}}\right)(A^{2}-A^{3})+{\rho}^{2}\,A^{3}\right]\,.
\nonumber \\
\en

\n Pour $\gamma =0$, cette expression se r\'eduit \`a celle de la
th\'eorie de Kaluza-Klein correspondante, (\ref{eq 77}).

Nous cherchons \`a construire une m\'etrique qui peut \^{e}tre
interpr\'et\'ee comme la g\'eom\'etrie ext\'erieure d'une corde
cosmique supraconductrice. Partons de la forme
normale de Jordan (\ref{eq A77}) pour la matrice $A$, o\`u nous
supposons ${\epsilon}_{1}{\epsilon}_{2}=1$ (le cas
${\epsilon}_{1}{\epsilon}_{2}=0$, donc $A^{3}=A^{2}$, a \'et\'e
\'etudi\'e \`a la fin de la sous-section \ref{sec-XaA}),

\be
A=\left( \begin{array}{cccc}
         1&0&0&0\\0&0&1&0\\0&0&0&1\\0&0&0&0
\end {array} \right)
\ee

\n et choisissons pour $C$ la matrice simple

\be
C=\left( \begin{array}{cccc}
         -{\alpha}^{2}&0&0&0\\0&0&0&-1\\0&0&-1&0\\0&-1&0&0
\end {array} \right)\,.
\ee

\n En reportant dans (\ref{eq met13}) on obtient

\be
\label{eq met14}
ds^{\,2}=-d{\rho}^{\,2}-{\alpha}^{2}\,{\rho}^{2}d{\theta}^{\,2}-dt^{\,2}
-2\left[(\ln\rho)^{2}-\frac{\gamma}{{\rho}^{2}}\right](dx^{5})^{2}-
2\,dz\,dx^{5}-4\,\ln\rho\,dt\,dx^{5}
\ee

\n o\`u on a identifi\'e les coordonn\'ees $x^{2}, x^{3}, x^{4}$ avec
respectivement $\theta, z, t$. Proc\'edons maintenant \`a
l'identification des coordonn\'ees physiques par la diagonalisation de
la 5-m\'etrique pour des valeurs de $\rho$ relativement grandes
(les m\^{e}mes coordonn\'ees pourront \'eventuellement avoir, pr\`es
de la source, une toute autre
signification). Pour all\'eger l'\'ecriture, posons

\bn
\label{eq pos1}
& &\frac{1}{{\alpha}_{3}} \equiv 2(\ln\rho)^{2}-\frac{2\gamma}{{\rho}^{2}}\\
\label{eq pos2}
& &\frac{{\alpha}_{4}}{{\alpha}_{3}} \equiv 2\ln\rho\,.
\en

\n Remarquons que, pour $\rho \rightarrow \infty$, ${\alpha}_{3}
\rightarrow 0_{\mbox{}+}$, d'o\`u en particulier

\bn
& &2\ln\rho = \sqrt{\frac{2}{{\alpha}_{3}}}\left(1+{\cal
O}({\alpha}_{3}^{3/2})\right)\nonumber \\
\label{eq pos3}
\\
& &{\alpha}_{4}=\sqrt{2{\alpha}_{3}}\left(1+{\cal O}({\alpha}_{3}^{3/2})
\right)\,.\nonumber
\en

\n Les valeurs propres de la m\'etrique (\ref{eq met14}) sont
$-1,\,-{\alpha}^{2}{\rho}^{2}$ et les trois autres sont solutions de

\be
n^{3}+\left(\frac{1}{{\alpha}_{3}}+1\right)n^{2}-\left(4(\ln\rho)^{2}+1-
\frac{1}{{\alpha}_{3}}\right)n+1=0\,.
\ee

\n En se servant de (\ref{eq pos3}), cette derni\`ere \'equation
s'\'ecrit pour $\rho \rightarrow \infty$
$$
n^{3}+\left(\frac{1}{{\alpha}_{3}}+1\right)n^{2}-
\left(\frac{1}{{\alpha}_{3}}+1\right)n-1 \simeq 0
$$
et admet pour solutions

\bn
& &n_{\mbox{}+}=1+{\cal O}({\alpha}_{3}^{3/2})\nonumber \\
\label{eq valeur}
& &n_{1}=-{\alpha}_{3}+{\cal O}({\alpha}_{3}^{3/2})\\
& &n_{2}=-\frac{1}{{\alpha}_{3}}-2+{\alpha}_{3}+{\cal O}({\alpha}_{3}^{3/2})\,.
\nonumber
\en

\n L'identification des coordonn\'ees $z,\,t,\,x^{5}$ se fait de la
mani\`ere suivante. D\'esignons par $N^{A}(\rho)$
les vecteurs propres; leurs composantes sont d\'etermin\'ees par

\bn
& &N^{1}=N^{2}=0\\
& &-N^{5}=n\,N^{3}\\
& &-N^{4}-\frac{{\alpha}_{4}}{{\alpha}_{3}}\,N^{3}=n\,N^{4}
\en

\n o\`u $n$ d\'esigne
une des 3 valeurs propres $n_{\mbox{}+},\,n_{1},\,n_{2}$.
Cherchons ensuite quelle est, parmi les coordonn\'ees
$z,\,t,\,x^{5}$, celle qui peut \^{e}tre associ\'ee \`a la valeur
propre positive $n_{\mbox{}+}$ et qui sera donc la coordonn\'ee temps;
autrement dit, l'axe de
la coordonn\'ee temps sera identifi\'e avec la direction dans laquelle
est orient\'e asymptotiquement le c\^{o}ne de lumi\`ere. Or pour
$n_{\mbox{}+}$ on a

\be
\frac{N^{4}(+)}{-N^{5}(+)}=\frac{N^{4}(+)}{N^{3}(+)}\sim
\sqrt{\frac{1}{2{\alpha}_{3}}}\rightarrow \infty
\ee

\n ce qui signifie que le c\^{o}ne de lumi\`ere est d\'ej\`a orient\'e
(pour $\rho \rightarrow \infty$) suivant l'axe d\'esign\'e par $t$;
$t$ est donc la coordonn\'ee temps. Pour $n=n_{1}$, on obtient

\be
\frac{N^{3}(1)}{N^{5}(1)}\sim \frac{1}{{\alpha}_{3}}\rightarrow
\infty\quad,\quad\frac{N^{3}(1)}{N^{5}(1)}\sim -\sqrt{\frac{1}{2{\alpha}_{3}}}
\rightarrow \infty
\ee

\n ce qui signifie que le vecteur propre associ\'e est orient\'e, pour
$\rho \rightarrow \infty$, vers le troisi\`eme axe, ce dernier peut
\^{e}tre confondu avec l'axe de la coordonn\'ee $z$ ou celui de la
coordonn\'ee $x^{5}$. Nous l'identifions avec l'axe de la coordonn\'ee
$z$ (la troisi\`eme coordonn\'ee cylindrique usuelle); dans le cas contraire,
$g_{55}$ serait \'egale asymptotiquement \`a $n_{1}$ qui tend vers
z\'ero quand $\rho$ tend vers l'infini et par cons\'equent la
projection 4-dimensionnelle, qui est une des id\'ees de base de la
th\'eorie de Kaluza-Klein et qui doit \^{e}tre r\'ealisable au moins
\`a grande distance de la source, deviendrait impossible. $n=n_{2}$
est bien la valeur propre associ\'ee \`a $x^{5}$; on v\'erifie que
dans ce cas les rapports $N^{5}(2)/N^{3}(2)$, $N^{5}(2)/N^{4}(2)$
divergent asymptotiquement. Finalement, les coordonn\'ees
$\rho,\,\theta,\,z,\,t,\,x^{5}$ de (\ref{eq met14}) sont bien les
coordonn\'ees physiques usuelles.

La solution (\ref{eq met14}), obtenue pour un choix particulier de $C$ et $A$,
peut \^{e}tre g\'en\'eralis\'ee. On peut
garder l'expression de $A$ et chercher \`a g\'en\'eraliser celle de
$C$. Une approche diff\'erente, mais \'equivalente, consiste \`a r\'ealiser
une transformation lin\'eaire de coordonn\'ees de la forme

\be
\left(\begin{array}{c} \theta\\z\\t\\x^{5} \end{array} \right)=B\,
\left(\begin{array}{c} {\theta}^{\prime}\\z^{\prime}\\t^{\prime}\\
{x^{5}}^{\prime} \end{array} \right)
\ee

\n de fa\c{c}on, d'une part, \`a ne pas m\'elanger la coordonn\'ee
$\theta$ avec les autres coordonn\'ees, et d'autre part \`a conserver
les directions principales asymptotiques du tenseur m\'etrique
d\'etermin\'ees ci-dessus. Les valeurs propres associ\'ees \'etant
telles qu'asymptotiquement $n_{3}\ll n_{4}\ll n_{5}$, cette
contrainte implique $(B^{-1})_{34}=(B^{-1})_{35}=(B^{-1})_{45}=0$. La
matrice $B$, qui ne d\'epend alors plus que de 6 param\`etres, ou 3
param\`etres $({\rho}_{0},\,p,\,q)$ en choisissant des \'echelles
arbitraires pour les coordonn\'ees $z,\,t,\,x^{5}$, peut s'\'ecrire

\be
\label{eq matB}
B=\left(\begin{array}{cccc}
1&0&0&0\\0&1&q-\ln{\rho}_{0}&p+(\ln{\rho}_{0})^{2}/2\\
0&0&1&-\ln{\rho}_{0}\\0&0&0&1 \end{array} \right)\,.
\ee

\n En revenant \`a la notation initiale --non prim\'ee-- on obtient la
m\'etrique

\bn
\label{eq met15}
ds^{\,2}&=&-d{\rho}^{\,2}-{\alpha}^{2}\,{\rho}^{2}\,d{\theta}^{\,2}-dt^{\,2}
-2\left[(\ln\frac{\rho}{{\rho}_{0}})^{2}+p-\frac{\gamma}{{\rho}^{2}}\right]
(dx^{5})^{2} \nonumber \\
 & &\mbox{}-2\,dz\,dx^{5}-2\left[2\ln\frac{\rho}{{\rho}_{0}}+q\right]dt\,dx^{5}
\en

\n qui s'obtient directement de (\ref{eq met13}) via les
transformations \mbox{$C\rightarrow B^{T}CB,\,A\rightarrow B^{-1}AB$} (voir
\'equation (\ref{eq bas}) plus bas).

\subsection{Interpr\'etation physique}
Les potentiels \'electromagn\'etiques sont donn\'es
d'apr\`es (\ref{eq potentiel}) par $A_{3}={\lambda}^{-1}g_{35}/g_{55}$,
$A_{4}={\lambda}^{-1}g_{45}/g_{55}$. Le comportement asymptotique
des champs magn\'etique --orthoradial-- et \'electrique --radial--
est donc ici

\be
B \sim \frac{1}{\rho(\ln\rho)^{3}}\qquad,\qquad
E \sim \frac{1}{\rho(\ln\rho)^{2}}\,.
\ee

\n Le champ est principalement \'electrique pour un observateur
situ\'e \`a grande distance de la source.

Comme nous allons le montrer en calculant le tenseur
d'impulsion-\'energie effectif, cette source s'interpr\`ete bien comme
une corde cosmique parcourue par un courant \'electrique. En effet,
les \'equations (\ref{eq GB}) peuvent \^{e}tre r\'einterpr\'et\'ees
comme des \'equations de Kaluza-Klein avec source

\be
\label{eq eff}
R^{A}_{\,B}-\frac{1}{2}\,R\,{\delta}^{A}_{\,B}=\kappa\,
T_{\mbox{\tiny eff}\,B}^{A}
\ee

\n o\`u le tenseur d'impulsion-\'energie effectif est la somme d'une
contribution distributionnelle, et d'une contribution proportionnelle
au tenseur de Lanczos. Les composantes $T_{\mbox{\tiny eff}\,5}^{\mu}$,
qui sont le second membre des \'equations de
Maxwell modifi\'ees de la th\'eorie de Kaluza-Klein (\'equations
(\ref{eq KK})), s'interpr\`etent comme les composantes de la
densit\'e de courant ($T_{\mbox{\tiny eff}\,5}^{\mu}={\lambda}^{-1}J^{\mu}$),
$T_{\mbox{\tiny eff}\,3}^{3}$ comme une densit\'e de
tension, $T_{\mbox{\tiny eff}\,4}^{4}$ comme une densit\'e de masse et enfin
$T_{\mbox{\tiny eff}\,5}^{5}$ comme une densit\'e de charge scalaire.
Pour calculer le premier membre de (\ref{eq eff}), \'ecrivons la m\'etrique
(\ref{eq met15}) sous la forme

\be
\label{eq met16}
ds^{\,2}={\lambda}_{ab}\,dx^{a}\,dx^{b}-d{\rho}^{\,2}
-{\alpha}^{2}\,{\rho}^{2}d{\theta}^{\,2}
\ee

\n ($a, b=3, 4, 5$), avec

\bn
\label{eq 174}
\lambda&=&{\cal C}\,\mbox{\large e}^{2(\ln\rho\,{\cal A}-
\gamma\,{\cal A}^{2}/{\rho}^{2})} \\
 &=&{\cal C}+2\ln\rho\,{\cal CA}+2\left[(\ln\rho)^{2}-
\frac{\gamma}{{\rho}^{2}}\right]{\cal CA}^{2}
\en

\n o\`u

\be
\label{eq bas}
{\cal C}=\left(\begin{array}{ccc} 0&0&-1\\
0&-1&2\ln{\rho}_{0}-q\\-1&2\ln{\rho}_{0}-q&-2p-2(\ln{\rho}_{0})^{2}
\end{array} \right)\quad;\quad
{\cal A}=\left(\begin{array}{ccc} 0&1&-q\\0&0&1\\0&0&0\end{array}
\right)\,.
\ee

\n D'o\`u, l'on calcule

\be
\mbox{Tr}{\cal A}=\mbox{Tr}{\cal A}^{2}=0\quad,\quad{\cal A}^{3}=0\,.
\ee

\n La m\'etrique (\ref{eq met16}) est de la forme 2-statiques (\ref{eq n+p})
avec

\be
\label{eq 2-stat}
h_{ij}\,dx^{i}\,dx^{j}=-d{\rho}^{\,2}
-{\alpha}^{2}\,{\rho}^{2}\,d{\theta}^{\,2}=-\left(\frac{\alpha {\rho}_{0}}
{r_{0}^{\alpha}}\right)^{2}r^{2\,\alpha -2}\,{\delta}_{ij}
\,dx^{i}\,dx^{j}
\ee

\n en posant $\rho ={\rho}_{0}(r/r_{0})^{\alpha}$
($r^{2}=x^{2}+y^{2}$). On obtient \`a partir des \'equations (\ref{abcd})
et (\ref{aijb})

\bnn
R^{a}_{\,b}&=&\frac{1}{2}\,\left(\frac{1}{\rho}\,
(\rho\,{\lambda}^{-1}{\lambda}_{,\,\rho})_{,\,\rho}\right)^{a}_{\,b}\\
 &=&{\cal A}^{a}_{\,b}\,\Delta\ln\rho -\gamma\,({\cal A}^{2})^{a}_{\,b}\,
\Delta\frac{1}{{\rho}^{2}}
\enn

\n o\`u $\Delta$ est le laplacien covariant dans la m\'etrique
(\ref{eq 2-stat}). On d\'emontre \`a l'aide du th\'eor\`eme de Green
que
$$
\Delta\ln\rho =\alpha\,\Delta\ln r = 2\,\pi\,\alpha\,|h|^{-1/2}\,
{\delta}^{2}(\vec{x})
$$
($|h|^{-1/2}{\delta}^{2}(\vec{x})$ est la distribution de Dirac
covariante). D'o\`u

\be
R^{a}_{\,b}=2\,\pi\,\alpha\,{\cal A}^{a}_{\,b}\,|h|^{-1/2}\,{\delta}^{2}
(\vec{x})
-\frac{4\gamma}{{\rho}^{4}}\,({\cal A}^{2})^{a}_{\,b}\,.
\ee

\n D'autre part,

\be
R^{1}_{\,1}=R^{2}_{\,2}=|h|^{-1}\,R_{1212}=2\,\pi\,(\alpha -1)\,
|h|^{-1/2}\,{\delta}^{2}(\vec{x})
\ee

\n pour la m\'etrique conique (\ref{eq 2-stat}).

Finalement les contributions distributionnelles au tenseur
d'impulsion-\'energie effectif (\ref{eq eff}) sont, pour la
matrice ${\cal A}$ donn\'ee par (\ref{eq bas})

\bn
& &T_{\mbox{\tiny eff}\,3}^{3}=T_{\mbox{\tiny eff}\,4}^{4}=
T_{\mbox{\tiny eff}\,5}^{5}=\frac{1-\alpha}{4G}\,|h|^{-1/2}\,
{\delta}^{2}(\vec{x})\,,\\
& &T_{\mbox{\tiny eff}\,4}^{3}=T_{\mbox{\tiny eff}\,5}^{4}=-\frac{1}{q}\,
T_{\mbox{\tiny eff}\,5}^{3}=\frac{\alpha}{4G}\,|h|^{-1/2}\,
{\delta}^{2}(\vec{x})
\en

\n (corde cosmique charg\'ee, anim\'ee d'un mouvement de translation
le long de son axe, et parcourue par un courant \'electrique si $q\neq
0$). La contribution \'etendue

\be
T_{\mbox{\tiny eff}\,5}^{3}=-\frac{\gamma}{2\pi G}\,\frac{1}{{\rho}^{4}}
\ee

\n correspondant \`a un courant \'electrique concentr\'e pr\`es de la
corde $\rho =0$. On peut donc dire qu'il s'agit d'une corde cosmique
supraconductrice.

\section{\'Etude des g\'eod\'esiques}
\setcounter{equation}{0}
Apr\`es avoir examin\'e dans la sous-section \ref{sec-sup} la nature de
la singularit\'e en $\rho =0$ d'un point de vue structural, examen au
terme duquel cette derni\`ere a \'et\'e interpr\'et\'ee comme une
corde cosmique
supraconductrice (\'etendue), nous l'examinerons dans cette section
d'un point de vue comportemental en \'etudiant le mouvement libre des
particules d'\'epreuve soumises au champ de force cr\'e\'e par la
corde. Nous verrons qu'une \'etude analytique permet de conclure que\\

${\bullet}$ pour $\gamma <0$ les particules d'\'epreuve peuvent
atteindre la singularit\'e, tandis que pour $\gamma \geq 0$ les particules
d'\'epreuve dont les 5-g\'eod\'esiques sont de genre temps ou
lumi\`ere n'atteignent jamais la singularit\'e;\\

${\bullet}$ seules les particules pour lesquelles les
5-g\'eod\'esiques sont de genre espace peuvent, ind\'ependamment de la
valeur de $\gamma$, aller \`a l'infini.\\

La loi du mouvement libre est r\'egie par l'\'equation (\ref{eq geod})
des g\'eod\'esiques que nous allons remplacer par un syst\`eme
d'int\'egrales premi\`eres. Tout le long
de ce chapitre nous n'avons consid\'er\'e que les m\'etriques
4-statiques admettant 4 vecteurs de Killing, donn\'es par ${\xi}_{a}\,^{A}=
{\delta}_{a}\,^{A}$, qui engendrent un groupe d'isom\'etrie
ab\'elien. Associ\'ees \`a ces isom\'etries, quatre int\'egrales
premi\`eres de mouvement ${\Pi}_{a}$ se d\'efinissent lors d'un
mouvement libre de particules d'\'epreuve et expriment
l'homog\'en\'eit\'e de l'espace-temps le long des 4 axes d\'efinis par
les 4 vecteurs ${\xi}_{a}\,^{A}$; elles sont donn\'ees par (voir, par
exemple, \cite{theseA})
$$
g_{AB}\,\frac{dx^{A}}{dp}\,{\xi}_{a}\,^{B}={\Pi}_{a}
$$
qui ne sont autres que les composantes covariantes des vitesses
$dx^{A}/dp$ le long des axes d\'efinis par les vecteurs de Killing;
elles peuvent \^{e}tre reli\'ees \`a des grandeurs physiques comme le
moment angulaire $j$ et la charge $q$ qui sont respectivement
proportionnels \`a ${\Pi}_{2},\,{\Pi}_{5}$ (voir aussi
\cite{theseA}). Or $g_{1a}=0$ et $g_{ab}={\Lambda}_{ab}$, d'o\`u par inversion

\be
\label{eq integ1}
\frac{dx^{a}}{dp}={\Lambda}^{ab}\,{\Pi}_{b}
\ee

\n qui est \'equivalente aux 4 \'equations g\'eod\'esiques pour $A=a$;
l'\'equation pour $A=1$ s'\'ecrit avec ${\Gamma}^{1}_{11}\equiv 0,\,
{\Gamma}^{1}_{ab}={\Lambda}_{ab\,,\rho}/2$

\be
\label{eq integ2}
2\,\frac{d^{2}\rho}{dp^{2}}+{\Pi}_{a}\,{\Lambda}^{ac}\,
{\Lambda}_{cd\,,\rho}\,{\Lambda}^{db}\,{\Pi}_{b}=0
\ee

\n o\`u les $dx^{a}/dp$ ont \'et\'e \'elimin\'es \`a l'aide de
(\ref{eq integ1}). En multipliant (\ref{eq integ2}) par $d\rho/dp$
cette \'equation s'int\`egre, en tenant compte de
${\Lambda}^{ac}{\Lambda}_{cd\,,p}{\Lambda}^{db}=-{\Lambda}^{ab}\,_{,p}\,,$
par

\be
\label{eq integ3}
\left(\frac{d{\rho}}{dp}\right)^{2}-{\Pi}_{a}\,{\Lambda}^{ab}\,{\Pi}_{b}
+const.=0\,.
\ee

\n Le param\`etre affine $p$ de la g\'eod\'esique peut \^{e}tre
choisi tel que $ds^{\,2}=\epsilon dp^{\,2}$ avec $\epsilon =-1, 0$
ou $1$, respectivement pour les 5-g\'eod\'esiques de genre espace,
lumi\`ere ou temps; dans ce cas la constante d'int\'egration
dans (\ref{eq integ3})
est \'egale \`a $\epsilon$. Finalement, l'\'equation des
g\'eod\'esiques (\ref{eq geod}) est \'equivalente au syst\`eme d'\'equation

\bn
\label{eq integ4}
& &\left(\frac{d{\rho}}{dp}\right)^{2}-{\Pi}_{a}\,{\Lambda}^{ab}(\rho)\,
{\Pi}_{b}+\epsilon=0\\
\label{eq integ5}
& &\frac{dx^{a}}{dp}={\Lambda}^{ab}(\rho)\,{\Pi}_{b}\,.
\en

\n Pour une \'etude qualitative du mouvement g\'eod\'esique, nous nous
contentons de reconsid\'erer la m\'etrique (\ref{eq met14})
\'equivalente de ce point de vue \`a (\ref{eq met15}):

\be
\label{eq integ6}
\epsilon\,dp^{\,2}=-d{\rho}^{\,2}-{\alpha}^{2}\,{\rho}^{2}\,d{\theta}^{\,2}-
dt^{\,2}-\frac{1}{{\alpha}_{3}}\,(dx^{5})^{2}-2\,dz\,dx^{5}-2\,
\frac{{\alpha}_{4}}{{\alpha}_{3}}\,dt\,dx^{5}
\ee

\n o\`u ${\alpha}_{3}(\rho),\,{\alpha}_{4}(\rho)$ sont donn\'es par
(\ref{eq pos1}), (\ref{eq pos2}). Les \'el\'ements ${\Lambda}^{ab}$
non nuls s'expriment par

\be
{\Lambda}^{22}=-\frac{1}{{\alpha}^{2}{\rho}^{2}}\;,\;
{\Lambda}^{33}=\frac{1}{{\alpha}_{3}}-\left(\frac{{\alpha}_{4}}{{\alpha}_{3}}
\right)^{2}\;,\;{\Lambda}^{34}=\frac{{\alpha}_{4}}{{\alpha}_{3}}\;,\;
{\Lambda}^{35}={\Lambda}^{44}=-1
\ee

\n et permettent de d\'evelopper (\ref{eq integ4}), (\ref{eq integ5})
par

\bn
\label{eq geod1}
&
&\left(\frac{d{\rho}}{dp}\right)^{2}+\left(\frac{{\Pi}_{2}^{2}}{{\alpha}^{2}}+2\gamma {\Pi}_{3}^{2}\right)\frac{1}{{\rho}^{2}}+2{\Pi}_{3}^{2}(\ln\rho)^{2}-
4{\Pi}_{3}{\Pi}_{4}\ln\rho
+2{\Pi}_{3}{\Pi}_{5}+{\Pi}_{4}^{2}+\epsilon=0
\nonumber\\\\
\label{eq geod2}
& &\frac{d\theta}{dp}=-\frac{{\Pi}_{2}}{{\alpha}^{2}{\rho}^{2}}\\
\label{eq geod3}
& &\frac{dz}{dp}=-\frac{2\gamma {\Pi}_{3}}{{\rho}^{2}}-2{\Pi}_{3}(\ln\rho)^{2}
+2{\Pi}_{4}\ln\rho-{\Pi}_{5}\\
\label{eq geod4}
& &\frac{dt}{dp}=2{\Pi}_{3}\ln\rho -{\Pi}_{4}\\
\label{eq geod5}
& &\frac{dx^{5}}{dp}=-{\Pi}_{3}\,.
\en

\n Ind\'ependamment de la valeur de la charge (proportionnelle \`a
${\Pi}_{5}$), le mouvement sur le cylindre
de Klein, \'equation (\ref{eq geod5}), se fait \`a vitesse constante
ind\'ependante de la position de la particule dans l'espace, tandis
que l'\'equation (\ref{eq geod2}) assure la conservation du momemt
angulaire ordinaire $L_{z}$. Examinons de plus pr\`es le comportement
de l'\'equation (\ref{eq geod1}) quand $\rho \rightarrow 0$; on d\'eduit:\\

\n 1) Si ${\Pi}_{3}\neq 0$ nous distinguons:\\
a) si

\be
2\,\gamma < -\frac{{\Pi}_{2}^{2}}{{\alpha}^{2}{\Pi}_{3}^{2}}
\ee

\n les termes logarithmiques sont n\'egligeables devant le terme en
$1/{\rho}^{2}$ qui est n\'egatif; l'\'equation (\ref{eq geod1}) admet
toujours des solutions, et les particules peuvent atteindre la
singularit\'e, quel que soit $\epsilon$. Ainsi pour $\gamma <0$, toutes
les particules peuvent, pour un rapport ${\Pi}_{2}^{2}/{\Pi}_{3}^{2}$
suffisamment petit, atteindre la singularit\'e au bout d'un temps
propre fini,\\
b) si

\be
2\,\gamma = -\frac{{\Pi}_{2}^{2}}{{\alpha}^{2}{\Pi}_{3}^{2}}
\ee

\n le terme en $1/{\rho}^{2}$ est absent alors que le premier membre
de (\ref{eq geod1}) reste positif ($\forall\,\epsilon$) d\^{u} au
terme en $(\ln\rho)^{2}$. La particule ne peut donc pas atteindre la
singularit\'e,\\
c) si

\be
2\,\gamma > -\frac{{\Pi}_{2}^{2}}{{\alpha}^{2}{\Pi}_{3}^{2}}
\ee

\n les termes logarithmiques sont toujours n\'egligeables devant le terme en
$1/{\rho}^{2}$ qui est maintenant positif ainsi que tout le premier
membre de (\ref{eq geod1}) qui n'admet donc pas de solutions
($\forall\,\epsilon$). En particulier, pour $\gamma >0$ aucune
particule ne peut atteindre --m\^{e}me radialement-- la
singularit\'e.\\

\n 2) Si ${\Pi}_{3}=0$ ($\gamma$ quelconque), les particules ne
peuvent atteindre la singularit\'e que radialement (${\Pi}_{2}=0$); ce
sont les particules pour lesquelles $\epsilon =-1$, ${\Pi}_{4}^{2}<1$.\\

L'\'etude du comportement de (\ref{eq geod1}) pour $\rho \rightarrow
\infty$ peut \^{e}tre men\'ee de la m\^{e}me fa\c{c}on; une fois de
plus, le mouvement d\'epend de lavaleur de ${\Pi}_{3}$. Si
${\Pi}_{3}\neq 0$, le premier membre de (\ref{eq geod1}), \'egal
asymptotiquement \`a
$$
\left(\frac{d{\rho}}{dp}\right)^{2}+2{\Pi}_{3}^{2}(\ln\rho)^{2}
$$
est positif, et donc (\ref{eq geod1}) n'admet pas de
solutions avec $\rho \rightarrow \infty$. Ind\'ependamment de la
valeur de $\gamma$, aucune particule
ne peut donc dans ce cas aller \`a l'infini. Si ${\Pi}_{3}=0$,
l'\'equation (\ref{eq geod1}) n'admet des solutions avec $\rho
\rightarrow \infty$ que pour $\epsilon = -1$
et ${\Pi}_{4}^{2}<1$. Finalement, et de fa\c{c}on ind\'ependante de la
valeur de $\gamma$, seules les particules pour lesquelles les
5-g\'eod\'esiques sont de genre espace peuvent aller \`a l'infini (si
${\Pi}_{4}^{2}<1$); le fait que les 5-g\'eod\'esiques de genre temps
ou lumi\`ere ne s'\'etendent pas \`a l'infini est une pathologie due
\`a l'hypoth\`ese --non physique-- de la sym\'etrie axiale.

\section{Conclusion}
\setcounter{equation}{0}
Nous avons trouv\'e --dans l'hypoth\`ese d'existence de 4 vecteurs de
Killing ${\xi}_{a}\,^{A}={\delta}_{a}\,^{A}$-- des solutions cordes
cosmiques charg\'ees (sans champ magn\'etique) et neutres \`a la
th\'eorie de Kaluza-Klein (sans terme de Gauss-Bonnet; section 3.2)
qui ont \'et\'e identifi\'ees dans la sous-section 3.3.2, leurs seules
sources effectives \'etant purement distributionnelles.

Dans la section 3.3 nous avons obtenu les m\^{e}mes solutions en
incorporant le terme de Gauss-Bonnet dans l'action d'Einstein-Hilbert
($\gamma \neq 0$), pour lesquelles le tenseur de Lanczos est nul et
qui constituent donc un cas particulier des solutions de Kaluza-Klein
($\gamma = 0$) avec une source effective distributionnelle; la
consid\'eration du terme de Gauss-Bonnet \'etant donc, de ce point de
vue, formelle.

Les solutions expos\'ees dans la section 3.3 ne sont pas toutes
acceptables physiquement, en omettant les solutions de la forme $X=A$
--ce qui constitue une premi\`ere s\'election-- le d\'eveloppement men\'e
dans la sous-section 3.4.1 n'a concern\'e, au d\'ebut, que les solutions
de Kaluza-Klein de la forme $X=(2/\rho)A$, puis enfin, par une
deuxi\`eme s\'election, que les solutions cordes cosmiques. La
m\'ethode a permis ensuite de construire une m\'etrique admettant
quatre, et {\em seulement quatre}, vecteurs de Killing (voir Annexe
4), d\'ependant de 4 param\`etres et pour laquelle le tenseur de
Lanczos --une fois contravariant-- proportionnel \`a la source
effective \'etendue est non nul. Le fait que la densit\'e effective de
courant $T_{\mbox{\tiny eff}\,5}^{3}$ ait une contribution continue
nous a conduit \`a qualifier la solution (exacte) corde cosmique de
supraconductrice.

Le tenseur de Lanczos se manifeste dans les \'equations du mouvement
g\'eod\'esique,
en particulier dans l'int\'egrale premi\`ere (\ref{eq geod1}), par un
terme qui ne se distingue du terme centrifuge que par le signe de
$\gamma$; si ce dernier est n\'egatif les particules sont en g\'en\'eral
attir\'ees vers la singularit\'e ind\'ependamment du genre de leurs
 g\'eod\'esiques. Inversement, si $\gamma$ est
positif, les g\'eod\'esiques de genre temps ou lumi\`ere ne
s'\'etendent pas jusqu'\`a la singularit\'e. Ces m\^{e}mes
g\'eod\'esiques ne s'\'etendent pas non plus jusqu'\`a l'infini
($\forall\;\gamma$); ce comportement pathologique pourrait
dispara\^{\i}tre dans le cas plus physique de cordes ferm\'ees.

\newpage

\chapter*{Annexe 1: Formulaire math\'ematique}
\markboth{Annexe 1}{}
\addcontentsline{toc}{section}{{\bf Annexe 1:} Formulaire math\'ematique}

{\bf ${\bullet}$ M\'etrique {\boldmath $g_{AB}$}}\\
L'inverse de $g_{AB}$ est not\'ee $g^{AB}$:

\be
g^{AB}\,g_{BC} = {\delta}^{A}_{C}
\ee

\be
g_{AB} = g_{BA} \;\;\; ; \;\;\; g_{AB;C} = g^{AB}\,_{;C} = 0
\ee

\noindent o\`u (;) d\'enote la d\'erivation covariante (voir
(\ref{eq dc})).\\

{\bf ${\bullet}$ Connexions affines {\boldmath ${\Gamma}^{A}_{BC}$}}

\bn
\label{eq connexion}
{\Gamma}^{A}_{BC} & \equiv & \frac{1}{2}\,g^{AD}\,\left(g_{BD,C}
                                      + g_{CD,B} - g_{BC,D}\right)\\
{\Gamma}^{A}_{BC} & = & {\Gamma}^{A}_{CB}
\en

\noindent o\`u (,) d\'enote une d\'erivation partielle ordinaire.\\

{\bf ${\bullet}$ D\'erivation covariante}

\bn
T^{A}\,_{;B} & \equiv & T^{A}_{,B} + {\Gamma}^{A}_{BC}\,T^{C}\nonumber\\
\label{eq dc}
T_{A;B} & \equiv & T_{A,B} - {\Gamma}^{C}_{AB}\,T_{C} \\
T^{AB}_{C}\,_{;D} & \equiv & T^{AB}_{C}\,_{,D} +
{\Gamma}^{A}_{DE}\,T^{BE}_{C} + {\Gamma}^{B}_{DE}\,T^{AE}_{C}
- {\Gamma}^{E}_{CD}\,T^{AB}_{E} \nonumber
\en \\

{\bf ${\bullet}$ Tenseur de Riemann {\boldmath $R^{A}\,_{BCD}$}}

\be
\label{eq Riemann}
R^{A}\,_{BCD} \equiv {\Gamma}^{A}_{BD,C} - {\Gamma}^{A}_{BC,D} +
{\Gamma}^{E}_{BD}\,{\Gamma}^{A}_{CE} -
{\Gamma}^{E}_{BC}\,{\Gamma}^{A}_{DE}
\ee\\

\noindent Propri\'et\'es alg\'ebriques de $R_{ABCD} \equiv g_{AE}\,
R^{E}\,_{BCD}$:
$$
R_{ABCD} = R_{CDAB}
$$
\be
R_{ABCD} = -R_{BACD} = -R_{ABDC} = +R_{BADC}
\ee
$$
R_{ABCD} + R_{ADBC} + R_{ACDB} = 0
$$ \\

{\bf ${\bullet}$ Tenseur de Ricci {\boldmath $R_{AB}$}}

\be
\label{eq Ricci}
R_{AB} \equiv R^{C}\,_{ACB} = R_{BA}
\ee \\

{\bf ${\bullet}$ Scalaire de courbure {\boldmath $R$}}

\be
\label{eq scalaire}
R \equiv g^{AB}\,R_{AB}
\ee \\

{\bf ${\bullet}$ Identit\'es de Bianchi}

\be
\label{eq Bianchi}
R_{A}\,^{B}\,_{;B} - \frac{1}{2}\,R_{;A} \equiv 0\,.
\ee\\

\newpage
{\bf ${\bullet}$ Formule d'interpolation de Lagrange pour la fonction
{\boldmath $f$}}

\be
\label{eq Lagrange}
f(A)=\sum_{i}f(p_{i})\prod_{j\neq i}\,\frac{A-p_{j}}{p_{i}-p_{j}}
\ee

\n o\`u les $p_{i}$ sont les valeurs propres distinctes de la matrice $A$.

\newpage

\chapter*{Annexe 2: Unicit\'e de la solution de Gauss-Bonnet
{\boldmath $X=A$}}
\markboth{Annexe 2}{}
\addcontentsline{toc}{section}{{\bf Annexe 2:} Unicit\'e de la solution de
  Gauss-Bonnet {\boldmath $X=A$}}
On se propose de montrer dans cette section l'unicit\'e de la solution
$X=A$ des \'equation de Gauss-Bonnet pour laquelle $a=b=0$. En ayant
montr\'e la non existence de solutions dans les cas ($a=0,\;b\neq0$),
($a\neq0,\;b=0$), il nous reste \`a traiter le cas g\'en\'erique
$a\neq0,\;b\neq0$. Dans ce cas l'\'equation (\ref{eq matr3}) devient

\be
\label{eq annulateur}
\psi(A)=0
\ee

\n avec

\be
\label{eq minimal3}
\psi(p) \equiv p^{3}-\frac{b}{a}\,p^{2}+\left(a^{2}-b+\frac{2}{\gamma}\right)p
+\frac{2(a^{2}+b)}{a\gamma}
\ee

\n et --on pr\'ef\`ere recopier les \'equations (\ref{eq cab}),
(\ref{eq dab})--

\bn
\label{eq cab1}
& &\gamma c=\gamma(-a^{3}+ab+\frac{b^{2}}{a})-10a-8\,\frac{b}{a}  \\
\label{eq dab1}
& &24\gamma d=\gamma(-a^{4}-a^{2}b+2b^{2})-8a^{2}+8b\,.
\en

$\psi(p)$ \'etant un polyn\^{o}me annulateur de $A$, la m\'ethode de
r\'esolution consiste \`a chercher, pour quelles valeurs des
param\`etres $a, b, c, d$, il est le polyn\^{o}me minimal de $A$ avec
les conditions (\ref{eq cab1}), (\ref{eq dab1}); sinon \`a en
recommencer avec des polyn\^{o}mes annulateurs d'ordre inf\'erieur et
ainsi de suite. $\psi(p)$ \'etant d'ordre 3, la d\'emonstration se fait
donc en 3 \'etapes.\\

\n \underline{\'Etape 1}

Supposons que $\psi(p)$ est le polyn\^{o}me minimal de $A$. Et soit
$\Delta(p)$ le polyn\^{o}me caract\'eristique de $A$

\be
\label{eq poca}
\Delta(p)=p^{4}-ap^{3}+\frac{1}{2}\,(a^{2}-b)p^{2}+\left(-\frac{1}{3}\,
c+\frac{1}{2}\,ab-\frac{1}{6}\,a^{3}\right)p+d \,.
\ee

\n Le reste de division de $\Delta(p)$ par $\psi(p)$ doit \^{e}tre
identiquement nul \cite{L} \'etant donn\'e que $\Delta(p)$ est aussi
un polyn\^{o}me annulateur de $A$ d'ordre sup\'erieur. Le reste de la
division analytique est un polyn\^{o}me d'ordre 2, en annulant
identiquement les coefficients de $p^{2},\;p^{1},\;p^{0}$ on obtient
respectivement

\bn
\label{eq rel1}
& &\gamma(-a^{4}-a^{2}b+2b^{2})-4a^{2}=0\\
\label{eq rel2}
& &\gamma(-7a^{4}+11a^{2}b-4b^{2})-20a^{2}+8b=0\\
\label{eq rel3}
& &d+\frac{2(a^{4}-b^{2})}{a^{2}\gamma}=0
\en

\n o\`u l'\'equation (\ref{eq cab1}) a \'et\'e utilis\'ee dans
(\ref{eq rel2}). En comparant (\ref{eq rel1}) et (\ref{eq dab1}), on
obtient
$$
d=\frac{-a^{2}+2b}{6\gamma}
$$
qui, combin\'ee avec (\ref{eq rel3}), donne
$$
11a^{4}+2ba^{2}-12b^{2}=0\,,
$$
(toutes les constantes sont r\'eelles), d'o\`u
$$
a^{2}=\frac{{\epsilon}_{b}\sqrt{133}-1}{11}\,b
$$
(${\epsilon}_{b}$ est le signe de $b$). Cette relation de
proportionnalit\'e entre $a^{2}$ et $b$ ne satisfait
pas l'\'equation obtenue en \'eliminant $\gamma$ entre (\ref{eq
  rel1}), (\ref{eq rel2}), qui s'\'ecrit
$$
\left(\frac{a^{2}}{b}\right)^{3}-7\left(\frac{a^{2}}{b}\right)^{2}+\frac{11}{2}\,\frac{a^{2}}{b}-2=0\,.
$$
Il est donc impossible que $\psi(p)$ soit le polyn\^{o}me minimal de
$A$ avec les conditions (\ref{eq cab1}), (\ref{eq dab1}).\\

\n \underline{\'Etape 2}

Cherchons s'il existe des polyn\^{o}mes annulateurs de $A$ d'ordre 2
$$
Q(p) \equiv p^{2}-sp+q\,.
$$
Supposons que $Q(p)$ soit le polyn\^{o}me minimal de $A$. D'o\`u en
prenant les traces de $Q(A)$ et de $AQ(A)$

\bn
\label{eq rel4}
& &b=as-4q\\
\label{eq rel5}
& &c=a(s^{2}-q)-4qs\,.
\en

\n Le reste de division de $\psi(p)$ --un annulateur de $A$-- par
$Q(p)$ est un polyn\^{o}me \mbox{d'ordre 1} qui doit \^{e}tre identiquement
nul, d'o\`u les relations suivantes annulant ses coefficients

\bn
\label{eq rel6}
& &0=a^{3}-a^{2}s+3aq+4qs+\frac{2a}{\gamma}\\
\label{eq rel7}
& &0=a^{2}+as-4q-2\gamma q^{2}
\en

\n o\`u (\ref{eq rel4}) a \'et\'e utilis\'ee. La formule (\ref{eq
  cab1}) devient apr\`es substitution de (\ref{eq rel4})

\be
\label{eq rel8}
c=-a^{3}+a^{2}s+as^{2}-4aq+\frac{16q^{2}}{a}-8qs-\frac{10a}{\gamma}-
\frac{8s}{\gamma}+\frac{32q}{\gamma a}\,.
\ee

\n On peut montrer que (\ref{eq rel5}) et (\ref{eq rel8}) sont
compatibles avec (\ref{eq rel6}), (\ref{eq rel7}). En effet, la
combinaison (\ref{eq rel5})$-(8/\gamma a)$(\ref{eq rel7}) donne en
sommant membre \`a membre

\be
\label{eq rel9}
c=as^{2}-aq-4qs+\frac{8}{\gamma a}(-a^{2}-as+4q+2\gamma q^{2})\,.
\ee

\n En r\'ealisant ensuite la combinaison (\ref{eq rel9})$-$(\ref{eq
  rel6}) et en sommant membre \`a membre, on retrouve (\ref{eq rel8}).\\
Pour le reste, nous posons ($s\neq0$)

\be
\alpha \equiv \frac{a}{s}\quad,\quad\beta \equiv \frac{q}{s^{2}}
\quad,\quad\delta \equiv \gamma s^{2}\,.
\ee

\n Les relations (\ref{eq rel6}), (\ref{eq rel7}) s'\'ecrivent
maintenant \footnote{Si $s=0$, les
  relations (\ref{eq rel6}), (\ref{eq rel7}) fusionnent en
\bn
& &2{\mu}^{2}+7\mu +2=0\quad \mbox{avec}\quad \mu \equiv \gamma q\\
\label{eq rel01}
& &\gamma a^{2}=-3\mu-2
\en
d'o\`u ${\mu}^{2}=(41\pm7\sqrt{33})/8$. De $Q(A)=A^{2}+q=0$ on
d\'eduit

\bn
\label{eq rel02}
& &\gamma b=-4\mu \\
\label{eq rel03}
& &24{\gamma}^{2}d=\pm24{\mu}^{2}\,.
\en
En substituant (\ref{eq rel01}), (\ref{eq rel02}) dans (\ref{eq
  dab1}), on obtient
$$
24{\gamma}^{2}d=19{\mu}^{2}+20
$$
qui n'est pas compatible avec (\ref{eq rel03}).}

\bn
\label{eq rel10}
& &{\alpha}^{2}-\alpha+4\,\frac{\beta}{\alpha}+3\beta +\frac{2}{\delta}=0\\
\label{eq rel11}
& &{\alpha}^{2}+\alpha-4\beta-2\delta {\beta}^{2}=0\,.
\en

\n Le reste de division de $\Delta(p)$ par $Q(p)$ doit, lui aussi,
\^{e}tre identiquement nul, d'o\`u les deux relations suivantes

\bn
\label{eq rel12}
& &4\beta (\alpha-2)=-(\alpha-1)(\alpha-2)(\alpha-3)\\
\label{eq rel13}
& &24\beta \delta \left(\frac{{\alpha}^{2}}{2}-\frac{3\alpha}{2}+\beta
+1\right)=\frac{24\gamma d}{s^{2}}
\en

\n o\`u $\gamma d/s^{2}$ est donn\'e en fonction de $\alpha, \beta,
\delta$ par (\ref{eq dab1}). Si\\

\n \underline{$\alpha \neq 2$} on a alors

\be
\label{eq rel14}
4\beta =-(\alpha-1)(\alpha-3)\,.
\ee

\n En \'eliminant $\delta$ entre (\ref{eq rel10}), (\ref{eq rel11}),
on obtient

\be
\label{eq rel15}
4\alpha{\beta}^{2}=({\alpha}^{2}+\alpha-4\beta)(-{\alpha}^{3}+{\alpha}^{2}-
3\alpha \beta-4\beta)
\ee

\n qui se transforme \`a l'aide de (\ref{eq rel14}) en

\be
\label{eq rel16}
-\alpha(\alpha-1)(\alpha-3)^{2}=(2{\alpha}^{2}-3\alpha+3)
({\alpha}^{2}+5\alpha+12)
\ee

\n dont le second membre est toujours positif; le premier membre n'est
positif que pour $0<\alpha <1$. Les racines r\'eelles de (\ref{eq
  rel16}) doivent donc appartenir \`a cet intervalle dans lequel le
premier membre de (\ref{eq rel16}) est strictement inf\'erieur
\`a 9/4 et le second membre est strictement sup\'erieur \`a $345/32
\simeq 10,8$; ce qui est impossible. Si\\

\n \underline{$\alpha = 2$} la relation (\ref{eq rel15}) devient
$$
8{\beta}^{2}-11\beta-6=0
$$
d'o\`u
$$
\beta =\frac{11\pm \sqrt{313}}{16}
$$
et
$$
\delta =-2(5\beta +2)\,.
$$
Or, pour ces valeurs, la relation (\ref{eq rel13}) devient
$$
-1692,67 \neq -1350,08
$$
si on prend le signe $\mbox{}(+)$ dans $\beta$, ou
$$
0,76 \neq -3,30
$$
si on prend le signe $\mbox{}(-)$ dans $\beta$.

Finalement, on conclut que la matrice $A$ n'admet pas de polyn\^{o}me
minimal d'ordre 2. \\

\n \underline{\'Etape 3}

Si $A-x$ est un polyn\^{o}me annulateur --minimal-- de $A$ avec les
conditions (\ref{eq cab1}), (\ref{eq dab1}), on a alors de $A=x$
$$
\begin{array}{lcl} a=4x&,&b=4x^{2}\\
                   c=4x^{3}&,&d=x^{4}\,. \end{array}
$$
Les \'equations (\ref{eq cab1}), (\ref{eq dab1}) donnent
respectivement
$$
\gamma x^{2}=-1
$$
$$
\gamma x^{2}=-4
$$
qui ne sont pas compatibles.

On a ainsi montr\'e que les relations (\ref{eq annulateur}), (\ref{eq
  cab1}), (\ref{eq dab1}) ne peuvent pas \^{e}tre satisfaites
simultan\'ement pour $a\neq0$ et $b\neq0$.

\newpage

\chapter*{Annexe 3: Unicit\'e de la solution {\boldmath $X=(2/\rho)A$}
de Gauss-Bonnet}
\markboth{Annexe 3}{}
\label{sec-unicite2}
\addcontentsline{toc}{section}{{\bf Annexe 3:} Unicit\'e de la solution
{\boldmath $X=(2/\rho)A$} de Gauss-Bonnet}
On se propose de montrer que $X=(2/\rho)A$ avec $a=b=c=1,\;d=0$ et
$A^{3}-A^{2}=0$ est l'unique solution \`a sym\'etrie cylindrique sous
la forme $X(\rho)=\alpha(\rho)A$ ($\alpha(\rho)$ non constante) de
la th\'eorie de Kaluza-Klein avec
terme de Gauss-Bonnet. Partons de l'expression

\be
\label{eq exp1}
X(\rho)=\alpha(\rho)\,A
\ee

\n et sans perdre de g\'en\'eralit\'e, on prend

\be
\label{eq exp2}
a=1
\ee

\n ($a, b, c, d$ sont d\'efinis par (\ref{eq defA})). Les \'equations
(\ref{eq scal2}), (\ref{eq matr2}) s'\'ecrivent respectivement

\be
\label{eq scal4}
18{\alpha}_{,\,\rho}+3(1+2b){\alpha}^{2}+\frac{\gamma}{2}\left(
3(1-3b+2c){\alpha}^{2}{\alpha}_{,\,\rho}+\frac{1}{2}\,(1-3b+2c-24d)
{\alpha}^{4}\right)=0
\ee
\bn
\label{eq matr4}
L(A)&\equiv &\gamma(6{\alpha}_{,\,\rho}+{\alpha}^{2}){\alpha}^{2}A^{3}-
\gamma(6{\alpha}_{,\,\rho}+b{\alpha}^{2}){\alpha}^{2}A^{2}\nonumber \\
 & &\mbox{}+\left[[\gamma(1-b){\alpha}^{2}+1](4{\alpha}_{,\,\rho}+
{\alpha}^{2})+{\alpha}^{2}\right]A+2[4{\alpha}_{,\,\rho}+(b+1){\alpha}^{2}]=0
\nonumber \\
\en

\n o\`u la relation (\ref{eq A4d}) a \'et\'e utilis\'ee dans (\ref{eq
  scal4}). Le polyn\^{o}me $L(p)$ est un polyn\^{o}me annulateur de la
matrice \underline{constante} $A$, or, ses coefficients sont fonctions
de $\rho$, pour qu'il en soit ainsi, ces derniers doivent \^{e}tre
proportionnels \footnote{On verra qu'ils ne peuvent pas s'annuler
  identiquement.} entre eux. On distingue deux cas:\\

\n 1) \underline{Cas $6{\alpha}_{,\,\rho}+{\alpha}^{2}\neq0$}\\

\n a) $b\neq1$. On doit avoir

\bn
\label{eq avoir1}
& &\frac{6{\alpha}_{,\,\rho}+b{\alpha}^{2}}{6{\alpha}_{,\,\rho}+{\alpha}^{2}}
=c_{1}\\
\label{eq avoir2}
& &\frac{[\gamma(1-b){\alpha}^{2}+1](4{\alpha}_{,\,\rho}+{\alpha}^{2})+
{\alpha}^{2}}
{\gamma(6{\alpha}_{,\,\rho}+{\alpha}^{2}){\alpha}^{2}}=c_{2}\\
\label{eq avoir3}
& &\frac{4{\alpha}_{,\,\rho}+(b+1){\alpha}^{2}}
{(6{\alpha}_{,\,\rho}+{\alpha}^{2}){\alpha}^{2}}=c_{3}
\en

\n ($c_{1}, c_{2}, c_{3}$ sont des constantes r\'eelles avec, dans ce
cas, $c_{1}\neq1$; sinon $b=1$). D'o\`u de (\ref{eq avoir1})

\be
\label{eq ded1}
{\alpha}_{,\,\rho}=\frac{c_{1}-b}{6(1-c_{1})}\,{\alpha}^{2}\,.
\ee

\n En substituant (\ref{eq ded1}) dans (\ref{eq avoir2}) pour supprimer
${\alpha}_{,\,\rho}$, on obtient --en particulier-- en comparant les
puissances de ${\alpha}^{4}$

\be
2\,c_{1}=3-b
\ee

\n (l'autre relation d\'etermine $c_{2}$ en fonction de $c_{1}$). Or,
${\alpha}_{,\,\rho}$ \'etant proportionnel \`a ${\alpha}^{2}$, la
relation (\ref{eq avoir3}) ne peut \^{e}tre satisfaite que si
$c_{3}=0$, d'o\`u $b=1$. On d\'eduit alors que pour $b\neq1$, $L(p)$ ne
peut \^{e}tre annulateur de $A$.\\

\n b) $b=1$. Les \'equations (\ref{eq avoir1}), (\ref{eq avoir2})
restent valables et (\ref{eq avoir3}) est maintenant \'equivalente \`a
(\ref{eq avoir2}) ($c_{3}=\gamma c_{2}$). D'o\`u de (\ref{eq avoir1}),
$c_{1}=1$. La relation (\ref{eq avoir2}) conduit \`a

\be
\label{eq ded3}
{\alpha}_{,\,\rho}=\frac{(c_{2}\gamma{\alpha}^{2}-2){\alpha}^{2}}
{2(2-3c_{2}\gamma{\alpha}^{2})}\,.
\ee

\n L'\'equation (\ref{eq matr4}) se ram\`eme \`a

\be
\label{eq matr5}
A^{3}-A^{2}+c_{2}A+2c_{2}=0
\ee

\n d'o\`u en prenant la trace

\be
\label{eq ded4}
c=1-5c_{2}\,.
\ee

\n L'\'equation caract\'eristique de $A$ est maintenent (voir (\ref{eq
  carac}))

\be
A^{4}-A^{3}+\frac{5c_{2}}{3}A+d=0\,,
\ee

\n qui, combin\'ee avec (\ref{eq matr5})$\times A$, donne

\be
\label{eq ded5}
3c_{2}A^{2}+c_{2}A-3d=0
\ee

\n d'o\`u en prenant la trace

\be
\label{eq ded6}
d=\frac{c_{2}}{3}\,.
\ee

\n En substituant (\ref{eq ded3}), (\ref{eq ded6}) dans le premier
membre de (\ref{eq matr4}), on obtient au num\'erateur --apr\`es
r\'eduction au m\^{e}me d\'enominateur-- un polyn\^{o}me en $\alpha$ qui
doit \^{e}tre identiquement nul. Le coefficient du terme le plus
\'elev\'e --${\alpha}^{6}$-- \'etant proportionnel \`a
$(\gamma c_{2})^{2}$, d'o\`u $c_{2}=0$, et par cons\'equent:
$c=1,\;d=0,\;A^{3}-A^{2}=0$ et ${\alpha}_{,\,\rho}=-{\alpha}^{2}/2$,
qui s'int\`egre par $\alpha=2/\rho$. On retrouve ainsi la solution
discut\'ee dans la sous-section \ref{sec-XaA}.\\

\n 2) \underline{Cas $6{\alpha}_{,\,\rho}+{\alpha}^{2}=0$}\\

\n a) $b\neq1$. Dans ce cas $L(A)$ est un polyn\^{o}me d'ordre 2. De
${\alpha}_{,\,\rho}=-{\alpha}^{2}/6$ on d\'eduit la valeur de
$\gamma(6{\alpha}_{,\,\rho}+b{\alpha}^{2}){\alpha}^{2}=
\gamma(b-1){\alpha}^{4}$ coefficient de $A^{2}$ dans $L(A)$, alors que
celui de $A$ contient un terme en ${\alpha}^{2}$ en plus, et par
cons\'equent ne peuvent pas \^{e}tre proportionnels. Le polyn\^{o}me
$L(p)$ n'est donc pas annulateur de $A$ dans ce cas.\\

\n b) $b=1$. Dans ce cas $L(A)$ est un polyn\^{o}me d'ordre 1 qui
s'\'ecrit

\be
\label{eq ded8}
A+2=0
\ee

\n d'o\`u en prenant la trace: Tr$A=-8$, qui en d\'esaccord avec
l'hypoth\`ese Tr$A=a=1$.\\

On a ainsi montr\'e l'unicit\'e de la solution $\alpha=2/\rho$ avec
$a=b=c=1,\;d=0,$ $A^{3}-A^{2}=0$.

\newpage

\chapter*{Annexe 4: Vecteurs de Killing de la corde cosmique supraconductrice}
\markboth{Annexe 4}{}
\addcontentsline{toc}{section}{{\bf Annexe 4:} Vecteurs de Killing de
  la corde cosmique supraconductrice}
On se propose d'int\'egrer --sans faire trop de commentaires--
l'\'equation de Killing

\be
\label{eq K0}
K_{A\,;B}+K_{B\,;A}=0
\ee

\n pour la m\'etrique corde cosmique supraconductrice:
$$
ds^{\,2}=-d{\rho}^{\,2}-{\alpha}^{2}\,{\rho}^{2}\,d{\theta}^{\,2}-
dt^{\,2}-\frac{1}{{\alpha}_{3}}\,(dx^{5})^{2}-2\,dz\,dx^{5}-2\,
\frac{{\alpha}_{4}}{{\alpha}_{3}}\,dt\,dx^{5}
$$
\'equivalente \`a (\ref{eq met15}). L'\'equation (\ref{eq K0}) est
\'equivalente \`a

\be
\label{eq K00}
K_{A\,,B}+K_{B\,,A}=2{\Gamma}^{C}_{AB}K_{C}\,.
\ee

\n \underline{$A=B=1$}
\bn
& &K_{1\,,\rho}=0 \;\Rightarrow \nonumber \\
\label{eq K1}
& &K_{1}=F(\theta, z, t, x^{5})\,.
\en

\n \underline{$A=1,\;B=2$}
\bn
& &K_{1\,,\theta}+K_{2\,,\rho}=\frac{2}{\rho}\,K_{2}\;\Rightarrow \nonumber \\
& &K_{2\,,\rho}-\frac{2}{\rho}\,K_{2}=-F_{,\,\theta}\;\Rightarrow \nonumber \\
\label{eq K2}
& &K_{2}={\rho}^{2}G(\theta, z, t, x^{5})+\rho F_{,\,\theta}(\theta, z,
t, x^{5})\,.
\en

\n \underline{$A=1,\;B=3$}
\bn
& &K_{1\,,z}+K_{3\,,\rho}=0\;\Rightarrow \nonumber \\
\label{eq K3}
& &K_{3}=-\rho F_{,\,z}(\theta, z, t, x^{5})+H(\theta, z, t, x^{5})\,.
\en

\n \underline{$A=1,\;B=4$}
\bn
&
&K_{1\,,t}+K_{4\,,\rho}=\left(\frac{{\alpha}_{4}}{{\alpha}_{3}}\right)_{,\rho}
K_{3}\;\Rightarrow \nonumber \\
& &K_{4\,,\rho}=-\rho \left(\frac{{\alpha}_{4}}{{\alpha}_{3}}\right)_{,\rho}
F_{,\,z}+\left(\frac{{\alpha}_{4}}{{\alpha}_{3}}\right)_{,\rho}H-F_{,\,t}
\nonumber \\
& &{\qquad}=\left(\frac{{\alpha}_{4}}{{\alpha}_{3}}\right)_{,\rho}H-2F_{,\,z}
-F_{,\,t}\;\Rightarrow \nonumber \\
\label{eq K4}
& &K_{4}=\frac{{\alpha}_{4}}{{\alpha}_{3}}\,H(\theta, z, t, x^{5})-
\rho(2F_{,\,z}+F_{,\,t})+L(\theta, z, t, x^{5})\,.
\en

\n \underline{$A=1,\;B=5$}
\bn
& &K_{1\,,5}+K_{5\,,\rho}={\Lambda}^{ab}{\Lambda}_{5b\,,\rho}K_{a}\nonumber\\
& &{\qquad\quad\qquad\,}=\left[-\frac{{\alpha}_{4}}{{\alpha}_{3}}
\left(\frac{{\alpha}_{4}}{{\alpha}_{3}}\right)_{,\rho}+
\left(\frac{1}{{\alpha}_{3}}\right)_{,\rho}\right]K_{3}+
\left(\frac{{\alpha}_{4}}{{\alpha}_{3}}\right)_{,\rho}K_{4}\nonumber\\
& &{\qquad\quad\qquad\,}=\frac{4\gamma}{{\rho}^{3}}\,K_{3}+\frac{2}{\rho}\,
K_{4}\;\Rightarrow \nonumber \\
& &K_{5\,,\rho}=\left(\frac{4\gamma}{{\rho}^{3}}+4\,\frac{\ln\rho}{\rho}\right)
H+\frac{2}{\rho}\,L-(4F_{,\,z}+2F_{,\,t}+F_{,\,5})-
4\gamma\,\frac{1}{{\rho}^{2}}\,F_{,\,z}\;\Rightarrow \nonumber \\
\label{eq K5}
& &K_{5}=\frac{1}{{\alpha}_{3}}\,H(\theta, z, t, x^{5})+
\frac{{\alpha}_{4}}{{\alpha}_{3}}\,L(\theta, z, t, x^{5})-
\rho (4F_{,\,z}+2F_{,\,t}+F_{,\,5})+4\gamma\,\frac{1}{\rho}\,F_{,\,z}
\nonumber \\
& &{\qquad\;}+M(\theta, z, t, x^{5})\,.
\en

\n Nous avons ainsi d\'etermin\'e $K_{1},\,K_{2},\,K_{3},\,K_{4},\,K_{5}$
en fonction de $\rho$ et \`a des fonctions $F(\theta, z, t, x^{5}),\,
G(\theta, z, t, x^{5}),\,H(\theta, z, t, x^{5}),\,L(\theta, z, t,
x^{5}),\,M(\theta, z, t, x^{5})$ pr\`es qui restent \`a d\'eterminer:\\

\n \underline{$A=B=2$}
\bnn
& &K_{2\,,\theta}=-{\alpha}^{2}\rho K_{1}\;\Rightarrow \\
& &G_{,\,\theta}\,{\rho}^{2}+(F_{,\,\theta\theta}+{\alpha}^{2}F)\,\rho=0
\;\Rightarrow \\
\enn
\be
\label{1}
\left\{\begin{array}{l}
\mbox{a)}\;G_{,\,\theta}=0\;\Rightarrow\;G\equiv G(z, t, x^{5})\\
\mbox{b)}\;F_{,\,\theta\theta}+{\alpha}^{2}F=0\;\Rightarrow\;F=
a(z, t, x^{5})\cos\alpha\theta +b(z, t, x^{5})\sin\alpha\theta \,.
\end{array} \right.
\ee\\

\n \underline{$A=2,\;B=3$}
\bnn
& &K_{2\,,z}+K_{3\,,\theta}=0\;\Rightarrow\\
& &G_{,\,z}\,{\rho}^{2}+H_{,\,\theta}=0\;\Rightarrow
\enn
\be
\label{2}
\left\{\begin{array}{l}
\mbox{a)}\;G_{,\,z}=0\;\Rightarrow\;G\equiv G(t, x^{5})\\
\mbox{b)}\;H_{,\,\theta}=0\;\Rightarrow\;H\equiv H(z, t, x^{5})\,.
\end{array} \right.
\ee\\

\n \underline{$A=2,\;B=4$}
\bnn
& &K_{2\,,t}+K_{4\,,\theta}=0\;\Rightarrow\\
& &G_{,\,t}\,{\rho}^{2}-2F_{,\,\theta z}\rho+L_{,\,\theta}=0\;\Rightarrow
\enn
\be
\label{3}
\left\{\begin{array}{l}
\mbox{a)}\;G_{,\,t}=0\;\Rightarrow\;G\equiv G(x^{5})\\
\mbox{b)}\;F_{,\,\theta z}=0\;\Rightarrow\;
F=a(t, x^{5})\cos\alpha\theta +b(t, x^{5})\sin\alpha\theta \\
\mbox{c)}\;L_{,\,\theta}=0\;\Rightarrow\;L\equiv L(z, t, x^{5})\,.
\end{array} \right.
\ee\\

\n \underline{$A=2,\;B=5$}
\bnn
& &K_{2\,,5}+K_{5\,,\theta}=0\;\Rightarrow\\
& &G_{,\,5}\,{\rho}^{2}-2F_{,\,\theta t}\rho +M_{,\,\theta}=0\;\Rightarrow
\enn
\be
\label{4}
\left\{\begin{array}{l}
\mbox{a)}\;G_{,\,5}=0\;\Rightarrow\;G\equiv -{\alpha}^{2}\,C^{2}\\
\mbox{b)}\;F_{,\,\theta t}=0\;\Rightarrow\;
F=a(x^{5})\cos\alpha\theta +b(x^{5})\sin\alpha\theta \\
\mbox{b)}\;M_{,\,\theta}=0\;\Rightarrow\;M\equiv M(z, t, x^{5})\,.
\end{array} \right.
\ee\\

\n \underline{$A=B=3$}
\bn
& &K_{3\,,z}=0\;\Rightarrow \nonumber  \\
\label{5}
& &H_{,\,z}=0\;\Rightarrow\;H\equiv H(t, x^{5})\,.
\en

\n \underline{$A=3,\;B=4$}
\bn
& &K_{3\,,t}+K_{4\,,z}=0\;\Rightarrow \nonumber \\
\label{6}
& &H_{,\,t}+L_{,\,z}=0\,.
\en

\n \underline{$A=3,\;B=5$}
\bnn
& &K_{3\,,5}+K_{5\,,z}=0\;\Rightarrow\\
& &H_{,\,5}+M_{,\,z}+\frac{{\alpha}_{4}}{{\alpha}_{3}}\,L_{,\,z}=0\;\Rightarrow
\enn
\bn
\label{7}
& &L_{,\,z}=0\Rightarrow\;L\equiv L(t, x^{5})\\
\label{8}
& &H_{,\,5}+M_{,\,z}=0\,.
\en
En substituant (\ref{7}) dans (\ref{6}) on a
\be
\label{9}
H_{,\,t}=0\;\Rightarrow\;H\equiv H(x^{5})\,.
\ee

\n \underline{$A=B=4$}
\bn
& &K_{4\,,t}=0\;\Rightarrow \nonumber \\
\label{10}
& &L_{,\,t}=0\;\Rightarrow\;L\equiv L(x^{5})
\en

\n \underline{$A=4,\;B=5$}
\bnn
& &K_{4\,,5}+K_{5\,,t}=-\left(\frac{{\alpha}_{4}}{{\alpha}_{3}}\right)
_{,\rho}K_{1}\;\Rightarrow\\
& &\frac{{\alpha}_{4}}{{\alpha}_{3}}\,H_{,\,5}+
\left(\frac{{\alpha}_{4}}{{\alpha}_{3}}\right)_{,\rho}F+L_{,\,5}+M_{,\,t}=0
\;\Rightarrow
\enn
\bn
\label{11}
& &F\equiv 0 \\
\label{12}
& &H_{,\,5}=0\;\Rightarrow\;H\equiv -C^{5} \\
\label{13}
& &L_{,\,5}+M_{,\,t}=0\,.
\en
En substituant (\ref{12}) dans (\ref{8}) on a
\be
\label{14}
M_{,\,z}=0\;\Rightarrow\;M\equiv M(t, x^{5})\,.
\ee

\n \underline{$A=B=5$}
\bnn
& &K_{5,\,5}=\frac{1}{2}\,\left(\frac{1}{{\alpha}_{3}}\right)
_{,\rho}K_{1}\;\Rightarrow\\
& &\frac{{\alpha}_{4}}{{\alpha}_{3}}\,L_{,\,5}+M_{,\,5}=0\;\Rightarrow
\enn
\bn
\label{15}
& &L_{,\,5}=0\;\Rightarrow\;L\equiv C^{4} \\
\label{16}
& &M_{,\,5}=0\;\Rightarrow\;M\equiv M(t)\,.
\en
En substituant (\ref{15}) dans (\ref{13}) on a
\be
\label{17}
M_{,\,t}=0\;\Rightarrow\;M\equiv -C^{3}\,.
\ee

\n Les fonctions $G,\,H,\,L,\,M$ sont donc des constantes r\'eelles
arbitraires et $F\equiv 0$. Finalement, on a
\bn
& &K_{1}\equiv 0\\
& &K_{2}=-{\alpha}^{2}\,C^{2}\,{\rho}^{2}\\
& &K_{3}=-C^{5}\\
& &K_{4}=-C^{5}\,\frac{{\alpha}_{4}}{{\alpha}_{3}}(\rho)+C^{4}\\
& &K_{5}=-C^{5}\,\frac{1}{{\alpha}_{3}}(\rho)+C^{4}\,
\frac{{\alpha}_{4}}{{\alpha}_{3}}(\rho)-C^{3}\,.
\en
et
\bn
& &K^{1}\equiv 0\\
& &K^{2}=C^{2}\\
& &K^{3}=C^{3}\\
& &K^{4}=C^{4}\\
& &K^{5}=C^{5}\,.
\en

On conclut ainsi que tout vecteur de Killing $K^{A}$, solution de
(\ref{eq K0}), est une combinaison lin\'eaire des 4: ${\xi}_{a}\,^{A}=
{\delta}_{a}\,^{A}$ avec des coefficients constants
($K^{A}=C^{a}{\xi}_{a}\,^{A}$).

\newpage
\chapter*{}
\markboth{}{}
\addcontentsline{toc}{section}{\bf Publications}
\vskip 6 cm
\centerline{\Huge \bf Publications}


\begin{thebibliography}{99}
\addcontentsline{toc}{chapter}{Bibliographie}
\bibitem{Kaluza} T. Kaluza, Sitzungsber. Preuss. Akad. Wiss. Phys. Math.
{\bf K1} (1921) 996.
\bibitem{Klein} O. Klein, Z. F. Physik {\bf 37} (1926) 895;\\
                O. Klein, Nature {\bf 118} (1926) 516.
\bibitem{Weinberg} S. Weinberg, ``Gravitation and Cosmology" (1972)
  Wiley, New York.
\bibitem{Jordan} P. Jordan, Ann. der. Phys. Lpz. {\bf 1} (1947) 219;\\
                 Y. Thiry, C. R. Acad. Sci. Paris {\bf 226} (1948) 216
                 et 1881.
\bibitem{KK} ``An introduction to Kaluza-Klein theories" ed. H. C. Lee
  (World Scientific\,, Singapore), (1984).
\bibitem{KS} R. Kerner, Ann. Inst. H. Poincar\'e {\bf 9} (1968) 143;\\
             A. Salam et J. Strahdee, Ann. Phys. (N. Y.) {\bf 141}
             (1982) 316.
\bibitem{Leutwyler} H. Leutwyler, Arch. Sci. {\bf 13} (1960) 549.
\bibitem{Dobiasch} P. Dobiasch et D. Maison, Gen. Rel. Grav. {\bf 14}
  (1982) 231.
\bibitem{Chodos} A. Chodos et S. Detweiler, Gen. Rel. Grav. {\bf 14}
  (1982) 979.
\bibitem{Sorkin} R. D. Sorkin, Phys. Rev. Lett. {\bf51} (1983) 87.
\bibitem{Gross} D. J. Gross et M. J. Perry, Nucl. Phys. B{\bf226}
  (1983) 29.
\bibitem{Clement131} G. Cl\'ement, Gen. Rel. Grav. {\bf 16} (1984) 131.
\bibitem{Xi} Z. M. Xi, Phys. Lett. B{\bf158} (1985) 215.
\bibitem{Clement137} G. Cl\'ement, Gen. Rel. Grav. {\bf 18} (1986) 137.
\bibitem{Belinsky} V. A. Belinsky et R. Ruffini, Phys. Lett. B{\bf89}
                   (1980) 195;\\
                   G. Gibbons et D. Wiltshire, Ann. Phys. (N. Y.)
                   {\bf167} (1986) 201;\\
                   G. Cl\'ement, Gen. Rel. Grav. {\bf 18} (1986) 861;\\
                   W. Bruckman, Phys. Rev. D{\bf36} (1987) 3674;\\
                   A. Levinson et A. Davidson,
                   Mod. Phys. Lett. A{\bf6} (1991) 2189.
\bibitem{Fer} J. A. Ferrari, Gen. Rel. Grav. {\bf22} (1990) 19.
\bibitem{Clement491} G. Cl\'ement, Gen. Rel. Grav. {\bf16} (1984) 491;\\
                     G. Cl\'ement, Ann. Physique coll. {\bf14} (1989) 55.
\bibitem{theseA} M. Azreg-A\"{\i}nou, ``Diffusion classique par un soliton
  wormhole de Kaluza-Klein", th\`ese de Magister, Universit\'e de
  Constantine (1988).
\bibitem{AAGC1} M. Azreg-A\"{\i}nou et G. Cl\'ement, Gen. Rel. Grav. {\bf22}
               (1990) 1119.
\bibitem{Karim} K. A\"{\i}t-Moussa et G. Cl\'ement, J. Math. Phys. {\bf32}
                (1991) 717; \\
                K. A\"{\i}t-Moussa,``Diffusion classique par un dyon
  de Kaluza-Klein", th\`ese de Magister, Universit\'e de Constantine (1989).
\bibitem{Belinskii} V. A. Belinskii et V. E. Zakharov, JETP
  (Sov. Phys.) {\bf48} (1978) 985.
\bibitem{Tom} A. Tomimatsu, Prog. Theor. Phys. {\bf74} (1985) 630.
\bibitem{AAGC2} M. Azreg-A\"{\i}nou et G. Cl\'ement, Gen. Rel. Grav. {\bf25}
               (1993) 881.
\bibitem{Vilenkin} A. Vilenkin, Phys. Rev. D{\bf23} (1981) 852.
\bibitem{Hiscock} W. A. Hiscock, Phys. Rev. D{\bf31} (1985) 3288.
\bibitem{Gott} J. Gott, Astrophys. J. {\bf288} (1985) 422.
\bibitem{Linet} B. Linet, Gen. Rel. Grav. {\bf17} (1985) 1109.
\bibitem{Witten} E. Witten, Nucl. Phys. B{\bf249} (1985) 557;\\
I. Moss et S. Poletti, Phys. Lett. B{\bf199} (1987) 34;\\
D. Harari et P. Sikivie, Phys. Rev. D{\bf37} (1988) 3438;\\
B. Linet, Class. Quant. Grav. {\bf6} (1989) 435;\\
T. M. Helliwel et D. A. Konkowski, Phys. Lett. A{\bf143} (1990) 438;\\
G. W. Gibbons, M. E. Ortiz et F. Ruiz Ruiz, Phys. Lett. B{\bf240} (1990) 50;\\
A. K. Raychaudhuri, Phys. Rev. D{\bf41} (1990) 3041.
\bibitem{Madore} J. Madore, Phys. Lett. A{\bf110} (1985) 289.
\bibitem{Zwiebach} B. Zwiebach, Phys. Lett. B{\bf156} (1985) 315.
\bibitem{Zumino} B. Zumino, Phys. Rep. {\bf137} (1986) 109.
\bibitem{WheelerB273} J. T. Wheeler, Nucl. Phys. B{\bf273} (1986) 732.
\bibitem{Wiltshire} D. L. Wiltshire, Phys. Lett. B{\bf169} (1986) 36;\\
H. Goenner, F. M\"{u}ller-Hoissen et R. Kerner, ``Spherically symmetric
solutions of a generalized Einstein-Maxwell theory", preprint
GOE-ITP-20/88 (1988).
\bibitem{Love} D. Lovelock, J. Math. Phys. {\bf 12} (1971) 498.
\bibitem{Leibowitz} E. Leibowitz et N. Rosen, Gen. Rel. Grav. {\bf4}
  (1973) 449.
\bibitem{Maison} D. Maison, Gen. Rel. Grav. {\bf10} (1979) 717.
\bibitem{ClementLett} G. Cl\'ement, Phys. Lett. A{\bf118} (1986) 11.
\bibitem{Clement3dim} L. Landau et E. Lifchitz, ``Th\'eorie des champs"
  (1989), 4$^{e}$ \'edition revue et compl\'et\'ee, \'Editions Mir Moscou;\\
                      G. Cl\'ement, Ann. of Phys. {\bf201} (1990) 241;\\
                      B. Linet, Gen. Rel. Grav. {\bf23} (1991) 15.
\bibitem{L} F. R. Gantmacher, ``Th\'eorie des matrices", tome {\bf1}
  (1966), Dunod, Paris.
\bibitem{Zouzou} G. Cl\'ement et I. Zouzou, Phys. Rev. D{\bf50} (1994)
  7271.
\bibitem{Kasner} E. Kasner, Amer. J. Math. {\bf43} (1921) 217.
\end{thebibliography}
\end{document}